\documentclass[3p,twocolumn]{elsarticle}
\usepackage{graphicx}
\usepackage{color}
\usepackage{amssymb,amsmath}
\usepackage{amsfonts}
\usepackage{stmaryrd}
\usepackage{bm}

\usepackage[colorlinks,citecolor=blue,urlcolor=blue,linkcolor=blue]{hyperref}

\journal{Physica D}

\begin{document}

\begin{frontmatter}

\title{Topological Dynamics of Volume-Preserving Maps Without an Equatorial Heteroclinic Curve}

\author[1]{Joshua G. Arenson}
\ead{jarenson@ucmerced.edu}
\author[1]{Kevin A. Mitchell}
\ead{kmitchell@ucmerced.edu}

\address[1]{School of Natural Sciences, University of California, Merced, California, 95343}

\begin{abstract}

Understanding the topological structure of phase space for dynamical systems in higher dimensions is critical for numerous applications, including the computation of chemical reaction rates and transport of objects in the solar system. Many topological techniques have been developed to study maps of two-dimensional (2D) phase spaces, but extending these techniques to higher dimensions is often a major challenge or even impossible. Previously, one such technique, homotopic lobe dynamics (HLD), was generalized to analyze the stable and unstable manifolds of hyperbolic fixed points for volume-preserving maps in three dimensions. This prior work assumed the existence of an equatorial heteroclinic intersection curve, which was the natural generalization of the 2D case. The present work extends the previous analysis to the case where no such equatorial curve exists, but where intersection curves, connecting fixed points may exist. In order to extend HLD to this case, we shift our perspective from the invariant manifolds of the fixed points to the invariant manifolds of the invariant circle formed by the fixed-point-to-fixed-point intersections. The output of the HLD technique is a symbolic description of the minimal underlying topology of the invariant manifolds. We demonstrate this approach through a series of examples.
 
\end{abstract}
%


\end{frontmatter}

\newpage

\section{Introduction}

The study of classical chemical reaction dynamics is at its core a question of transport in Hamiltonian phase space.  Initial studies of the phase space geometry of reaction dynamics were restricted to two active degrees of freedom~\cite{DeLeon81,Davis85,Davis86}. Already for two degrees of freedom, it was seen that chaos could play a critical role. The current frontier for understanding phase space structures governing reaction dynamics is systems with three or more degrees of freedom~\cite{Wiggins01,Uzer02,Waalkens04c,Gabern05,ChunBiu06,Waalkens07,Paskauskas08,Ezra09,Ciftci13,MacKay14,Naik19}. Such work is not solely relevant to reaction dynamics but to other transport problems in Hamiltonian systems as well, such as celestial dynamics. Much of the research on transport for three degree-of-freedom systems has focused on transition-state theory, based on normally hyperbolic invariant manifolds (NHIMs).  Less attention has been paid to the global structure of the stable and unstable manifolds attached to NHIMs.  These manifolds are co-dimension one and (in the best case) divide phase space into topologically distinct regions, unfortunately, the invariant manifolds need not define (finite-volume) resonance zones and lobes, and hence these manifolds need not partition phase space into finite domains~\cite{Beigie95, Toda95}. This was first realized by Wiggins~\cite{Wiggins90}, followed by an explicit chemical example by Gililan and Ezra~\cite{Gillilan91}. As an alternative approach, Jung, Montoya, and collaborators~\cite{Jung10,Drotos14,Gonzalez14,Drotos16,Montoya20b} have studied the topological structure of chaotic scattering functions for three-degree-of-freedom Hamiltonian systems. They have shown how symbolic dynamics can be extracted from the doubly-differential cross section and then related back to the fractal structure of the chaotic saddle itself.  

Three-degree-of-freedom Hamiltonian systems generate flows in a six-dimensional phase space. If one is fortunate, this flow can be reduced, via a good surface-of-section, to a symplectic map on a four-dimensional phase space. This paper considers volume-preserving maps of a three-dimensional (3D) phase space as an intermediate step to symplectic maps in four dimensions (4D).  As previous studies in 3D have shown, even these maps have a wealth of complex behavior, and many open questions about their dynamics remain~\cite{Lomeli98,Lomeli00,Dullin09,James10,James13}. We consider here the global structure of intersecting two-dimensional (2D) stable and unstable manifolds of hyperbolic fixed points in 3D.  Specifically, we use finite pieces of these manifolds to generate symbolic dynamics describing the forced subsequent evolution of the manifolds. We note that complications can occur in 3D that have no analogue in 2D namely, the invariant manifolds of fixed points may not specify well defined resonance zones and lobes. We illustrate two methods for circumventing such complications. 
 
While we are interested in 3D volume-preserving maps as a stepping stone to the study of higher dimensional phase spaces, 3D volume-preserving maps are an important area of research in their own right and exhibit a plethora of fascinating phenomena. Within the realm of 3D volume-preserving maps, one can study behavior as diverse as particle advection in incompressible fluid flows~\cite{Aref17,Meiss15}, mixing of granular media in a tumbler \cite{Christov14}, the motion of charged particles along magnetic field lines in a plasma~\cite{Bazzani98}, and circular swimmers in a 2D incompressible fluid~\cite{Khurana12, Berman20}. A deeper history of 2D and 3D chaotic transport can be found in reviews by Aref et al.~\cite{Aref17} and Meiss~\cite{Meiss15}. 

Our work on 3D volume-preserving maps is based on prior studies of 2D maps. These studies focused on the structure of one-dimensional invariant manifolds of hyperbolic fixed points and periodic orbits, and how these manifolds intersect one another~\cite{ Easton86, Rom-Kedar90, Rom-Kedar94, Easton98}. If these stable and unstable manifolds intersect, they force a complex series of subsequent intersections. One technique to study the complicated topology that arises is homotopic lobe dynamics (HLD)~\cite{Mitchell03a, Mitchell06, Mitchell09, Mitchell12a, Novick12, Byrd14, Sattari16, Sattari17}. The underlying goal of HLD is to reduce the complex networks of stable and unstable manifolds and their intersections to a set of symbolic equations that describe the minimal underlying topology. An alternative technique for understanding the underlying topology of invariant manifolds in 2D was developed by Collins~\cite{Collins99, Collins02a, Collins04, Collins05a, Collins05b}. Collins's approach is based on train tracks and the Bestvina-Handel algorithm~\cite{Bestvina95}. This approach was recently shown to be dual to HLD~\cite{Collins19}. The input to both techniques is finite-time information in the form of finite-length intervals of the stable and unstable manifolds and their intersections; the output is a set of symbolic equations that predicts the minimum forced evolution of the system arbitrarily far into the future. Said another way, the existence of finite-time topological structure forces future structure to exist in specific, predictable ways. The symbolic dynamics in 2D HLD describe the evolution of 1D curves. It also allows one to assign symbolic itineraries to trajectories, thereby classifying chaotic trajectories of 2D maps.

Based on the preceding studies of 2D maps, we have been motivated to study 3D volume-preserving maps by extending HLD to analyze the global structure of 2D stable and unstable manifolds~\cite{Maelfeyt17,Smith17}. The input to the 3D HLD technique is (finite-area) pieces of intersecting 2D stable and unstable manifolds attached to hyperbolic fixed points (or as we discuss in Sec.~\ref{Invariant manifolds attached to the invariant circle}, attached to an invariant circle), which we call a \textit{trellis}. The output of the technique is a set of graphical equations that encodes the minimum topological structure of the manifolds as they are mapped forwards. The unstable manifolds are broken up into submanifolds called \textit{bridges}. The mathematical underpinning of HLD is homotopy theory. We punch ring-shaped holes (\textit{obstruction rings}) in the 3D phase space adjacent to specific 1D intersection curves. These obstruction rings are carefully chosen to topologically force the dynamics of the unstable manifolds. Based on the obstruction rings, the bridges are grouped into homotopy classes (called \textit{bridge classes}). The bridge classes are the elements that make up the symbolic dynamics of the system. When the map is applied to each bridge class, it produces a set of concatenated bridge classes. Physically the resulting symbolic dynamics describes how a 2D sheet will be stretched by the map. 

The graphical equations mapping bridge classes forward can be reduced to a transition matrix, represented pictorially by a transition graph. The largest eigenvalue of the transition matrix is the topological entropy forced by the finite trellis, which is a lower bound to the topological entropy of the full tangle and of the map itself. This topological entropy describes the rate at which a 2D sheet is stretched by the system.

The prior work on 3D HLD~\cite{Maelfeyt17,Smith17} was restricted to a particular class of 3D maps that have a so-called \textit{equatorial} heteroclinic intersection curve; the union of the stable and unstable manifolds of two fixed points up to this intersection curve divides phase space into two distinct regions. This is entirely analogous to how a primary intersection point is used to define a \textit{resonance zone} for 2D maps. As noted above not all 2D stable and unstable manifolds of fixed points intersect in such a convenient way. The objective of the current paper is to explore cases where such an equatorial curve does not exist. One alternative is that the 2D invariant manifolds of the fixed points intersect along a curve whose endpoints coincide with the fixed points. We call this a \textit{pole-to-pole} heteroclinic intersection curve. In this case we are unable to define a bounded resonance zone using just the 2D stable and unstable manifolds of the fixed points. We will show how one may extend 3D HLD to analyze such systems.
 
The topology of the 2D stable and unstable manifolds can become very intricate for numerically defined models (see Ref.~\cite{Smith17}). Thus we have opted to explain our extension of 3D HLD using a series of ``toy" examples. These examples were chosen to illustrate the basic concept of 3D HLD and how a shift in perspective allows us to extend it to systems we could not previously work with. A motivated reader will be able to apply the same techniques to a wide class of trellises. 

Section~\ref{Invariant manifolds attached to fixed points} introduces a number of key definitions concerning the invariant manifolds attached to fixed points and their intersections. Section~\ref{Invariant manifolds attached to the invariant circle} discusses how a pole-to-pole intersection curve allows us to recast our definitions for invariant manifolds of an invariant circle. Section~\ref{Reversibility} briefly touches on time-reversibility of maps. Sections~\ref{Example 1}-\ref{Example 5} are a series of examples that demonstrate how we extend the 3D HLD technique. Section~\ref{Example 1} discusses a previously tractable 3D system where the invariant manifolds of the fixed points have an equatorial intersection curve. For the uninitiated reader this acts as a primer on the prior work on 3D HLD.  Section~\ref{Example 2} presents a simple case where an equatorial intersection curve does not exist. We show that we can extract rules for topological forcing in this example, but we cannot analyze the full trellis.  Section~\ref{Example 3} analyzes a basic trellis composed of invariant manifolds attached to the invariant circle constructed from pole-to-pole intersection curves. We are able to successfully apply HLD to this system. Section~\ref{Example 4} considers the trellis analyzed in Sec.~\ref{Example 2} and, extending it slightly, reanalyzes it as invariant manifolds of the invariant circle. With the slight modification, we successfully apply HLD to the full trellis and show that the graphical equations in Sec.~\ref{Example 4} reduce to the graphical equations in Sec.~\ref{Example 2}. Section~\ref{Example 5} is the culmination of our work. An equatorial intersection does not exist between the 2D manifolds of the fixed points, and we require two homoclinic curves to construct a resonance zone using the invariant manifolds of the invariant circle. The bridge dynamics are explicitly 3D in nature and include both 2D and 1D components. Concluding remarks are in Sec.~\ref{Conclusion}.

\section{Preliminaries}
\label{Preliminaries} 

\subsection{Invariant manifolds attached to fixed points}
\label{Invariant manifolds attached to fixed points}

\begin{figure}
\centering

\includegraphics[width = 1\columnwidth]{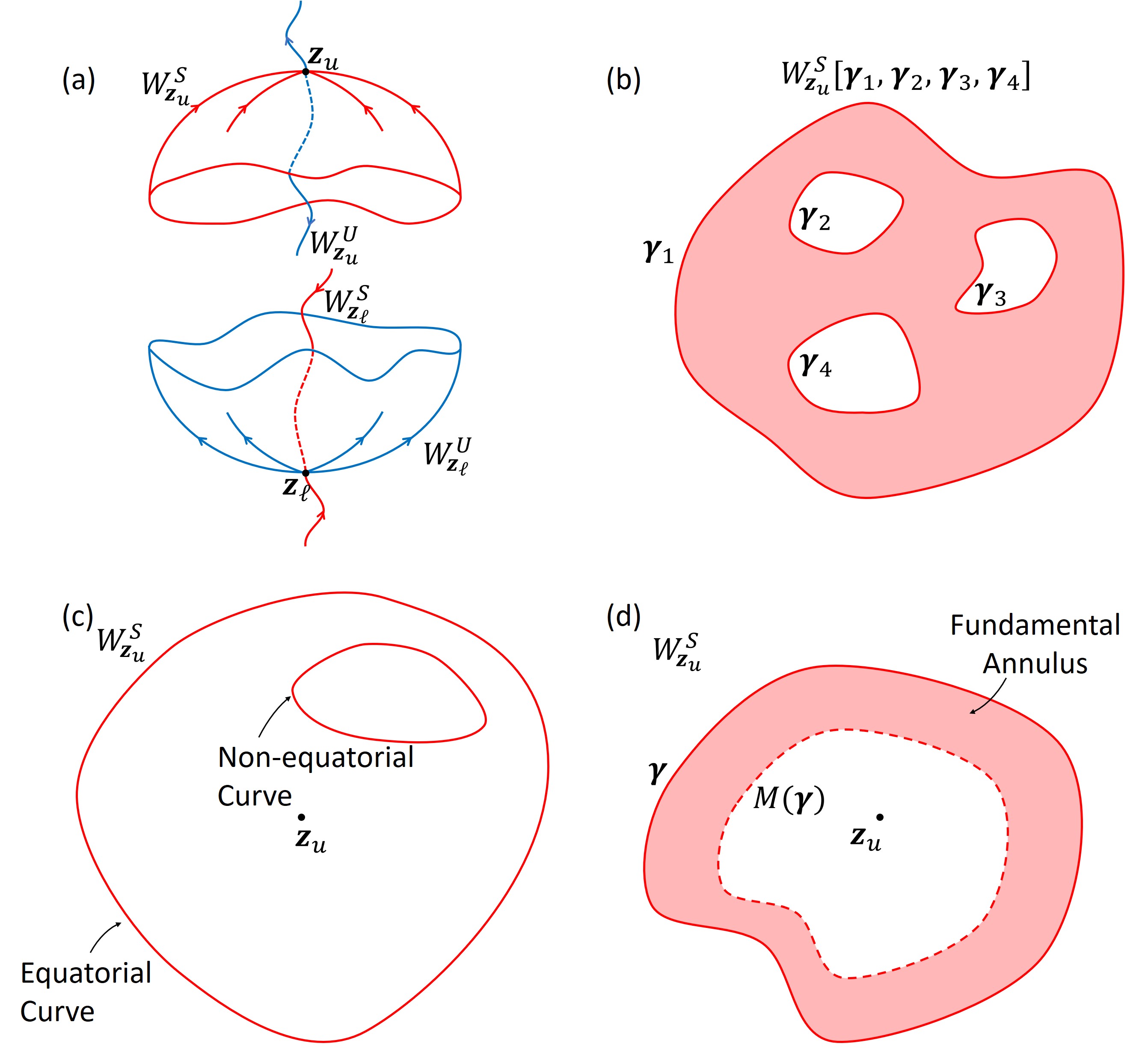}

\caption{(a) Two fixed points with their resultant invariant manifolds. The lower fixed point $\bm{z}_\ell$ has a 2D unstable manifold $W^U_{\bm{z}_\ell}$ and a 1D stable manifold $W^S_{\bm{z}_\ell}$. The upper fixed point $\bm{z}_u$ has a 2D stable manifold $W^S_{\bm{z}_u}$ and 1D unstable manifold $W^U_{\bm{z}_u}$. (b) A submanifold of a 2D stable manifold. It is uniquely defined by the boundary curves $\bm{\gamma}_1$, $\bm{\gamma}_2$, $\bm{\gamma}_3$, and $\bm{\gamma}_4$. (c) Two curves on the stable manifold. An equatorial curve encloses the fixed point. A non-equatorial curve does not. (d) A fundamental annulus defined by the curve $\bm{\gamma}$ (included in the annulus) and its iterate $M(\bm{\gamma})$ (not included in the annulus).}
\label{fig1}
\end{figure}

Suppose that we have a volume-preserving map $M$ in $\Bbb{R}^3$ and that $M$ has two hyperbolic fixed points, which we assume lie on the $z$-axis.  We assume the upper fixed point, denoted $\mathbf{z}_u$, has two stable directions and one unstable direction.  See Fig.~\ref{fig1}a.  The two stable directions point along the horizontal plane and the unstable direction points vertically. Similarly, we assume the lower fixed point, denoted $\mathbf{z}_\ell$, has two unstable directions, aligned horizontally, and one stable direction, aligned vertically.  The 2D stable manifold of $\mathbf{z}_u$ is denoted $W^S_{\mathbf{z}_u}$, and the 2D unstable manifold of $\mathbf{z}_\ell$ is denoted $W^U_{\mathbf{z}_\ell}$. Two-dimensional connected submanifolds of $W^S_{\mathbf{z}_u}$ can be specified by the set of $n$ curves ${\bm \gamma}_1, ..., {\bm \gamma}_n$ that form the boundary of the submanifold (Fig.~\ref{fig1}b).  We designate the (closed) submanifold by the notation $W^S_{\mathbf{z}_u} [ {\bm \gamma}_1, ..., {\bm \gamma}_n ]$.  Similar notation applies to submanifolds of $W^U_{\mathbf{z}_\ell}$.  The 1D unstable manifold of $\mathbf{z}_u$ and stable manifold of $\mathbf{z}_\ell$ are denoted $W^U_{\mathbf{z}_u}$ and $W^S_{\mathbf{z}_\ell}$, respectively.  (Closed) subintervals of these manifolds, with endpoints $\mathbf{a}$ and $\mathbf{b}$, are denoted by $W^U_{\mathbf{z}_u}[ \mathbf{a}, \mathbf{b} ]$ and similarly for $W^S_{\mathbf{z}_\ell}$.

We focus first on the 2D invariant manifolds $W^S_{\mathbf{z}_u}$ and $W^U_{\mathbf{z}_\ell}$.  Following Lomeli and Meiss~\cite{Lomeli00}, we define fundamental annuli and primary intersections of stable and unstable manifolds.  We first define an \textit{equatorial} curve of either invariant manifold as a non-self-intersecting curve that winds once around the fixed point, i.e. an equatorial curve bounds a topological disk (within the invariant manifold) that includes the fixed point in its interior.  See Fig.~\ref{fig1}c.  Next, we define a \textit{proper loop} ${\bm \gamma}$ as an equatorial curve that does not intersect its own iterate, i.e. $M({\bm \gamma}) \cap {\bm \gamma} = \varnothing$.  A \textit{fundamental domain}, or \textit{fundamental annulus}, is then the region of an invariant manifold between a given proper loop and its iterate. See Fig.~\ref{fig1}d. One edge of a fundamental annulus is open and the other closed, which we denote by $W^S_{\mathbf{z}_u}({\bm \gamma}, M({\bm \gamma}) ]$ if ${\bm \gamma}$ is omitted and $W^S_{\mathbf{z}_u}[{\bm \gamma}, M({\bm \gamma}) )$ if $M({\bm \gamma})$ is omitted, and similarly for $W^U_{\mathbf{z}_\ell}$.  Note that each trajectory within the invariant manifold passes through a given fundamental annulus exactly once.  The collection of all fundamental annuli in $W^U_{\mathbf{z}_\ell}$ is denoted $\mathcal{F}^U_{\mathbf{z}_\ell}$ and all fundamental annuli in $W^S_{\mathbf{z}_u}$ is denoted $\mathcal{F}^S_{\mathbf{z}_u}$.  Fundamental annuli are important because they can be used to generate the entire invariant manifold, and indeed this is often how invariant manifolds are computed in practice.  Heteroclinic intersections between the stable and unstable manifolds are often detected by fixing the stable fundamental annulus and then iterating the unstable annulus forward.  Each heteroclinic trajectory will then land exactly once within the fundamental stable annulus.

Lomeli and Meiss~\cite{Lomeli00} define the \textit{intersection index} $\kappa$ between two fundamental annuli $\mathcal{U} \in \mathcal{F}^U_{\mathbf{z}_\ell}$ and $\mathcal{S} \in \mathcal{F}^S_{\mathbf{z}_u}$ as the largest iterate of $\mathcal{S}$ that still intersects $\mathcal{U}$, i.e.
\begin{equation}
\kappa(\mathcal{U},\mathcal{S}) = \max \left\{ k \in \Bbb{Z} |
    M^k(\mathcal{S}) \cap \mathcal{U} \neq \varnothing \right\}.
\label{r1}
\end{equation}
We define an \textit{index-0 point} as a point that lies in the intersection between two fundamental annuli of intersection index 0,  i.e. $\mathbf{r}$ is an index-0 point if $\mathbf{r} \in \mathcal{S} \cap \mathcal{U}$ for some $\mathcal{U} \in \mathcal{F}^U_{\mathbf{z}_\ell}$ and $\mathcal{S} \in \mathcal{F}^S_{\mathbf{z}_u}$ satisfying $\kappa(\mathcal{U},\mathcal{S}) = 0$.  Similarly, an \textit{index-0 curve} is one that consists entirely of index-0 points. 

\begin{figure}
\centering
\includegraphics[width = 1\columnwidth]{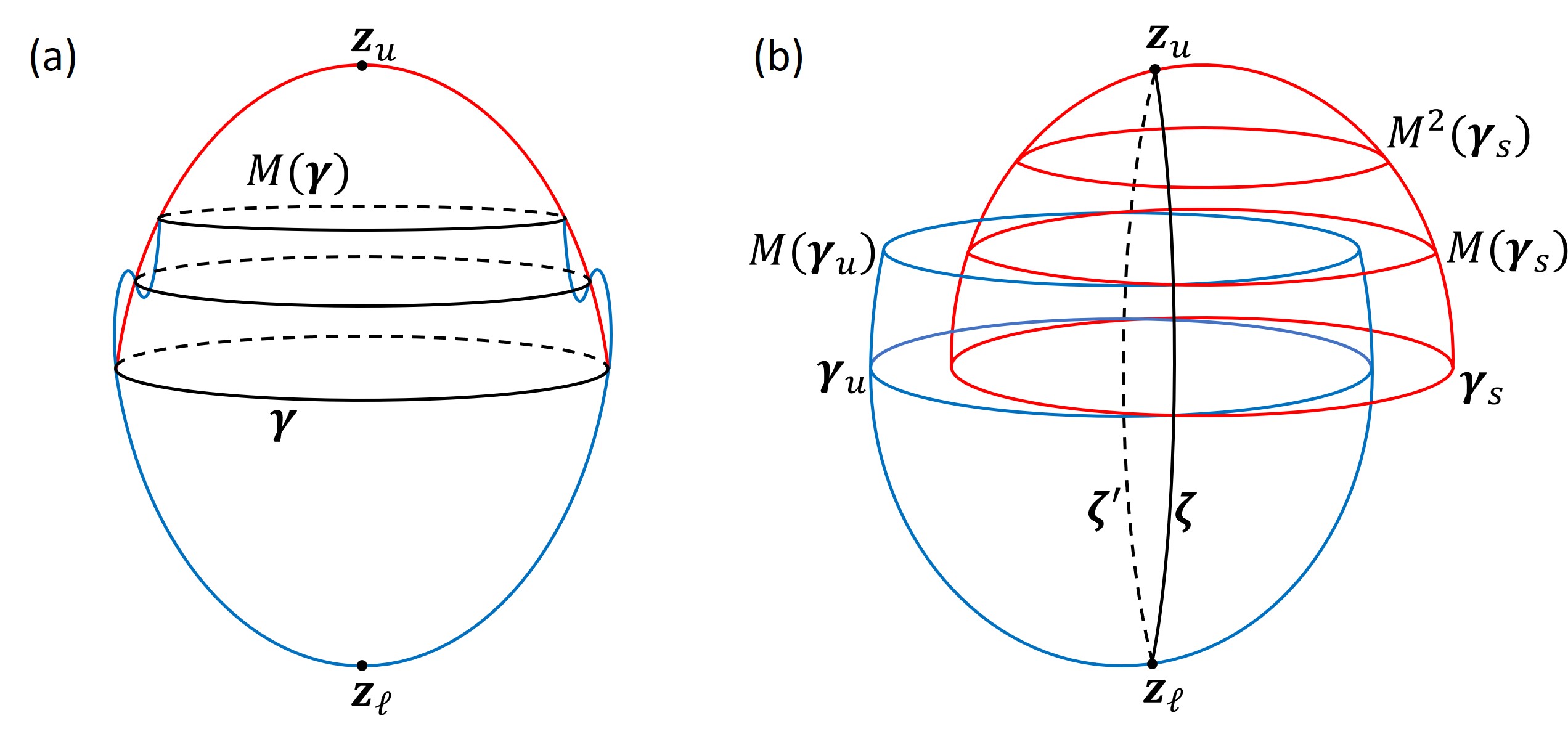}
\caption{(a) 2D stable (red) and unstable (blue) manifolds with a primary (equatorial) intersection curve $\bm{\gamma}$. (b) 2D stable and unstable manifolds without an equatorial intersection curve. $\bm{\gamma_s}$ and $\bm{\gamma_u}$ are equatorial but are not intersection curves. Two pole-to-pole intersection curves $\bm{\zeta}$ and $\bm{\zeta}^{'}$ exist connecting $\bm{z}_u$ and $\bm{z}_\ell$. }
\label{fig2}
\end{figure}

We define a \textit{primary intersection curve} $\bm{\gamma}$ as an equatorial curve on both the stable and unstable manifolds such that the stable disk $W^S_{\mathbf{z}_u}[{\bm \gamma} ]$ and the unstable disk $W^U_{\mathbf{z}_\ell}[{\bm \gamma} ]$ only intersect at their common boundary ${\bm \gamma}$.  See Fig.~\ref{fig2}a. It is clear that a primary intersection curve has index 0. Our definition of primary intersection curve reduces to the definition of a primary intersection point for 2D maps~\cite{Easton86}~\footnote{Our definition of a primary intersection curve differs from Lomeli and Meiss~\cite{Lomeli00}. Their primary intersection curve is what we call an index-0 curve.}. For such an intersection ${\bm \gamma}$, choose stable and unstable fundamental domains $\mathcal{S} = W^S[{\bm \gamma}, M({\bm \gamma}))$ and $\mathcal{U} = W^U[M^{-1}({\bm \gamma}), {\bm \gamma})$.  The set of all index-0 points is readily seen to be $M(\mathcal{U}) \cap \mathcal{S}$, plus all forward and backward iterates. In other words, we do not need to search over all possible pairs of fundamental annuli with index 0. We need only consider the pair $(M(\mathcal{U}),\mathcal{S})$, which has index 0. In Fig.~\ref{fig2}a, there are, up to iteration, two primary intersection curves ${\bm \gamma}$ and $\bm{\beta}$. Despite their convenience, primary intersection curves need not exist, and many important examples do not have them. For example, we have been unable to find primary intersection curves in the family of volume-preserving quadratic maps~\cite{Lomeli98,Dullin09,James10,James13}. Other kinds of index-0 curves may form loops that do not encircle the fixed point or curves that stretch from pole to pole, that is curves that converge upon $\mathbf{z}_u$ in one direction and upon $\mathbf{z}_\ell$ in the other.  Figure~\ref{fig2}b illustrates a pole-to-pole intersection curve of index 0.  Both non-equatorial index-0 loops and pole-to-pole index-0 curves exist in the family of 3D volume-preserving quadratic maps~\cite{James13}. 

We further refine our analysis of heteroclinic intersections by defining the \textit{index} $\sigma$ of a heteroclinic intersection $\mathbf{r}$.  This index is  the smallest intersection index of any two fundamental annuli that intersect at $\mathbf{r}$, i.e.
\begin{align}
\sigma(\mathbf{r}) = & \min \left\{ \kappa(\mathcal{U},\mathcal{S})  | \right. \nonumber \\
   & \left. \mathbf{r} \in  \mathcal{S} \cap \mathcal{U} \text{ for }
   \mathcal{U} \in \mathcal{F}^U_{\mathbf{z}_\ell}, \mathcal{S} \in \mathcal{F}^S_{\mathbf{z}_u} \right \}.
\label{r2}
\end{align}
 An equivalent characterization of heteroclinic intersection points is via the transition number.  For any two fundamental annuli $\mathcal{U} \in \mathcal{F}^U_{\mathbf{z}_\ell}$ and $\mathcal{S} \in \mathcal{F}^S_{\mathbf{z}_u}$, the \textit{transition   number} $\tau_{\mathcal{US}}$ of a heteroclinic trajectory $\mathbf{r}_i$ \textit{relative} to $(\mathcal{U},\mathcal{S})$ is defined as the number of iterates needed for the trajectory to map from $\mathcal{U}$ to $\mathcal{S}$, i.e.
\begin{align}
\tau_{\mathcal{U} \mathcal{S}}(\mathbf{r}_i) = &n, \text{ where } M^n(\mathbf{r}_j) \in \mathcal{S} \nonumber \\
 &\text{ when } \mathbf{r}_j \in \mathcal{U} \text{ for some }j. 
\end{align}
Typically one chooses the unstable fundamental annulus to ``precede'' the stable fundamental annulus so that the transition numbers are positive.  This is formalized by the concept of a properly ordered pair of fundamental annuli: $\mathcal{U} \in \mathcal{F}^U_{\mathbf{z}_\ell}$ and $\mathcal{S} \in \mathcal{F}^S_{\mathbf{z}_u}$ are said to be \textit{properly   ordered} if $\mathcal{U} \cap M^n(\mathcal{S}) = \varnothing$ for all $n \ge 0$.  We  then define \textit{the} transition number $\tau$ of a trajectory $\mathbf{r}_i$, independent of the choice of fundamental annuli, as
\begin{align}
\tau(\mathbf{r}_i)  & = \min \left \{ \tau_{\mathcal{U}
    \mathcal{S}}(\mathbf{r}_i) | \right. \nonumber \\ & \left. \mathcal{S} \in \mathcal{F}^S_{\mathbf{z}_u},\; \mathcal{U} \in \mathcal{F}^U_{\mathbf{z}_\ell} \text{ are properly
    ordered} \right \}. 
\end{align}
Note that $\tau$ is constant on a single connected intersection curve, whereas $\tau_{\mathcal{U} \mathcal{S}}$ need not be.  
%
%
It follows immediately from the above definitions that the transition number of a heteroclinic intersection equals its index plus one, i.e. $\tau = \sigma  + 1$.  

Primary intersection curves are again particularly useful for analyzing transition numbers. Assuming that such a curve ${\bm \gamma}$ exists, choose $\mathcal{S} = W^S[{\bm \gamma}, M({\bm \gamma}))$ and $\mathcal{U} = W^U[M^{-1}({\bm \gamma}), {\bm \gamma})$ as above. Then $\tau(\mathbf{r}) = \tau_{\mathcal{U} \mathcal{S}}(\mathbf{r})$ for all heteroclinic intersections $\mathbf{r}$ simultaneously.  There is no need to consider other fundamental annuli.  If no primary intersection exists, however, there may be no choice of $\mathcal{U}$ and $\mathcal{S}$ that simultaneously minimizes the relative transition number for all heteroclinic intersections.  This is true even when considering a single heteroclinic intersection curve; as noted previously, there may be no choice of $\mathcal{U}$ and $\mathcal{S}$ such that $\tau_{\mathcal{U} \mathcal{S}}$ is constant on the curve.

 For a primary intersection curve $\bm{\gamma}$, consider the two caps $W^S_{\mathbf{z}_u}[{\bm \gamma} ]$ and $W^U_{\mathbf{z}_\ell}[{\bm \gamma} ]$, as shown in Fig.~\ref{fig2}a. These caps bound a compact domain $R$, which we call the \textit{resonance zone}.  Specifying $\mathcal{S} = W^S[{\bm \gamma}, M({\bm \gamma}))$ and $\mathcal{U} = W^U[M^{-1}({\bm \gamma}), {\bm \gamma})$ as above, define the \textit{escape time} for every point in $\mathcal{U}$ as the number of iterates for it to map out of $R$.  (For simplicity, assume that no trajectory reenters $R$ after it has escaped.)  Then the set of points with a given escape time $n$ is divided into disconnected open \textit{escape domains}.  The boundary of these open domains are heteroclinic curves with transition number $n$.  \textit{Escape-time plots} (ETPs), i.e. 2D plots of the escape time, are an effective way to visualize the structure of heteroclinic intersection curves when primary intersections exist. See Fig.~\ref{Ex1ETP} in Sec.~\ref{Example 1}. In this paper, we use both \textit{forward} and \textit{backward} escape-time plots defined respectively on the unstable and stable fundamental annuli using the forward and backward maps.

\begin{figure*}
\centering
\includegraphics[scale=.195]{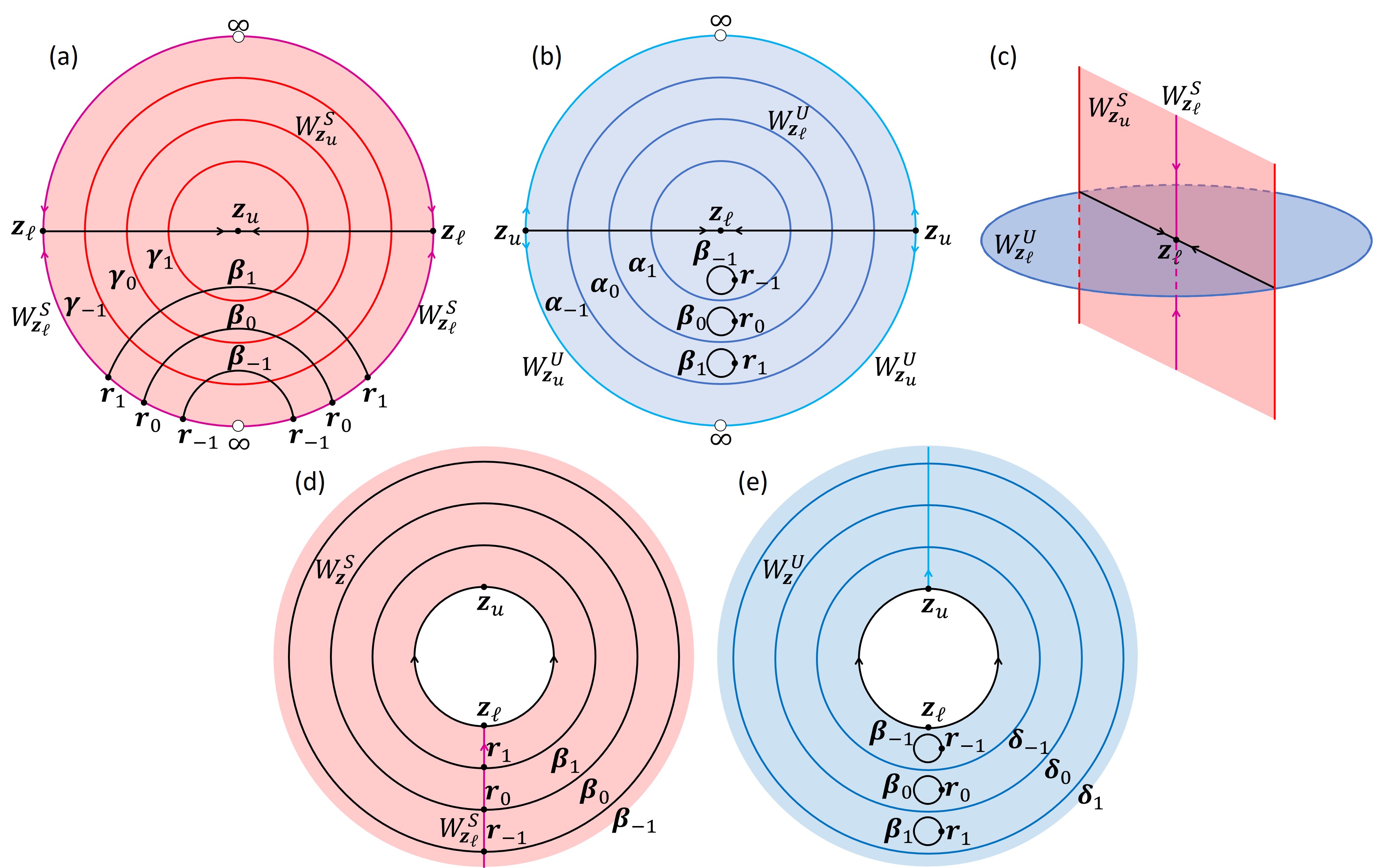}
\caption{(a) The union of the 2D stable manifold $W^S_{\bm{z}_u}$ (red disk) and the 1D stable manifold $W^S_{\bm{z}_\ell}$ (magenta). Two pole-to-pole intersection curves connect the fixed points $\bm{z}_u$ and $\bm{z}_\ell$. The red curves $\bm{\gamma}_n$ form a family of proper loops. The 1D intersection curves $\bm{\beta}_n$ (black) asymptotically approach the points  $\bm{r}_n$ on the 1D stable manifold. (b) The union of the 2D unstable manifold $W^U_{\bm{z}_\ell}$ (blue) and the 1D unstable manifold $W^U_{\bm{z}_u}$ (cyan). The intersection curves $\bm{\beta}_n$ from (a) together with the points $\bm{r}_n$ are closed circles. (c) A 3D view of the invariant manifolds in the vicinity of the fixed point $\bm{z}_\ell$. (d) One-half of the 2D stable manifold of the invariant circle formed by the pole-to-pole intersection curves from $\bm{z}_\ell$ to $\bm{z}_u$. The union of $\bm{\beta}_n$ and $\bm{r}_n$ forms a proper loop of the invariant circle. (e) One-half of the 2D unstable manifold of the invariant circle. The union of $\bm{\beta}_n$ and $\bm{r}_n$ is not an equatorial curve on the unstable manifold of the invariant circle $\bm{z}$.}
\label{fig3}
\end{figure*}

We now shift our focus to the 1D stable and unstable manifolds of $\mathbf{z}_\ell$ and $\mathbf{z}_u$, respectively, and their relationship to the 2D manifolds.  Many of the above definitions for the 2D manifolds have similar formulations for the 1D manifolds.  A \textit{fundamental domain} of either 1D manifold is simply a half-open interval between a point $\mathbf{r}$ and its iterate $M(\mathbf{r})$.  The collections of all such fundamental domains of the 1D manifolds are denoted $\mathcal{F}^U_{\mathbf{z}_u}$ and $\mathcal{F}^S_{\mathbf{z}_\ell}$. We then define the intersection index $\tau_{\mathcal{U} \mathcal{S}}$ between a 1D (stable/unstable) fundamental domain and a 2D (unstable/stable) fundamental domain using Eq.~(\ref{r1}) as before. The definition of an index-0 point between a 1D (stable/unstable) invariant manifold and 2D (unstable/stable) invariant manifold similarly generalizes: an index-0 point is a point that lies in the intersection of two fundamental domains of intersection index 0. More generally, the index $\sigma$ of a heteroclinic intersection between 1D invariant and 2D invariant manifolds is defined analogous to Eq.~(\ref{r2}).  Similarly the definition of the transition number $\tau_{\mathcal{U} \mathcal{S}}$ of a heteroclinic point $\mathbf{r}$ relative to fundamental domains $\mathcal{U}$ and $\mathcal{S}$ carries over analogously, as does the concept of properly ordered fundamental domains and the definition of the transition number $\tau$ of a heteroclinic point.

\subsection{Invariant manifolds attached to an invariant circle}
\label{Invariant manifolds attached to the invariant circle}

Suppose now that a pole-to-pole index-0 curve exists between $W^U_{\mathbf{z}_\ell}$ and $W^S_{\mathbf{z}_u}$, as in Fig.~\ref{fig2}b.  For simplicity, we assume that there are only two such pole-to-pole curves and that these curves are invariant, i.e. each curve maps to itself.  In general, there can be any even number of pole-to-pole intersection curves, and they may each be invariant or form periodic families of curves.  All of our results can easily be extended to this more general case.  

Now, let $\mathbf{r}_0$ be a heteroclinic intersection point between the 1D manifold $W^S_{\mathbf{z}_\ell}$ and the 2D manifold $W^U_{\mathbf{z}_\ell}$.  Given the presence of the two pole-to-pole curves, we assert that there must be a 1D intersection ${\bm \beta}_0$ between the 2D manifolds $W^S_{\mathbf{z}_u}$ and $W^U_{\mathbf{z}_\ell}$.  The set ${\bm \beta}_0$ has the topology of a curve with a single point removed at $\mathbf{r}_0$; that is, the union of ${\bm \beta}_0$ and $\mathbf{r}_0$ is a continuous curve.  Furthermore, the index of the set ${\bm \beta}_0$ equals the index of the point $\mathbf{r}_0$. These facts will be proved below.

Fig.~\ref{fig3}a shows a convenient way of visualizing heteroclinic
intersections when two pole-to-pole intersections exist.  The open
disk in Fig.~\ref{fig3}a represents the entirety of
$W^S_{\mathbf{z}_u}$.  The fixed point $\mathbf{z}_u$ is at the center
of the disk.  A proper loop ${\bm \gamma}_0$ encircles the fixed
point.  Its forward iterate ${\bm \gamma}_1$ is closer to
$\mathbf{z}_u$ and its backward iterate ${\bm \gamma}_{-1}$ is farther
away.  The regions between iterates of the proper loop are fundamental
stable domains.  In this representation, as ${\bm \gamma}_0$ is mapped
backward an arbitrary number of times it approaches, but never
reaches, the outer boundary of the disk.  At the left and rightmost
boundary points of the disk is the lower fixed point
$\mathbf{z}_\ell$.  Though represented twice in the figure, these two
points are geometrically the same and are thus identified with one
another.  The black horizontal line represents both pole-to-pole
intersection curves connecting $\mathbf{z}_u$ to $\mathbf{z}_\ell$.
The lower half of $W^S_{\mathbf{z}_u}$ in Fig.~\ref{fig3}a corresponds
to the left piece of $W^S_{\mathbf{z}_u}$ in Fig.~\ref{fig2}b,
which is in the ``interior'' region, whereas the upper half of
$W^S_{\mathbf{z}_u}$ in Fig.~\ref{fig3}a corresponds to the right piece of
$W^S_{\mathbf{z}_u}$ in Fig.~\ref{fig2}b, which remains
in the ``exterior'' region.

The 1D stable manifold of $\mathbf{z}_\ell$ is shown as the magenta
boundary of the disk in Fig~\ref{fig3}a.  It is divided into four separate curves, each beginning at $\mathbf{z}_\ell$ and terminating at the open circle at either the bottom or the top.  Just as the left copy of $\mathbf{z}_\ell$ is identified with the right copy of $\mathbf{z}_\ell$, the lower left branch of $W^S_{\mathbf{z}_\ell}$ is identified with the lower right branch of $W^S_{\mathbf{z}_\ell}$. Similarly, the upper left branch of of $W^S_{\mathbf{z}_\ell}$ is identified with the upper right branch.  These identifications mean that there is a single upper branch of $W^S_{\mathbf{z}_\ell}$, corresponding to the bottom half of $W^S_{\mathbf{z}_\ell}$ in Fig.~\ref{fig1}a, and a single lower branch, corresponding of the top half in Fig.~\ref{fig1}a.  Because the stable manifold $W^S_{\mathbf{z}_u}$ of the upper fixed point approaches the lower fixed point (via backward iteration) along the pole-to-pole intersection curve, the stable manifold $W^S_{\mathbf{z}_u}$ is eventually drawn away from $\mathbf{z}_\ell$ (via backward iteration) along the 1D curve $W^S_{\mathbf{z}_\ell}$, so that the 2D manifold $W^S_{\mathbf{z}_u}$ converges upon the 1D manifold $W^S_{\mathbf{z}_\ell}$. This geometry is shown in Fig.~\ref{fig3}c. For this reason, we have placed the red curve $W^S_{\mathbf{z}_\ell}$ along the boundary of the disk representing $W^S_{\mathbf{z}_u}$ in Fig.~\ref{fig3}a.  Finally, the open circles at the top and bottom of the disk are not points within $W^S_{\mathbf{z}_\ell}$, but can be thought of as points ``at infinity'' along the 1D stable manifold.

Recall that $W^S_{\mathbf{z}_\ell}$ transversely intersects the 2D unstable manifold $W^U_{\mathbf{z}_\ell}$ at the point $\mathbf{r}_0$. Then because $W^S_{\mathbf{z}_u}$ converges upon $W^S_{\mathbf{z}_\ell}$, $W^S_{\mathbf{z}_u}$ must also intersect $W^U_{\mathbf{z}_\ell}$ in the curve ${\bm \beta}_0$.  We have drawn ${\bm \beta}_0$ as a single arc in Fig.~\ref{fig3}a, though in fact ${\bm \beta}_0$ could intersect $W^S_{\mathbf{z}_\ell}$ multiple times.  Notice that ${\bm \beta}_0$ does not fit within a single fundamental domain, as defined by the curves ${\bm \gamma}_n$.  Indeed, there is no proper loop for which the resulting fundamental domain would include the entire curve ${\bm \beta}_0$, since ${\bm \beta}_0$ terminates at $W^S_{\mathbf{z}_\ell}$.

Fig.~\ref{fig3}b is a representation of $W^U_{\mathbf{z}_\ell}$ analogous to Fig.~\ref{fig3}a for $W^S_{\mathbf{z}_u}$.  Here the point $\mathbf{r}_0$ is not on the boundary, since it lies within the 2D manifold $W^U_{\mathbf{z}_\ell}$ where the 1D stable manifold $W^S_{\bm{z}_\ell}$ intersects it.  Thus, the curves ${\bm \beta}_n$ can lie within a single unstable fundamental domain, assuming the proper loops are chosen appropriately, as we have shown with the loops ${\bm \alpha}_n$ in Fig.~\ref{fig3}b.

Another convenient way of thinking about the invariant manifolds is to recognize that the two fixed points together with the two pole-to-pole intersection curves form an invariant circle, which we denote by $\mathbf{z}$ without a subscript.  Then the stable manifold $W^S_{\mathbf{z}}$ of the invariant circle is two-dimensional and equal to the union of $W^S_{\mathbf{z}_u}$ and $W^S_{\mathbf{z}_\ell}$.  The analogous statement is true for the unstable manifold $W^U_{\mathbf{z}}$.  The stable manifold $W^S_{\mathbf{z}}$ has two branches, corresponding to the upper and lower halves of the disk in Fig.~\ref{fig3}a.  Focusing on just the lower half-disk, we may wrap the horizontal black line into a circle, gluing the two points representing $\mathbf{z}_\ell$ together.  We similarly glue the two lines representing $W^S_{\mathbf{z}_\ell}$ together, forming the image in Fig.~\ref{fig3}d.  This branch of $W^S_{\mathbf{z}}$ begins on the interior black circle representing $\mathbf{z}$ and extends outward.  Note that each arc representing ${\bm \beta}_n$ in Fig.~\ref{fig3}a is now wrapped into a circle surrounding $\mathbf{z}$ in Fig.~\ref{fig3}d. We can similarly represent the lower branch of $W^U_{\mathbf{z}}$ in Fig.~\ref{fig3}b by the image in Fig.~\ref{fig3}e.

Note that all of the original definitions of equatorial curves, proper loops, fundamental domains, indices, transition numbers, primary intersection curves, etc. introduced above for the 2D manifolds $W^S_{\mathbf{z}_u}$ and $W^U_{\mathbf{z}_\ell}$ can now be directly applied to the branches of $W^S_{\mathbf{z}}$ and $W^U_{\mathbf{z}}$.  In Fig.~\ref{fig3}d, we then see that the curves ${\bm \beta}_n$, combined with their missing points $\mathbf{r}_n$, are homoclinic proper loops of $W^S_{\mathbf{z}}$.  This is an important realization for analyzing the topological structure of manifolds with pole-to-pole intersections.  It can be much easier and more natural to analyze these manifolds as invariant manifolds of the invariant circle $\mathbf{z}$ than as invariant manifolds of the two fixed points. This shall be explored in Sec.~\ref{Example 3} - Sec.~\ref{Example 5}.

\subsection{Reversibility}
\label{Reversibility}
A reversible map $M$ is defined as a map with a symmetry operator $S$ such that $M^{-1} = S \circ M \circ S$. Here $S$ must be idempotent, i.e. $S = S^{-1}$. A consequence of reversibility is that $W^S_{\bm{z}_u} = S(W^U_{\bm{z}_\ell})$. Assuming $S$ to be linear its eigenvalues must be either $+1$ or $-1$. In 3D there are only three possibilities: a single negative eigenvalue, two negative eigenvalues, or three negative eigenvalues. With appropriate rotations of phase space, we can express any $S$ as $S(x,y,z) = (x,y,-z)$, $S(x,y,z) = (-x,y,-z)$, or $S(x,y,z) = (-x,-y,-z)$.

Consider  $S(x,y,z) = (x,y-z)$. Under this operator every point on the $xy$-plane is invariant under $S$. As a consequence of $W^S_{\bm{z}_u} = S(W^U_{\bm{z}_\ell})$, any equatorial intersection of $W^U_{\bm{z}_\ell}$ with the $xy$-plane must be a primary intersection curve, as in Fig.~\ref{fig2}a. Thus this symmetry is a convenient way of forcing a primary intersection curve to exist.

\begin{figure*}
\centering
\includegraphics[width=1\linewidth]{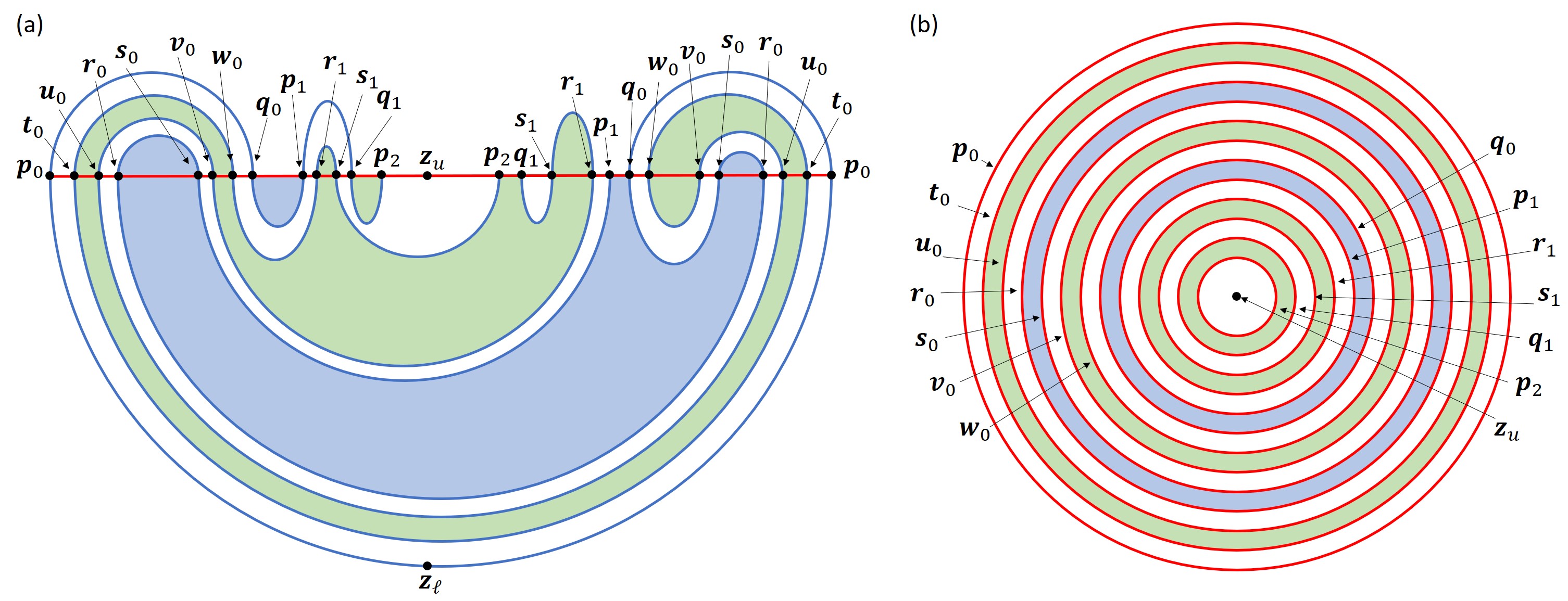}
\caption{(a) A cross-section of 2D stable and unstable manifolds growing from fixed points $\bm{z}_u$ and $\bm{z}_\ell$ respectively. The stable and unstable manifolds intersect at an primary intersection curve $\bm{p}_0$. Iterating the primary unstable cap forward twice results in a series of intersections with the stable manifold. (b) A top down view of the stable manifold showing the heteroclinic intersections between the stable and unstable manifolds.}
\label{Ex1Trel}
\end{figure*}

Now consider $S(x,y,z) = (-x,y,-z)$. In this case every point on the $y$-axis is invariant, and thus any intersection of the $y$-axis by $W^U_{\bm{z}_\ell}$ results in an intersection point with $W^S_{\bm{z}_u}$. These forced intersection points generically line on a heteroclinic intersection curve. However, this curve need not be equatorial. Thus this symmetry is convenient for exploring cases without primary intersection curves.  

Finally we consider $S(x,y,z) = (-x,-y,-z)$. In this case the only invariant point under $S$ is the origin. Systems with this symmetry operator do not generically have any forced intersection points. However, other advantages of reversibility still exist. 

Reversibility produces a number of advantages when computing manifolds and applying HLD. Applying the symmetry operator to the unstable manifold produces the stable manifold and vice versa. This is desirable when computing manifolds numerically as it cuts computation time in half, and computations for 2D (and higher dimensional) manifolds can be resource intensive. A second advantage is that the forward and backward ETPs are geometrically identical, requiring only a single computation. The examples in Sec.~\ref{Example 3} and Sec.~\ref{Example 5} use time-reversibility to simplify the analysis.

\section{Example 1} 
\label{Example 1}

\begin{figure}
\centering
  \includegraphics[width=1\linewidth]{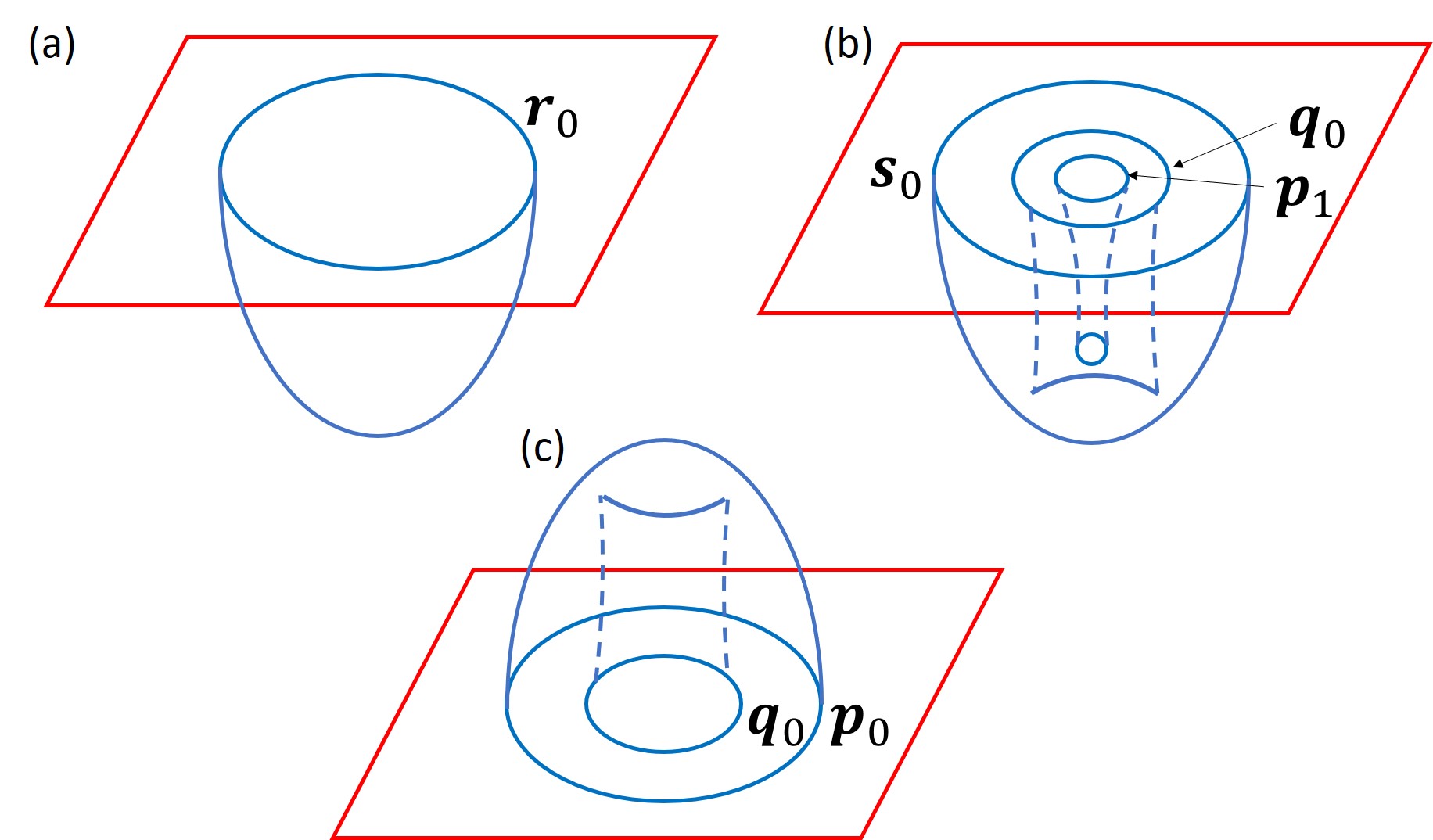}
\caption{(a) The bridge $W^U_{\bm{z}_\ell}[\bm{r}_0]$ which exists inside the resonance zone. With only one intersection curve, this bridge takes the form of a ``cap''. (b) The interior bridge $W^U_{\bm{z}_\ell}[\bm{s}_0,\bm{q}_0,\bm{p}_1]$. With three nested intersection curves, the bridge takes the form of a ``tridge''. (c) The exterior bridge $W^U_{\bm{z}_\ell}[\bm{p}_0,\bm{q}_0]$. This bridge has two intersection curves nested inside each other forming a ``bundt cake''. }
\label{Ex1BridgeExamples}
\end{figure}

We begin with an example of a system whose 2D stable and unstable manifolds of the fixed points $\bm{z}_u$ and $\bm{z}_\ell$ intersect at a primary intersection curve. The purpose of this example is to introduce the techniques used in Refs.~\cite{Maelfeyt17,Smith17}. Figure~\ref{Ex1Trel}a shows a cross section of the trellis, while Fig.~\ref{Ex1Trel}b shows a top down view of the trellis. The stable (red) and unstable (blue) manifolds intersect at the equatorial intersection $\bm{p}_0$. The trellis is made up of a series of iterates of the \textit{primary} unstable cap $W^U_{\bm{z}_\ell}[\bm{p}_0]$. Each iterate of the primary unstable cap produces a series of concatenated ``bridges''. A \textit{bridge} is defined as a 2D submanifold of the unstable manifold all of whose boundary circles lie within the stable cap $W^S_{\bm{z}_u}[\bm{p}_0]$ and which does not otherwise intersect the stable cap, i.e. bridges are the pieces one obtains when the unstable manifold is cut by the stable cap. Figure~\ref{Ex1BridgeExamples} shows three examples of bridges: a ``cap'' with a single boundary circle, a ``bundt cake'' with two nested boundary circles, and a ``tridge'' with three nested boundary circles. The first iterate of the unstable cap $W^U_{\bm{z}_\ell}[\bm{p}_0]$ produces three bridges interior to the resonance zone: the original cap $W^U_{\bm{z}_\ell}[\bm{p}_0]$, an interior cap $W^U_{\bm{z}_\ell}[\bm{r}_0]$, and an interior tridge $W^U_{\bm{z}_\ell}[\bm{q}_0,\bm{s}_0,\bm{p}_1]$. The first iterate of $W^U_{\bm{z}_\ell}[\bm{p}_0]$ also produces two bundt cake bridges exterior to the resonance zone: $W^U_{\bm{z}_\ell}[\bm{p}_0,\bm{q}_0]$ and $W^U_{\bm{z}_\ell}[\bm{r}_0,\bm{s}_0]$. The cap $W^U_{\bm{z}_\ell}[\bm{r}_0]$ can now be iterated forward producing an additional two interior caps, $W^U_{\bm{z}_\ell}[\bm{t}_0]$ and $W^U_{\bm{z}_\ell}[\bm{u}_0]$, an interior tridge $W^U_{\bm{z}_\ell}[\bm{v}_0,\bm{w}_0,\bm{r}_1]$, and produces two exterior bundt cakes, $W^U_{\bm{z}_\ell}[\bm{t}_0,\bm{w}_0]$ and $W^U_{\bm{z}_\ell}[\bm{u}_0,\bm{v}_0]$. The forward iterate of $W^U_{\bm{z}_\ell}[\bm{p}_0,\bm{q}_0]$ produces $W^U_{\bm{z}_\ell}[\bm{p}_1,\bm{q}_1]$, $W^U_{\bm{z}_\ell}[\bm{r}_0,\bm{s}_0]$ produces $W^U_{\bm{z}_\ell}[\bm{r}_1,\bm{s}_1]$, and $W^U_{\bm{z}_\ell}[\bm{q}_0,\bm{s}_0,\bm{p}_1]$ produces $W^U_{\bm{z}_\ell}[\bm{q}_1,\bm{s}_1,\bm{p}_2]$.

\begin{figure}
\centering
\includegraphics[width=1\linewidth]{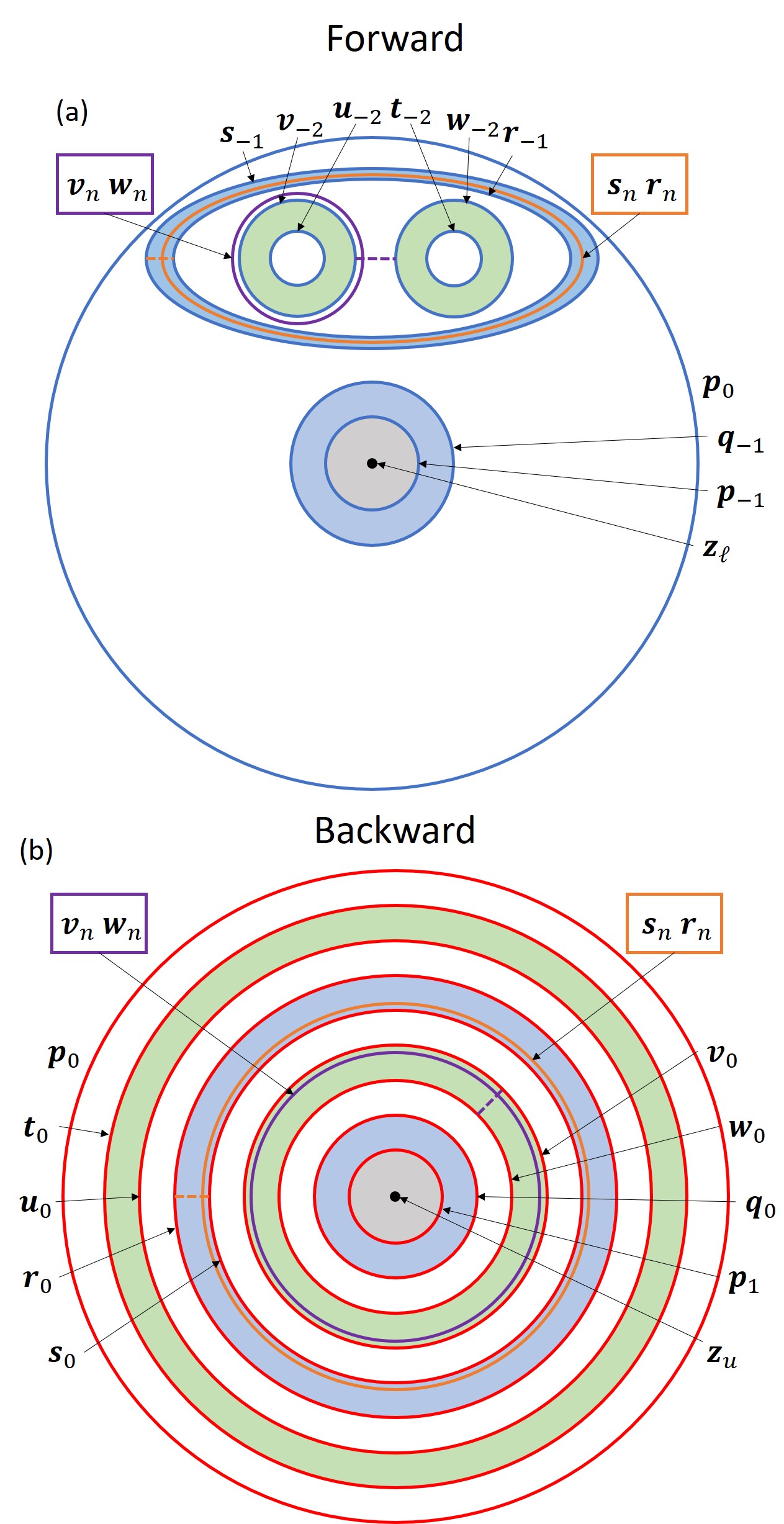}
\caption{(a) Forward and (b) backward escape-time plots. The color (blue or green) indicates the number of iterates (either forward or backward) needed to escape the resonance zone. The boundaries of escape domains are heteroclinic intersection curves, blue in (a) and red in (b). Dashed lines connect pseudoneighbor pairs; note that these lines do not intersection any other heteroclinic curves. The solid lines intersecting the dashed lines are the obstruction rings, which are labeled in boxes by their corresponding pseudoneighbor pair.}
\label{Ex1ETP}
\end{figure}

\begin{figure}
	\centering
  \includegraphics[width=1\linewidth]{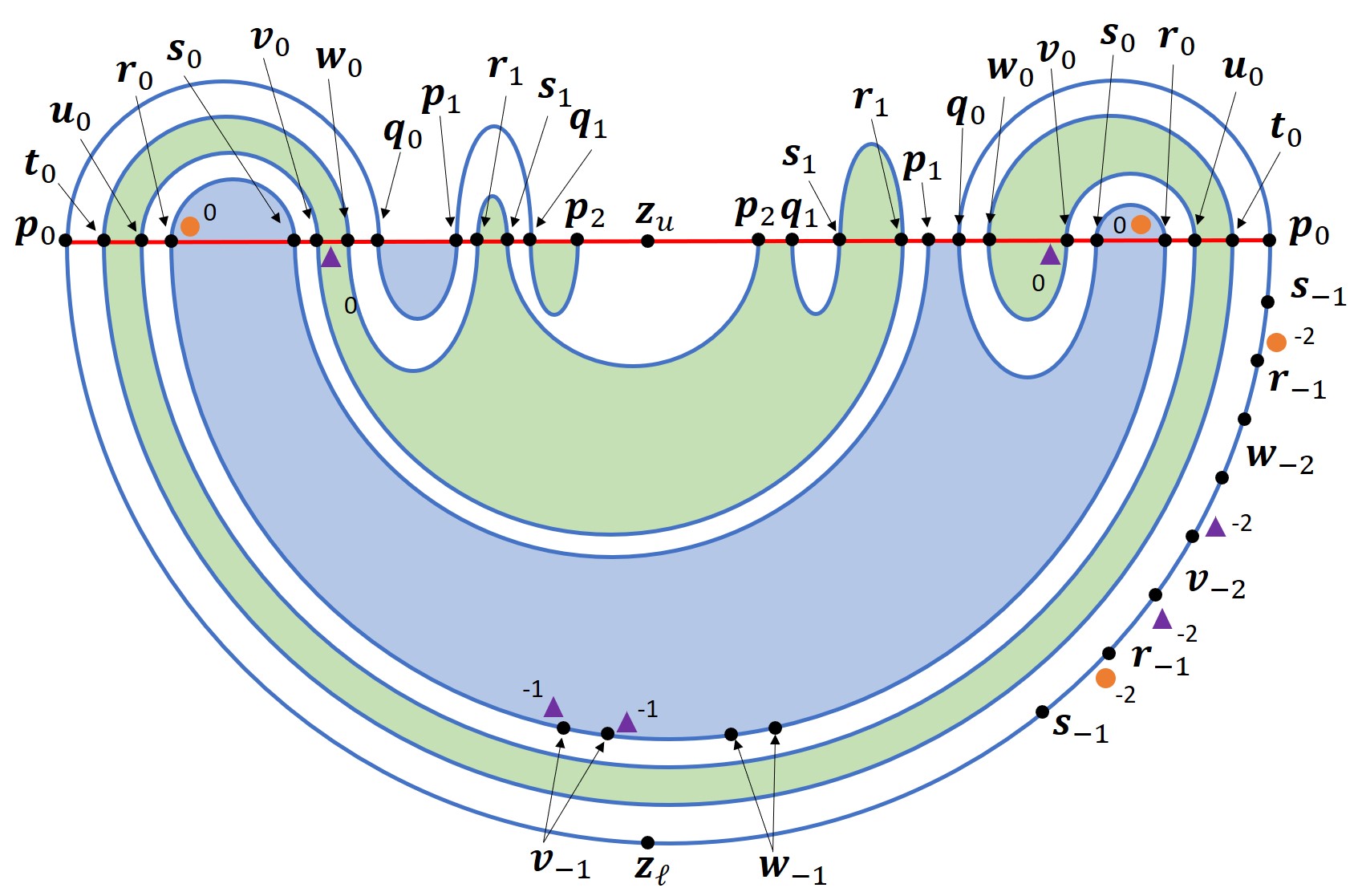}
  \caption{The trellis with the preiterates of the pseudoneighbors, $\bm{r}_0$, $\bm{s}_0$, $\bm{w}_0$, and $\bm{v}_0$ shown. The obstruction rings are placed slightly perturbed from one pseudoneighbor in a pair toward the other.  Each obstruction ring intersects the cross-section at two points, represented either by a pair of purple triangles or orange circles.}
\label{Ex1RingsPreIter}
\end{figure}

\begin{figure*}
\centering
  \includegraphics[width=1\linewidth]{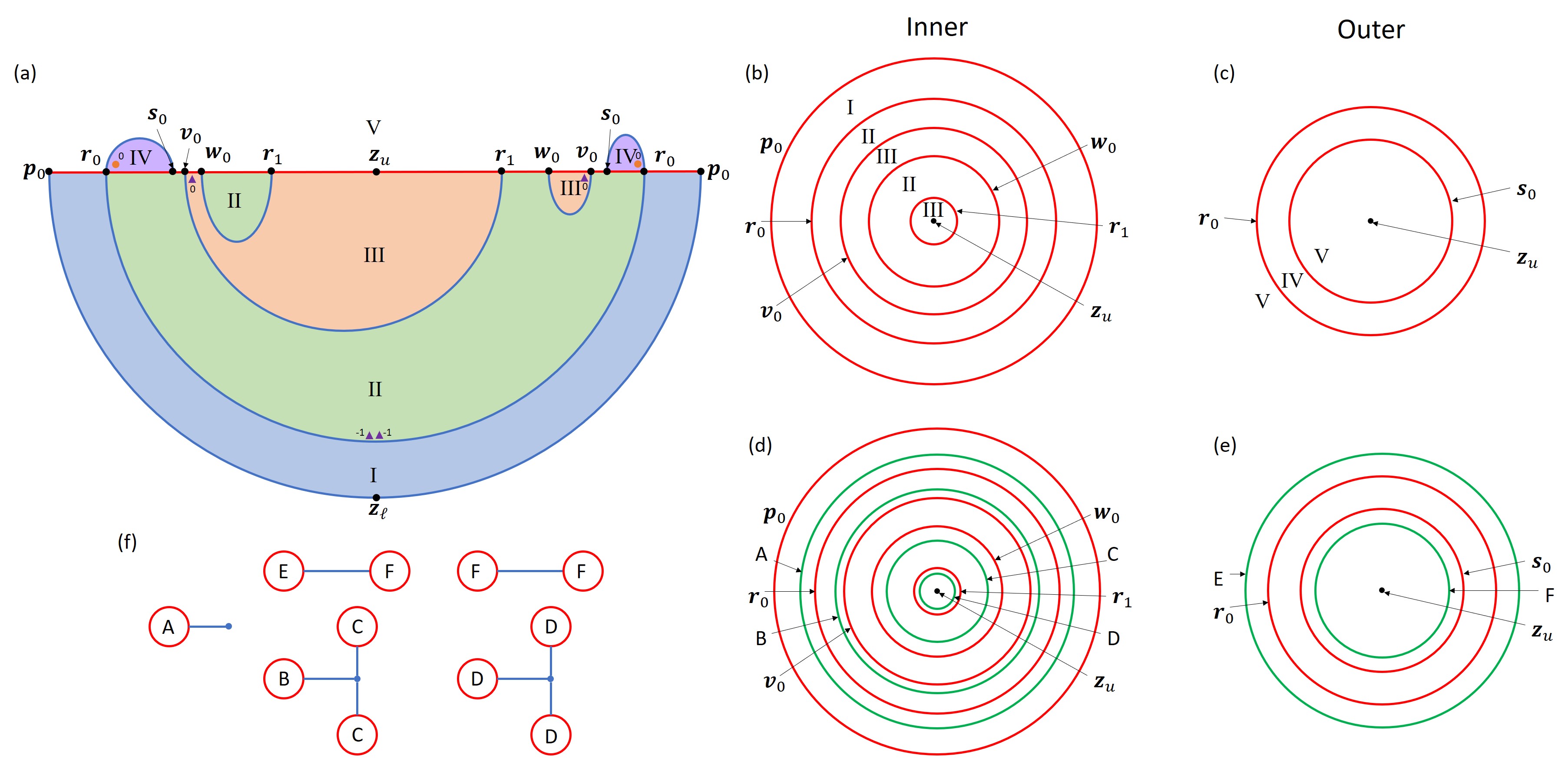}
  \caption{(a) A cross-section view of the primary division. Volumes partitioned by the bridges are labeled I-V. Each obstruction ring intersects the cross-section at two points, represented either by a pair of purple triangles or orange circles. (b) The inner stable division.  (c) The outer stable division. The heteroclinic intersection curves that appear in the inner or outer stable division are the boundary curves of inner or outer bridges in the primary division. (d) The inner stable division with boundary classes (green). (e) The outer stable division with boundary classes (green). (f) The complete set of bridge classes for the system. Each bridge class is uniquely specified by its boundary classes (red circles with letters) connected together by the unstable manifold (blue lines). }
\label{Ex1PrimaryDiv}
\end{figure*}

Next we place \textit{obstruction rings} in our system. These rings are obstructions in phase space designed to prevent bridges from being pulled back through the stable manifold. Placement of the rings are crucial to the topological distinction of different bridges. To identify the proper placement of the rings we need to investigate the forward and backward escape-time plots of the trellis as seen in Fig.~\ref{Ex1ETP}. ETPs record the number of iterates for points to exit the resonance zone. To construct the ETP we iterate points forward (or backward) from the fundamental unstable (or stable) annulus until they exit the resonance zone. Following Ref.~\cite{Maelfeyt17} we identify pairs of \textit{pseudoneighbor} intersection curves from the ETPs. Two heteroclinic curves $\bm{\alpha}_n$ and $\bm{\beta}_n$ form a pair of pseudoneighbors if $\bm{\alpha}_n$ and $\bm{\beta}_n$, or some iterate $\bm{\alpha}_m$ and $\bm{\beta}_m$, are adjacent on both the forward and backward ETPs, more precisely, if a line can be drawn between the two curves on both the forward and backward ETPs without intersecting any other heteroclinic curve. (An individual intersection curve can be a self-pseudoneighbor. See Ref.~\cite{Maelfeyt17}.) Note that the iterate of a pseudoneighbor pair is a pseudoneighbor pair. As seen in Fig.~\ref{Ex1ETP} there are two pseudoneighbor pairs $[\bm{v}_n,\bm{w}_n]$ and $[\bm{r}_n,\bm{s}_n]$. We draw the obstruction rings in the ETPs slightly perturbed from one of the pseudoneighbor intersections such that they lie between the two pseudoneighbors. The position of the rings in the ETPs dictate their placement in phase space as shown in Fig.~\ref{Ex1RingsPreIter}.

 We define homotopy classes of bridges with respect to the obstruction rings, which are viewed as ring-shaped holes in phase space. Two bridges are homotopically identified if one can be continuously distorted into the other without passing through an obstruction ring and while keeping all boundary circles attached to the stable cap. To determine these \textit{bridge classes}, we construct the \textit{primary division} of phase space. The primary division is a partitioning of phase space into a set of 3D domains. The primary division is obtained by cutting phase space along the following 2D manifolds:
\begin{enumerate}
 \item the stable component of the trellis, e.g. the stable cap $W^S_{\bm{z}_u}[\bm{p}_0]$;
 \item any bridge that includes a pseudoneighbor in its interior, i.e. within the bridge but not as a boundary circle;
 \item any bridge with a boundary circle that is a \textit{primary inert} pseudoneighbor---i.e. the first iterate of a pseudoneighbor to land on the stable component of the trellis---and for which the corresponding obstruction ring is nudged toward the interior of the bridge. 
\end{enumerate}
Figure.~\ref{Ex1PrimaryDiv}a shows the primary division of Example 1. By Cutting Rule 1 we include the stable cap $W^S_{\bm{z}_u}[\bm{p}_0]$. From Cutting Rule 2 we include the unstable cap $W^U_{\bm{z}_\ell}[\bm{p}_0]$ since every pseudoneighbor eventually maps into it in the backward-time direction. Furthermore the cap $W^U_{\bm{z}_\ell}[\bm{r}_0]$ is also included by Rule 2 since it contains the pseudoneighbors $\bm{v}_{-1}$ and $\bm{w}_{-1}$. Finally we include the bridges $W^U_{\bm{z}_\ell}[\bm{r}_0,\bm{s}_0]$ and $W^U_{\bm{z}_\ell}[\bm{v}_0,\bm{w}_0,\bm{r}_1]$ by Rule 3. In total this partitions phase space into five regions (Fig.~\ref{Ex1PrimaryDiv}a).

The stable cap is in turn partitioned by the boundary curves of the bridges that cut up phase space into the primary division. In fact, we define two primary divisions of the stable cap, one defined by the boundaries of bridges outside of the resonance zone and one by the boundaries of bridges inside the resonance zone. Figure~\ref{Ex1PrimaryDiv}b shows the inner stable division while Fig.~\ref{Ex1PrimaryDiv}c shows the outer stable division. The primary divisions of the stable cap define two sets of homotopy classes (inner and outer) for curves in the stable cap. We call these \textit{boundary classes}. Each bridge class can be uniquely specified by its boundary classes. The boundary classes for Example 1 are the green curves in Fig.~\ref{Ex1PrimaryDiv}d and Fig.~\ref{Ex1PrimaryDiv}e.

We denote a bridge class using a double bracket notation with the boundary classes that specify the bridge class enclosed. Example 1 has three inner bridge classes, the cap $\llbracket A \rrbracket$, the tridge $\llbracket B,C,C \rrbracket$, and the tridge $\llbracket D,D,D \rrbracket$, and two outer bridge classes, the bundt cakes $\llbracket D,E \rrbracket$ and $\llbracket F,F \rrbracket$. Bridge classes can also be represented in a graphic form as seen in Fig.~\ref{Ex1PrimaryDiv}f. This form is more convenient to represent the concatenation of bridge classes. Each boundary class is represented by a letter surrounded by a red circle, indicating its intersection with the stable manifold. These circles are connected with blue lines representing the connecting unstable surface.

\begin{figure*}
 \centering
\includegraphics[width=.75\linewidth]{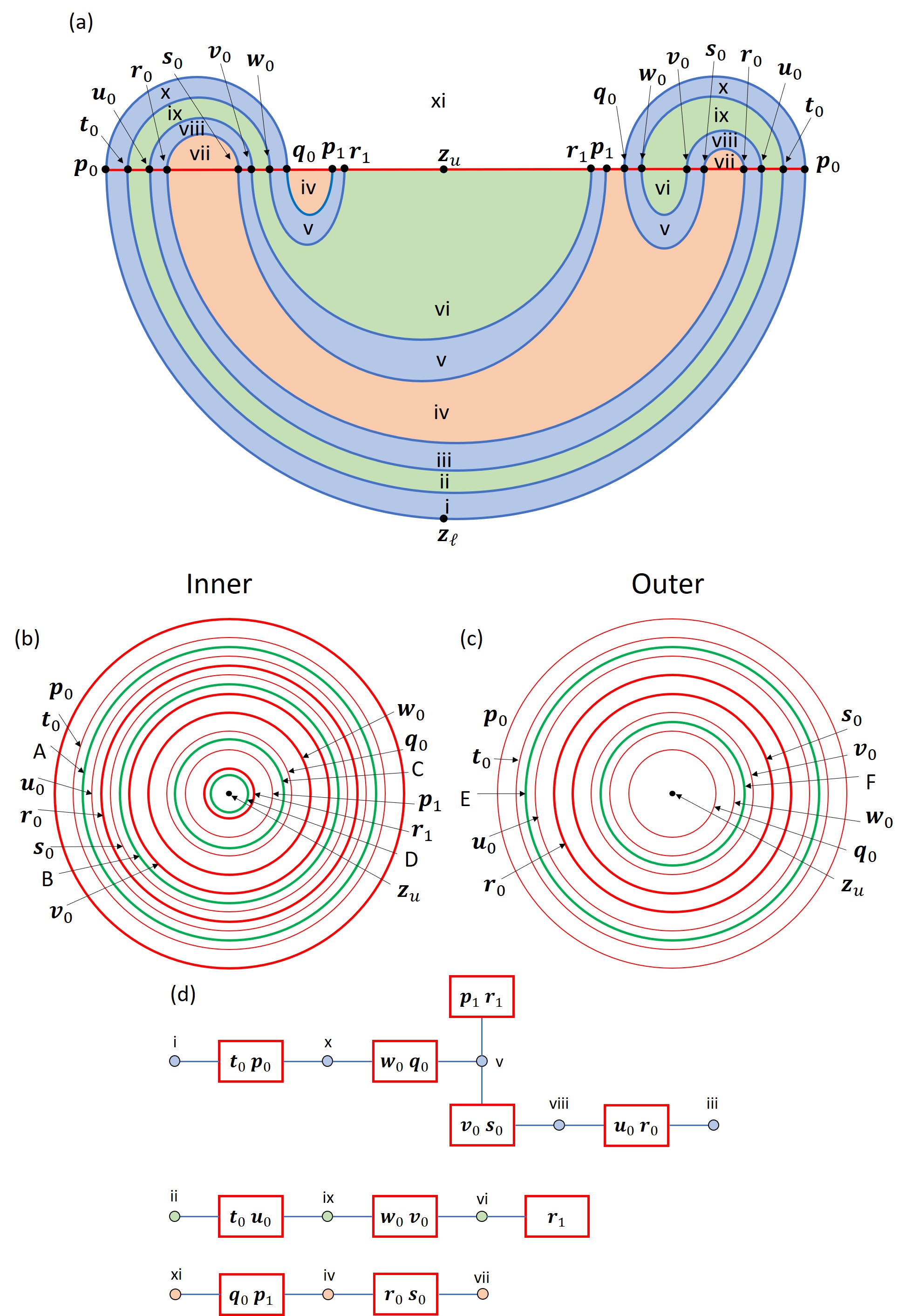}
\caption{(a) A cross section view of the secondary division of phase space. Volumes partitioned by bridges are labeled i-xi. (b) The inner secondary stable division. (c) The outer secondary stable division. Curves that also appear in the primary division are shown in bold while those that appear exclusively in the secondary division are not. Boundary classes are shown in green. (d) The connection graph of the secondary division. The connection graph identifies how the partitioned volumes of the secondary division are connected across the stable fundamental annulus. }
\label{Ex1SecondaryDiv}
\end{figure*}

To understand how the bridge classes are stretched and folded when they are iterated forward, we create a new division of phase space called the secondary division. The secondary division is constructed by cutting along the following surfaces:
\begin{enumerate}
 \item the stable component of the trellis, e.g. the cap $W^S_{\bm{z}_u}[\bm{p}_0 ]$;
 \item the forward iterate of every bridge with a pseudoneighbor in its interior, i.e. the iterate of those bridges included by Rule 2 of the primary division.
\end{enumerate}
Cutting phase space this way generates Fig.~\ref{Ex1SecondaryDiv}a. This division of phase space creates eleven domains labeled with lower-case Roman numerals. The boundary of each domain is made up of some number of bridges and some number of pieces of the stable manifold. For example, region i is bounded by the bridges $W^U_{\bm{z}_\ell}[\bm{p}_0]$, $W^U_{\bm{z}_\ell}[\bm{t}_0]$, and the stable piece $W^S_{\bm{z}_u}[\bm{p}_0,\bm{t}_0]$. Just like the primary division, the boundary curves of the bridges that make up the secondary division divide the stable cap in two ways. Figure~\ref{Ex1SecondaryDiv}b and Fig.~\ref{Ex1SecondaryDiv}c show the inner and outer secondary divisions of the stable cap. The bold red curves represent boundary curves that also occur in the primary division. Green curves are the boundary classes.

\begin{figure*}
\centering
\includegraphics[width=.65\linewidth]{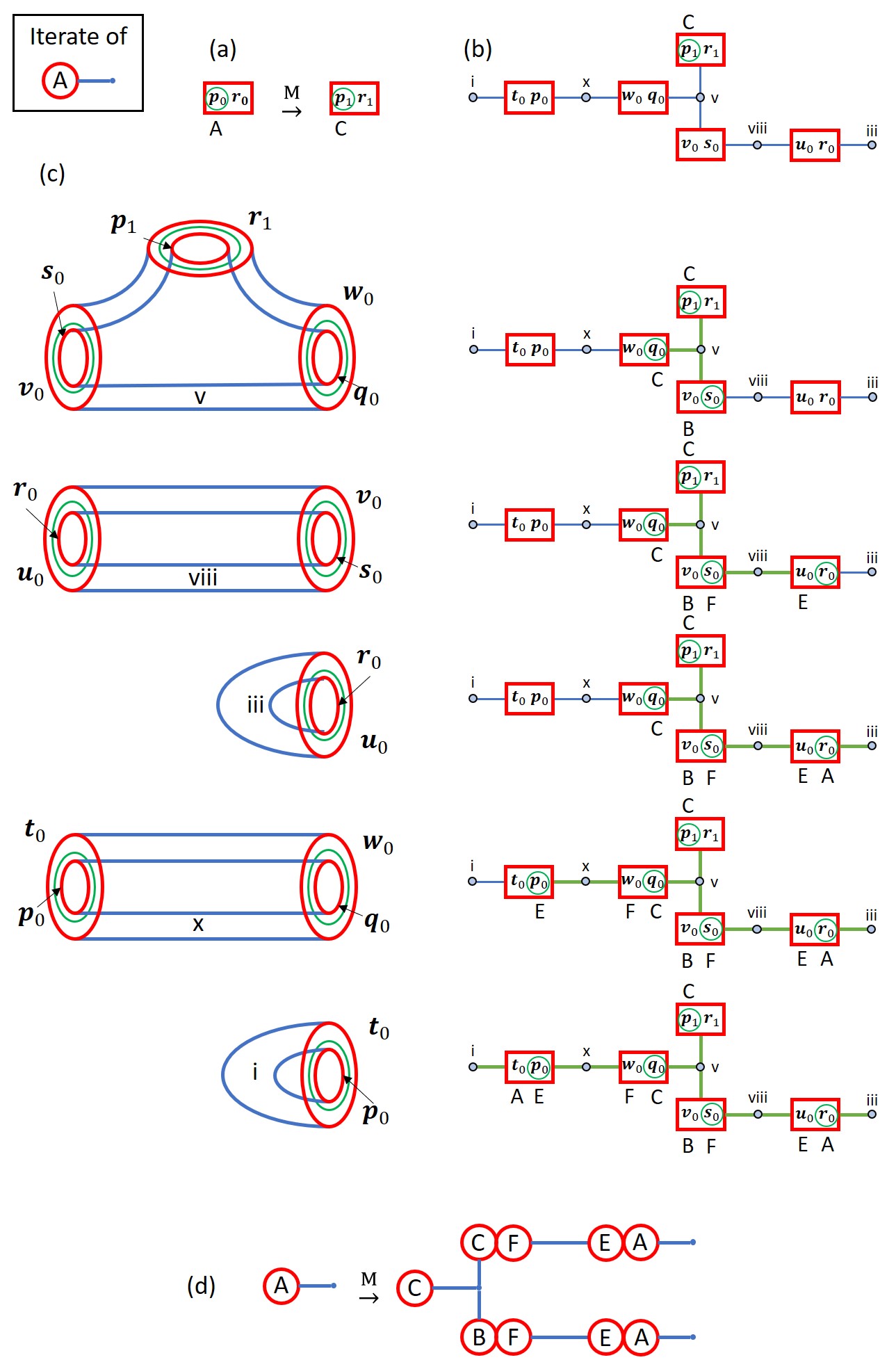}
\caption{A step-by-step illustration of the process to iterate $\llbracket A \rrbracket$ forward. (a) The forward iterate of the boundary class $A$, represented by the green circle, is the boundary class $C$, represented by the second green circle. (b) The component of the connection graph used to identify the forward iterate, showing boundary class $C$. (c) The step-by-step process identifying the forward iterate of $\llbracket A \rrbracket$. Each of the regions the forward iterate passes through is shown on the left, while the connections made are shown on the right. Boundary classes are labeled next to each box. (d) The final concatenation of bridge classes that make up the forward iterate of $\llbracket A \rrbracket$.   }
\label{Ex1StepThrough}
\end{figure*}

We specify that two domains of the secondary division are \textit{connected} if they share a common boundary along a piece of the stable fundamental annulus. This relationship is represented graphically by the \textit{connection graph}. See Fig.~\ref{Ex1SecondaryDiv}d. Every domain of the secondary division is represented as a circular node in the connection graph. Each circular node is connected to some number of red boxes, where each box represents one connected piece of the stable boundary for that domain. These pieces are labeled by their boundary curves. Two domains that are connected to one another are attached to a common red box, representing the mutual boundary between them. For example, the domains i and x are separated by the piece of the stable annulus $W^S_{\bm{z}_u}[\bm{t}_0,\bm{p}_0]$. Note that the connection graph for Example 1 has three connected components.

We map all of the bridge classes forward one iterate. We begin with the class $\llbracket A \rrbracket$, which includes the bridge $W^U_{\bm{z}_\ell}[\bm{p}_0]$. This class is specified by the boundary class $A$; a representative curve for $A$ can be chosen to lie within the domain $W^S_{\bm{z}_u}[\bm{p}_0,\bm{r}_0]$ (as seen in Fig.~\ref{Ex1PrimaryDiv}d). The forward iterate of this representative curve must lie within the domain $W^S_{\bm{z}_u}[\bm{p}_1,\bm{r}_1]$. This curve is of boundary class $C$, as seen by the inner secondary division (Fig.~\ref{Ex1SecondaryDiv}b). Note that even though the green curve $C$ is not between the curves $\bm{p}_1$ and $\bm{r}_1$ in Fig.~\ref{Ex1SecondaryDiv}b, it could be deformed to lie between these two curves without passing through a curve from the primary division (bold red curve). 

Figure~\ref{Ex1StepThrough} shows how to construct the forward iterate of $\llbracket A \rrbracket$. In Fig.~\ref{Ex1StepThrough}a we show how the boundary class $A$ maps forward.  First the forward iterate of the domain $W^S_{\bm{z}_u}[\bm{p}_0,\bm{r}_0]$ is $W^S_{\bm{z}_u}[\bm{p}_1,\bm{r}_1]$, both shown as red boxes in Fig.~\ref{Ex1StepThrough}a. Since $A$ is between $\bm{p}_0$ and $\bm{r}_0$, we circle $\bm{p}_0$ in green. Since $\bm{p}_0$ iterates to $\bm{p}_1$, we circle $\bm{p}_1$ in green as well. As discussed above this curve is of boundary class $C$. Figure~\ref{Ex1StepThrough}b shows the component of the connection graph containing $W^S_{\bm{z}_u}[\bm{p}_1,\bm{r}_1]$. As in Fig.~\ref{Ex1StepThrough}a we circle $\bm{p}_1$ and note that it is of boundary class $C$. We know from the connection graph that the forward iterate of $\llbracket A \rrbracket$ must enter region v. The first row of Fig.~\ref{Ex1StepThrough}c shows a topological representation of region v bounded by two nested tridges. Since the unstable manifold cannot intersect itself, the forward iterate of $\llbracket A \rrbracket$ is forced to intersect the domain $W^S_{\bm{z}_u}[\bm{w}_0,\bm{q}_0]$, whose intersection curve is of boundary class $C$, and the domain $W^S_{\bm{z}_u}[\bm{s}_0,\bm{v}_0]$, whose intersection curve is of boundary class $B$. We shade the connection graph lines in green to show how the iterate occupies region v, while placing green circles around $\bm{q}_0$ and $\bm{s}_0$, representing the intersection curves. Due to the intersection curve around $\bm{s}_0$, the forward iterate of $\llbracket A \rrbracket$ is forced to enter region viii. Since region viii is exterior, we note that a boundary curve between $\bm{s}_0$ and $\bm{v}_0$ is of the class $F$ with respect to the outer primary stable division (Fig.~\ref{Ex1SecondaryDiv}c). Examining the topology of region viii on the second row of Fig.~\ref{Ex1StepThrough}c, we see that the forward iterate of $\llbracket A \rrbracket$ is forced to intersect the domain $W^S_{\bm{z}_u}[\bm{r}_0,\bm{u}_0]$. We place a green circle around $\bm{r}_0$, which represents boundary class $E$. Next the iterate of $\llbracket A \rrbracket$ passes through the inner region iii. From the inner perspective the curve around $\bm{r}_0$ has boundary class $A$. The topological representation of region iii on row three of Fig.~\ref{Ex1StepThrough}c consists of two nested caps. The minimal topological form for the iterate of $\llbracket A \rrbracket$ within region iii is a terminating cap. Returning to the curve around $\bm{q}_0$, we see it generates a similar process as the curve around $\bm{s}_0$, occupying regions x and i as seen in rows four and five of Fig.~\ref{Ex1StepThrough}c. Putting all this together gives the forward iterate of $\llbracket A \rrbracket$ seen in Fig.~\ref{Ex1StepThrough}d.

\begin{figure}
\centering
\includegraphics[width=1\linewidth]{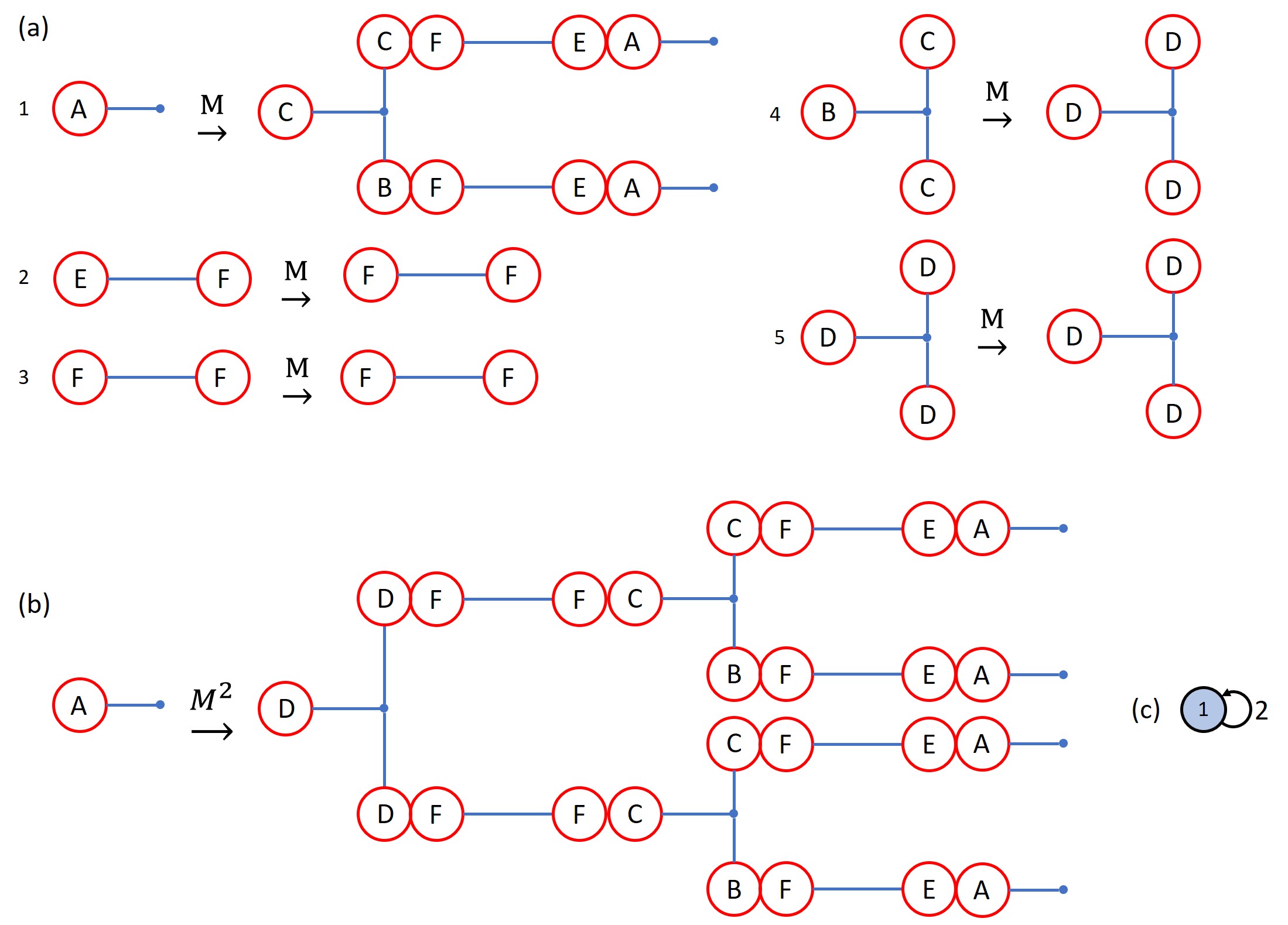}
\caption{(a) The complete set of bridge classes for the dynamics and their forward iterates. (b) Bridge class $\llbracket A \rrbracket$ iterated forward twice. After two iterations, $\llbracket A \rrbracket$ produces four copies of itself. (c) The transition graph for the single active bridge class.}
\label{Ex1BridgeIters}
\end{figure}

The iterates of the remaining four bridge classes in Fig.~\ref{Ex1PrimaryDiv}f are easier to construct. In the case of bridge class $\llbracket E,F \rrbracket$, we look at the representative bridge $W^U_{\bm{z}_\ell}[\bm{p}_0,\bm{q}_0]$. The forward iterate of $W^U_{\bm{z}_\ell}[\bm{p}_0,\bm{q}_0]$ is $W^U_{\bm{z}_\ell}[\bm{p}_1,\bm{q}_1]$, which is a single bridge, of class $\llbracket F,F \rrbracket$. The forward iterate of $W^U_{\bm{z}_\ell}[\bm{p}_1,\bm{q}_1]$ is $W^U_{\bm{z}_\ell}[\bm{p}_2,\bm{q}_2]$, also of class $\llbracket F,F \rrbracket$. Thus $\llbracket F, F \rrbracket$ maps to itself. Since all iterates of $\llbracket E,F \rrbracket$ consist of a single bridge class, we say that $\llbracket E,F \rrbracket$ is \textit{inert}. By the same logic, $\llbracket F,F \rrbracket$ is also inert. The same process can be applied to the remaining two bridge classes, which are also inert. Figure~\ref{Ex1BridgeIters}a summarizes the complete set of dynamics for Example 1. Note that $\llbracket A \rrbracket$ is not inert because it produces multiple bridge classes upon iteration. A bridge class that is not inert is called \textit{active}.

Having determined the iterates of all the bridge classes of the trellis, we know the forced topology of the unstable manifold. For example, suppose we wanted to iterate the bridge class of the unstable cap $W^U_{\bm{z}_\ell}[\bm{p}_0]$ forward twice. We would first iterate the bridge class $\llbracket A \rrbracket$, to which $W^U_{\bm{z}_\ell}[\bm{p}_0]$ belongs, and then iterate each resultant bridge class forward, resulting in Fig.~\ref{Ex1BridgeIters}b. [Note that the dynamics produced by iterating $\llbracket A \rrbracket$ forward twice is exactly the trellis in Fig.~\ref{Ex1Trel}.] Additionally, a lower bound of the topological entropy can be determined from the symbolic dynamics. We first create the \textit{transition matrix} $\mathsf{T}$, where the component $T_{ij}$ records the number of times that bridge class number $i$ appears in the iterate of bridge class number $j$. The log of the largest eigenvalue of $\mathsf{T}$ is a lower bound of the topological entropy. It is sufficient to only consider the submatrix of the active bridge classes, since inert bridge classes do not contribute to the topological entropy. In Example 1 there is a single active bridge class, $\llbracket A \rrbracket$, whose iterate contains two copies of $\llbracket A \rrbracket$. This produces the one-by-one matrix $\mathsf{T} = (2)$, which generates $h = \ln{2}$ as a lower bound to the topological entropy of the original map $M$. Note that a transition matrix can be represented as a transition graph, as in Fig.~\ref{Ex1BridgeIters}c. In future examples we will just show the transition graph. 

\begin{figure*}
\centering
\includegraphics[width=.85\linewidth]{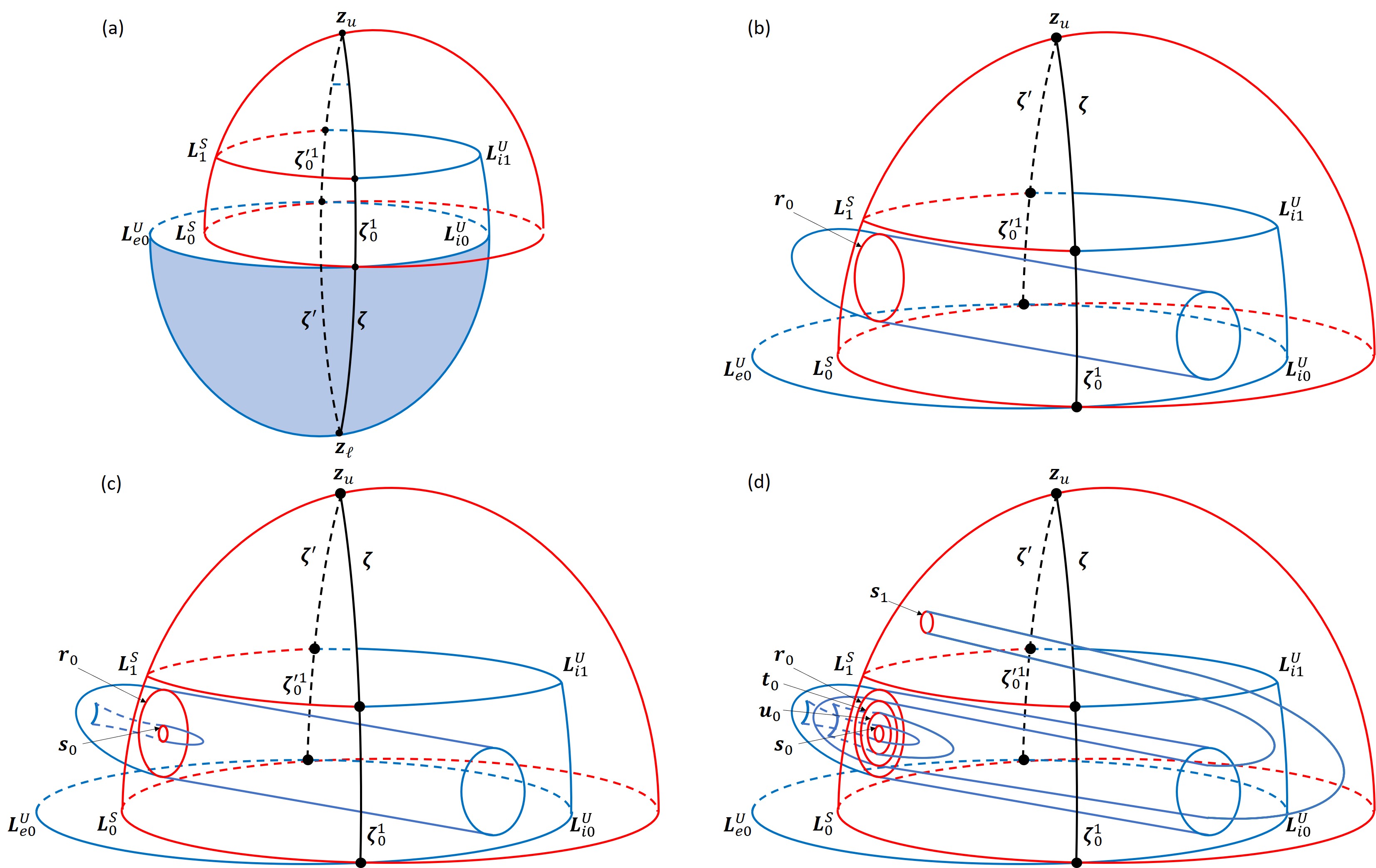}
\caption{A sequence of images representing trellises with increasing complexity. (a) A trellis with no additional intersections beyond the pole-to-pole intersections $\bm{\zeta}$ and $\bm{\zeta}^{'}$. (b) A trellis where the unstable manifold inside the stable cap is stretched across to intersect the stable cap. (c) A trellis where the cap $W^U_{\bm{z}_\ell}[\bm{r}_0]$ from (b) is pulled back through the stable cap. (d) The trellis in (c) with the forward iterate of $W^U_{\bm{z}_\ell}[\bm{s}_0]$ shown. }
\label{Ex2Build}
\end{figure*}

\section{Example 2}
\label{Example 2}

Example 1 examined the case where there is primary intersection curve $\bm{p}_0$ between the stable cap and primary unstable cap. Example 2  explores a case where there is no primary intersection curve between the stable and unstable caps, but there are pole-to-pole intersection curves, as in Fig.~\ref{fig2}b. Despite the lack of a well defined resonance zone, we can still extract homotopic lobe dynamics on a subset of the full trellis.

Figure~\ref{Ex2Build} constructs the topology of the trellis in Example 2. We define the stable cap and primary unstable cap as the submanifolds of the stable and unstable manifolds up to their first intersections with the $xy$-plane (Fig.~\ref{Ex2Build}a). We label these intersections $\bm{L}_0^S$ and $\bm{L}_0^U$ respectively, and we assume that they are proper loops. We use these curves to form the stable and unstable fundamental annuli $W^S_{\bm{z}_u}[\bm{L}_{0}^S,\bm{L}_1^S]$ and $W^U_{\bm{z}_\ell}[\bm{L}_{-1}^U,\bm{L}_0^U]$. The curve $\bm{L}_0^U$ is  broken into an interior segment $\bm{L}_{i0}^U$ and exterior segment $\bm{L}_{e0}^U$.  Figure~\ref{Ex2Build}a shows the first iterate of $\bm{L}_{i0}^U$ and the unstable manifold between them. The curves $\bm{L}_0^U$ and $\bm{L}_1^U$ intersect the pole-to-pole intersection curves $\bm{\zeta}$ and $\bm{\zeta}^{'}$ and define the segments $\bm{\zeta}^1_0$ and $\bm{\zeta}_0^{'1}$ between them. This allows us to define the interior half-annulus $W^U_{\bm{z}_\ell}[\bm{\zeta}_0^1,\bm{L}_{i0}^U,\bm{\zeta}_0^{'1},\bm{L}_{i1}^U]$.

\begin{figure}
\centering
\includegraphics[width=.55\linewidth]{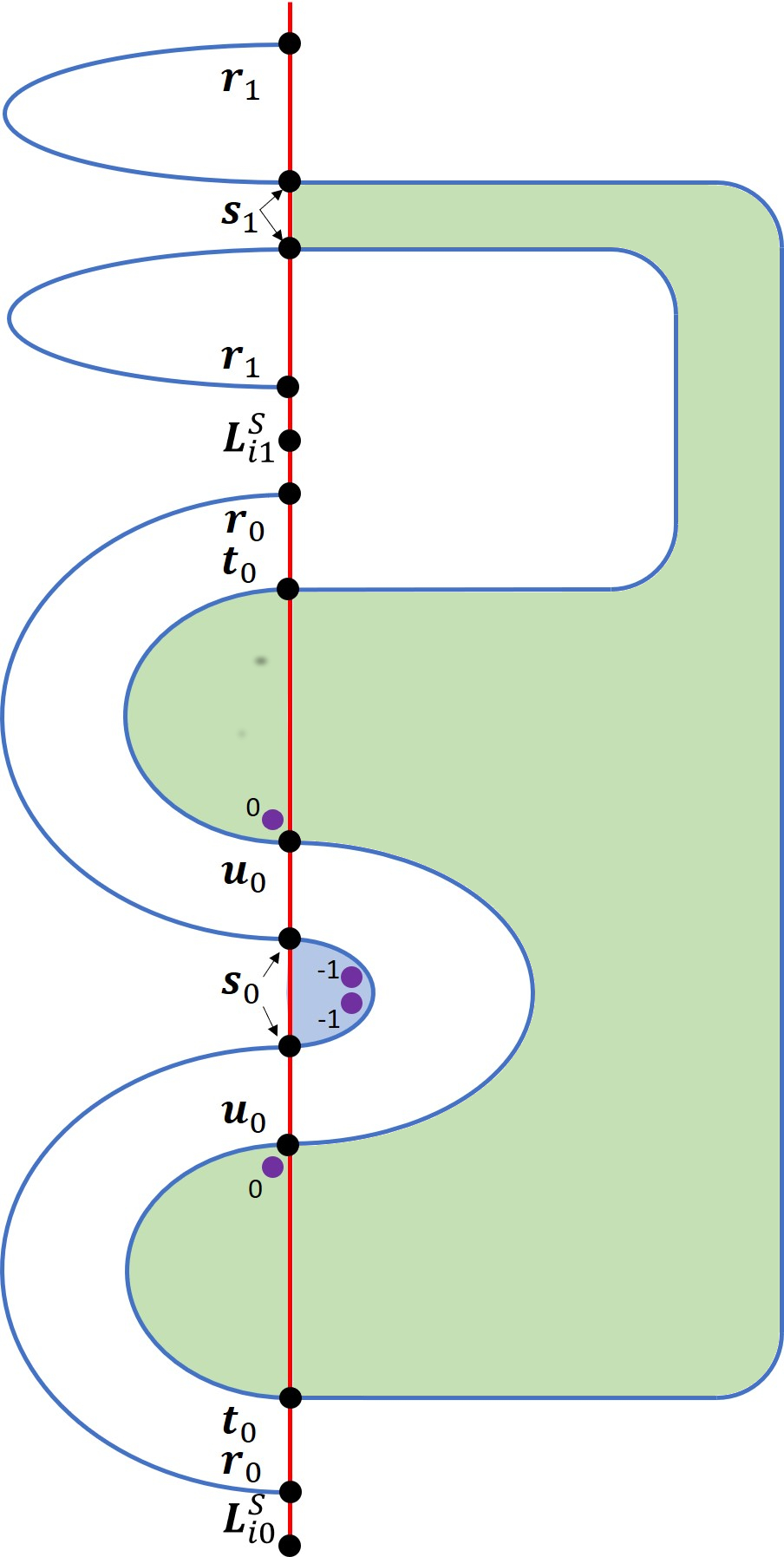}
\caption{A cross section of Fig.~\ref{Ex2Build}d showing only the bridges, all of whose boundary curves lie entirely within the stable cap. Only the part of the stable cap containing these boundary curves is shown.}
\label{Ex2Reduced}
\end{figure}

In Fig.~\ref{Ex2Build}b, we modify the dynamics of Fig.~\ref{Ex2Build}a by pushing a piece of the half-annulus $W^U_{\bm{z}_\ell}[\bm{\zeta}_0^1,\bm{L}_{i_0}^U,\bm{\zeta}_0^{'1},\bm{L}_{i_1}^U]$ across the interior region until it intersects the stable cap at $\bm{r}_0$. This creates the unstable submanifold $W^U_{\bm{z}_\ell}[\bm{\zeta}_0^1,\bm{L}_{i_0}^U,\bm{\zeta}_0^{'1},\bm{L}_{i_1}^U,\bm{r}_0]$. The intersection $\bm{r}_0$ lies on the fundamental stable annulus $W^S_{\bm{z}_u}[\bm{L}_{0}^S,\bm{L}_{1}^S]$. The trellis in Fig.~\ref{Ex2Build}b still does not generate any topological entropy as no new heteroclinic intersections are forced to exist at any finite iterate. In Fig.~\ref{Ex2Build}c we again modify the dynamics by introducing additional structure. We take the cap $W^U_{\bm{z}_\ell}[\bm{r}_0]$ in Fig.~\ref{Ex2Build}b and push a part of it back through the stable cap $W^S_{\bm{z}_u}[\bm{L}_0^S]$ forming the intersection $\bm{s}_0$, as seen in Fig.~\ref{Ex2Build}c. While this forms an interior cap $W^U_{\bm{z}_\ell}[\bm{s}_0]$ and exterior bundt cake $W^U_{\bm{z}_\ell}[\bm{r}_0,\bm{s}_0]$, this still does not produce any topological entropy. In Fig.~\ref{Ex2Build}d we add to the trellis in Fig.~\ref{Ex2Build}c by iterating the interior cap $W^U_{\bm{z}_\ell}[\bm{s}_0]$ forward. This iterate is stretched back to the fundamental stable annulus producing the interior macaroni $W^U_{\bm{z}_\ell}[\bm{s}_1,\bm{t}_0]$, the exterior bundt cake $W^U_{\bm{z}_\ell}[\bm{t}_0,\bm{u}_0]$, and the interior cap $W^U_{\bm{z}_\ell}[\bm{u}_0]$. The trellis in Fig.~\ref{Ex2Build}d has forced dynamics with non-zero topological entropy as we shall show.

\begin{figure}
\centering
\includegraphics[width=.5\linewidth]{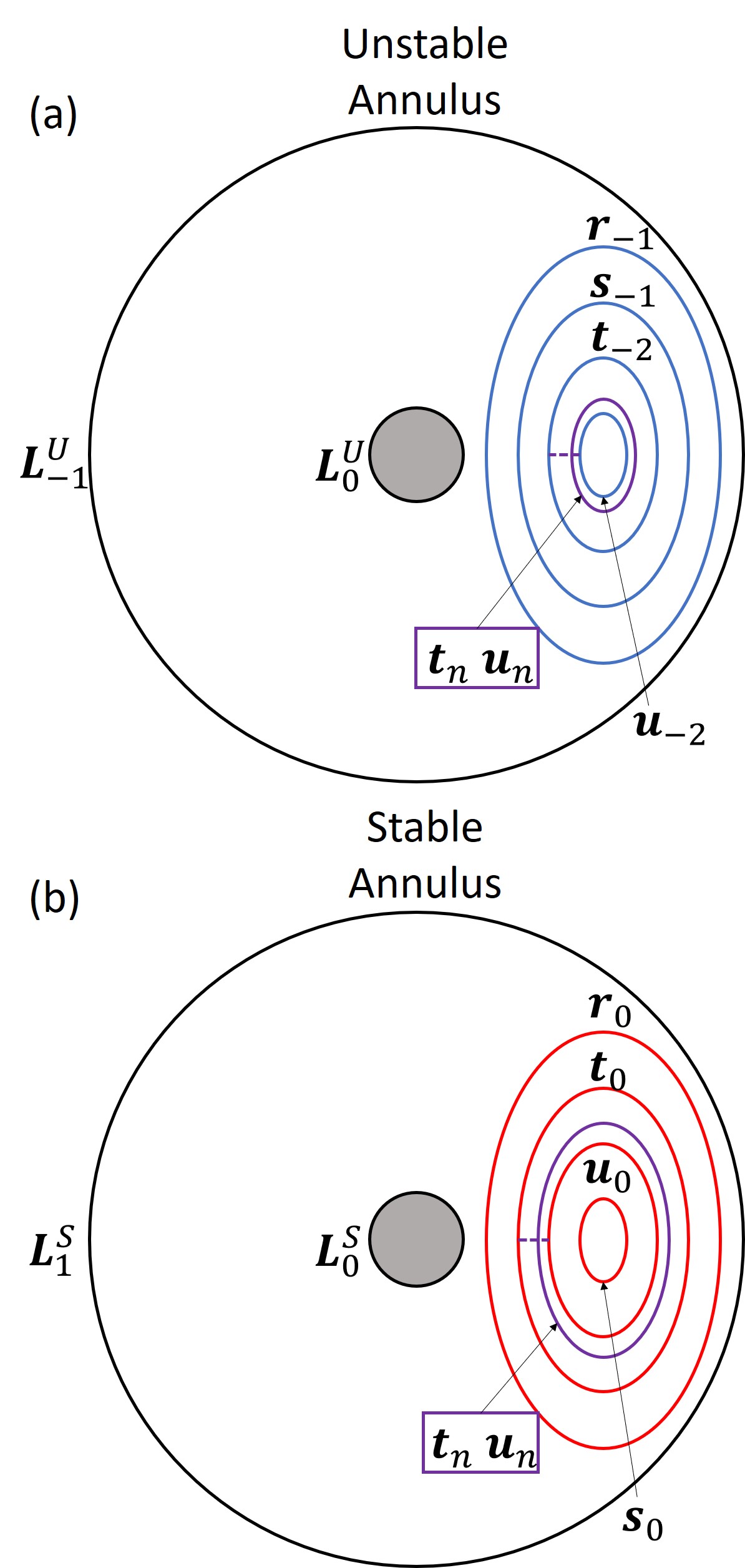}
\caption{(a) The heteroclinic intersections that lie on the unstable fundamental annulus $W^U_{\bm{z}_\ell}[\bm{L}_{-1}^U,\bm{L}_0^U]$. (b) The heteroclinic intersections that lie on the stable fundamental annulus $W^S_{\bm{z}_u}[\bm{L}_{0}^S,\bm{L}_1^S]$. Pseudoneighbor pairs are identified based on curves that are adjacent in both the forward and backward fundamental annuli.}
\label{Ex2Annuli}
\end{figure}

Some unstable submanifolds such as $W^U_{\bm{z}_\ell}[\bm{L}_0^U]$ and $W^U_{\bm{z}_\ell}[\bm{\zeta}_0^1,\bm{L}_{i0}^U,\bm{\zeta}_{0}^{'1},\bm{L}_{i1}^U,\bm{r}_0]$ are not bridges because their boundaries do not solely lie on the stable cap. This makes defining escape times, primary divisions, and secondary divisions awkward. Therefore in this example we focus solely on the unstable submanifolds that are true bridges and apply HLD to those submanifolds only. Fig~\ref{Ex2Reduced} shows the relevant bridges.

Since we do not have a well defined resonance zone, we avoid ETPs and work solely with the heteroclinic intersection curves directly. Figure~\ref{Ex2Annuli} shows the stable and unstable fundamental annuli with the heteroclinic intersections present in Fig.~\ref{Ex2Reduced}. We use these plots the same way we use the ETPs to identify pseudoneighbor pairs and place obstruction rings. Looking at both the annuli in Fig.~\ref{Ex2Annuli}a and Fig.~\ref{Ex2Annuli}b, we see that $[\bm{t}_n,\bm{u}_n]$ form the sole pseudoneighbor pair. We place an obstruction ring slightly perturbed from $\bm{u}_n$ toward $\bm{t}_n$ in Fig.~\ref{Ex2Annuli}. Two of these obstruction rings are present in Fig.~\ref{Ex2Reduced} represented by two pairs of purple dots labeled with their iterate number. 

\begin{figure*}
\centering
\includegraphics[width=.85\linewidth]{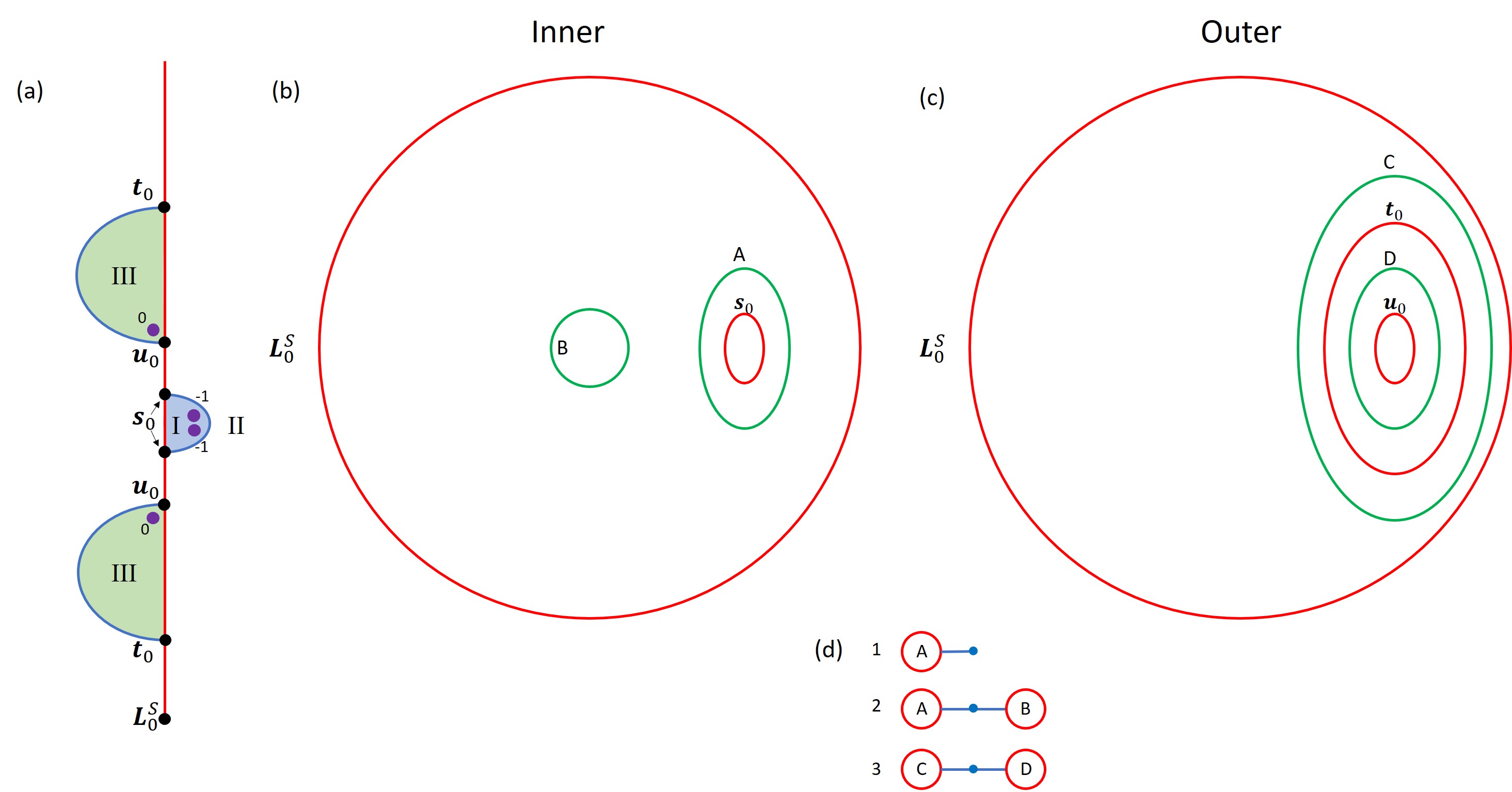}
\caption{ (a) The primary division of phase space. (b) The inner stable division. (c) The outer stable division. (d) The bridge classes that make up the system.}
\label{Ex2PrimaryDiv}
\end{figure*}

With the obstruction rings placed we construct the primary division of phase space in Fig~\ref{Ex2PrimaryDiv}a. We include the bridge $W^U_{\bm{z}_\ell}[\bm{s}_0]$ based on Rule 2 of constructing the primary division in Sec.~\ref{Example 1} and $W^U_{\bm{z}_\ell}[\bm{t}_0,\bm{u}_0]$ based on Rule 3. We omit the unstable cap $W^U_{\bm{z}_\ell}[\bm{L}_0^U]$ since it is not a bridge. Using the primary division, we construct the inner and outer stable divisions in Fig.~\ref{Ex2PrimaryDiv}b and Fig.~\ref{Ex2PrimaryDiv}c. We identify the boundary classes A through D from boundary curves in Fig.~\ref{Ex2Reduced}. Examining Fig.~\ref{Ex2Reduced} we find bridge classes $\llbracket A \rrbracket$, $\llbracket A,B \rrbracket$ and $\llbracket C,D \rrbracket$ as seen in Fig.~\ref{Ex2PrimaryDiv}d. 

\begin{figure*}
\centering
\includegraphics[width=1\linewidth]{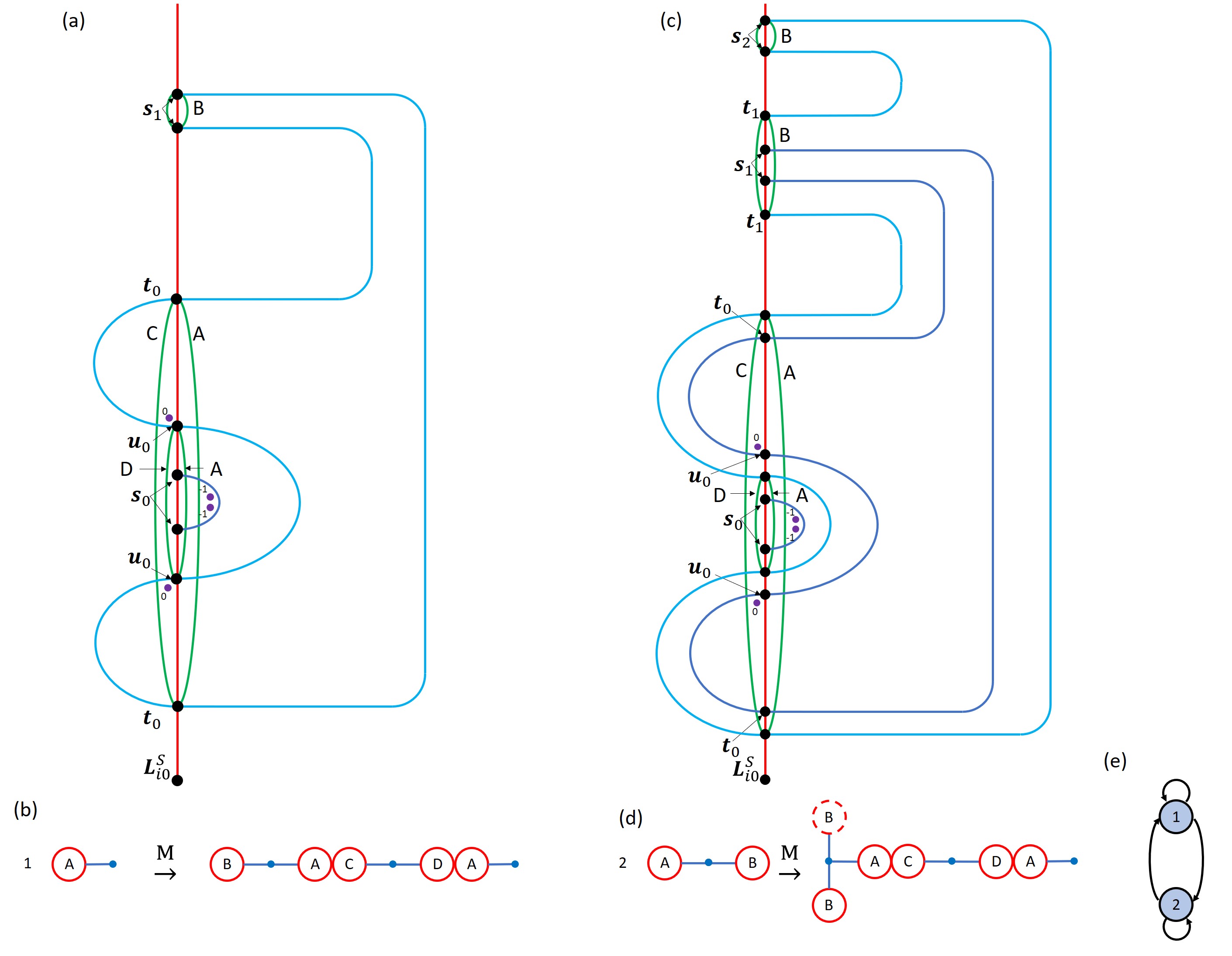}
\caption{ We show the construction of the forward iterate of each of the active bridge classes. (a) and (c) shows the active bridge being iterated forward in blue. The forward iterate of the bridge class is shown in cyan. The heteroclinic intersection curves of the forward iterate are in green and the boundary class they belong to labeled for both the inner and outer stable division. (b) and (d) show the symbolic equation for the two active bridge classes. (e) is the transition graph of the two active classes. }
\label{Ex2Iterates}
\end{figure*}

To determine the iterate of the bridge class $\llbracket A \rrbracket$, let us consider the iterate of the representative bridge $W^U_{\bm{z}_\ell}[\bm{s}_0]$. We iterate the boundary curve $\bm{s}_0$ forward to $\bm{s}_1$, as seen in Fig.~\ref{Ex2Iterates}a. Curve $\bm{s}_1$ is of boundary class $B$ (Fig~\ref{Ex2PrimaryDiv}b). Following Fig.~\ref{Ex2Iterates}a, $\bm{s}_1$ is connected to $\bm{t}_0$ by the interior macaroni $W^U_{\bm{z}_\ell}[\bm{s}_1,\bm{t}_0]$. Curve $\bm{t}_0$ has boundary class $A$ on the interior so that $W^U_{\bm{z}_\ell}[\bm{s}_1,\bm{t}_0]$ is of bridge class $\llbracket B,A \rrbracket$. To the left of the stable manifold, boundary curve $\bm{t}_0$ is connected to $\bm{u}_0$ by a bundt cake forming the bridge $W^U_{\bm{z}_\ell}[\bm{t}_0,\bm{u}_0]$. On the exterior $\bm{t}_0$ and $\bm{u}_0$ have boundary classes $C$ and $D$ respectively. This means $W^U_{\bm{z}_\ell}[\bm{t}_0,\bm{u}_0]$ belongs to bridge class $\llbracket C,D \rrbracket$. To the right of the stable manifold, $\bm{u}_0$ is terminated by a cap. $\bm{u}_0$ also has interior boundary class $A$ so that this cap $W^U_{\bm{z}_\ell}[\bm{u}_0]$ belongs to bridge class $\llbracket A \rrbracket$. Putting all this together, the iterate of $\llbracket A \rrbracket$ is the concatenation of $\llbracket B,A \rrbracket$, $\llbracket C,D \rrbracket$, and $\llbracket A \rrbracket$ as shown in Fig.~\ref{Ex2Iterates}b.

We determine the iterate of the bridge class $\llbracket A,B \rrbracket$ by examining the iterate of the representative bridge $W^U_{\bm{z}_\ell}[\bm{t}_0,\bm{s}_1]$. The forward iterates of $\bm{t}_0$ and $\bm{s}_1$ are $\bm{t}_1$ and $\bm{s}_2$, each placed as in Fig.~\ref{Ex2Iterates}c. The curves $\bm{t}_1$ and $\bm{s}_2$ each have boundary class $B$. Curve $\bm{t}_1$ cannot be directly connected to curve $\bm{s}_2$ by a macaroni, because this macaroni would then intersect the bridge $W^U_{\bm{z}_\ell}[\bm{t}_0,\bm{s}_1]$. To properly connect $\bm{t}_1$ to $\bm{s}_2$, the manifold is forced to have an additional intersection around $\bm{t}_0$ as in Fig.~\ref{Ex2Iterates}c. This new boundary has internal boundary class $A$. Yielding a bridge class $\llbracket B,B,A \rrbracket$ to the right of the stable manifold. On the left this new intersection curve has boundary class $C$. Following Fig.~\ref{Ex2Iterates}c this intersection curve is connected by a bundt cake to a new intersection curve between $\bm{u}_0$ and $\bm{s}_0$. This new intersection has outer boundary class $D$ meaning the bundt cake has bridge class $\llbracket C,D \rrbracket$. To the right of the stable manifold, the intersection curve between $\bm{u}_0$ and $\bm{s}_0$ is terminated by a cap of bridge class $\llbracket A \rrbracket$. In summary the iterate of bridge class $\llbracket A,B \rrbracket$ is the concatenation of the bridge classes in Fig.~\ref{Ex2Iterates}d. In the iterate of $\llbracket A,B \rrbracket$ a new bridge class $\llbracket A,B,B \rrbracket$ appears. The iterate of $\llbracket A,B,B \rrbracket$ is identical to the iterate of $\llbracket A,B \rrbracket$ except that $\llbracket A,B,B \rrbracket$ is replaced by the new bridge $\llbracket A,B,B,B \rrbracket$. This pattern repeats itself with each iterate of $\llbracket A,B,B,... \rrbracket$ producing a new bridge class with an additional $B$ boundary class. All of these additional $B$ boundary classes are unimportant to the symbolic dynamics. (They are inert in the sense of Ref.~\cite{Maelfeyt17}.) We therefore indicate the additional $B$ boundary class with a dashed circle in Fig.~\ref{Ex2Iterates}d. Furthermore, we identify all of these classes as one symbol in the symbolic dynamics. The result is that the system has two active bridge classes $\llbracket A \rrbracket$ and $\llbracket A,B \rrbracket$ each of which produces one copy of itself and one copy of the other. 

\begin{figure}
\centering
\includegraphics[width=1\linewidth]{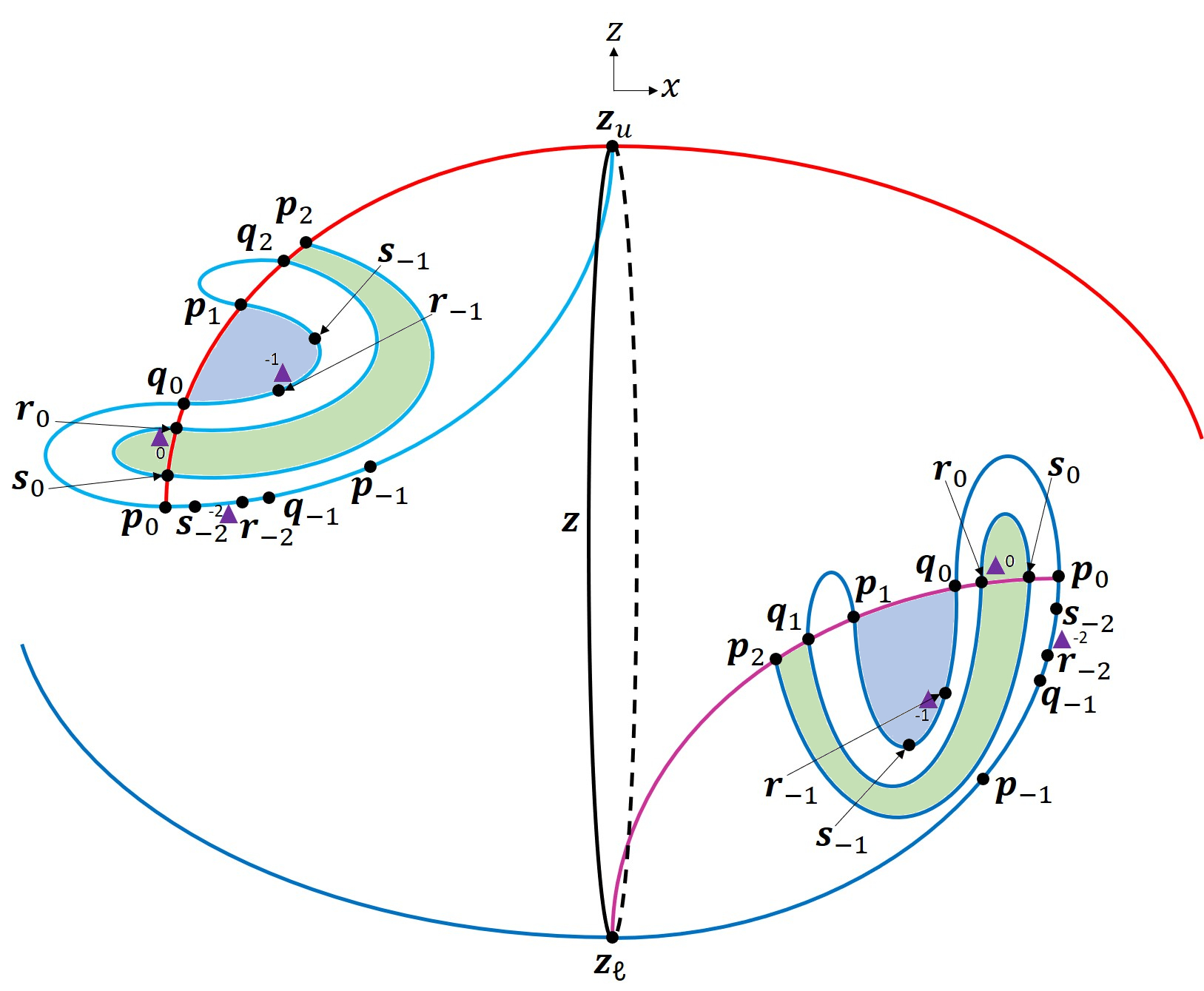}
\caption{An example of invariant manifolds attached to the invariant circle $\bm{z}$. The stable manifold of the invariant circle $\bm{z}$ is the union of the 1D stable manifold (magenta) extending from the lower fixed point and the 2D stable manifold (red) extending from the upper fixed point. Similarly the unstable manifold of $\bm{z}$ is the union of the 1D unstable manifold (cyan) extending from the upper fixed point and the 2D unstable manifold (blue) extending from the lower fixed point. The stable and unstable manifolds of $\bm{z}$ intersect at several 1D curves marked with dots. The obstruction ring (purple triangles) are placed slightly perturbed from $\bm{r}_n$ toward $\bm{s}_n$.}
\label{Ex3Trel}
\end{figure}

From the iterates of the active classes, we get the transition graph shown in Fig.~\ref{Ex2Iterates}(e). Each iterate of bridge class 1, i.e. $\llbracket A \rrbracket$, produces one copy of class 1 and class 2, i.e. $\llbracket A,B \rrbracket$. Bridge class 2  also produces a copy of both bridge classes 1 and 2. From the corresponding transition matrix, we find a topological entropy of $h = \ln{2}$.

We have shown here that HLD can be applied to manifolds in a localized region of phase space without needing a well defined resonance zone. In Example 4 we revisit this geometry in the context of a well defined resonance zone and provide an alternative analysis.

\section{Example 3} 
\label{Example 3}

\begin{figure}
\centering
\includegraphics[width=.65\linewidth]{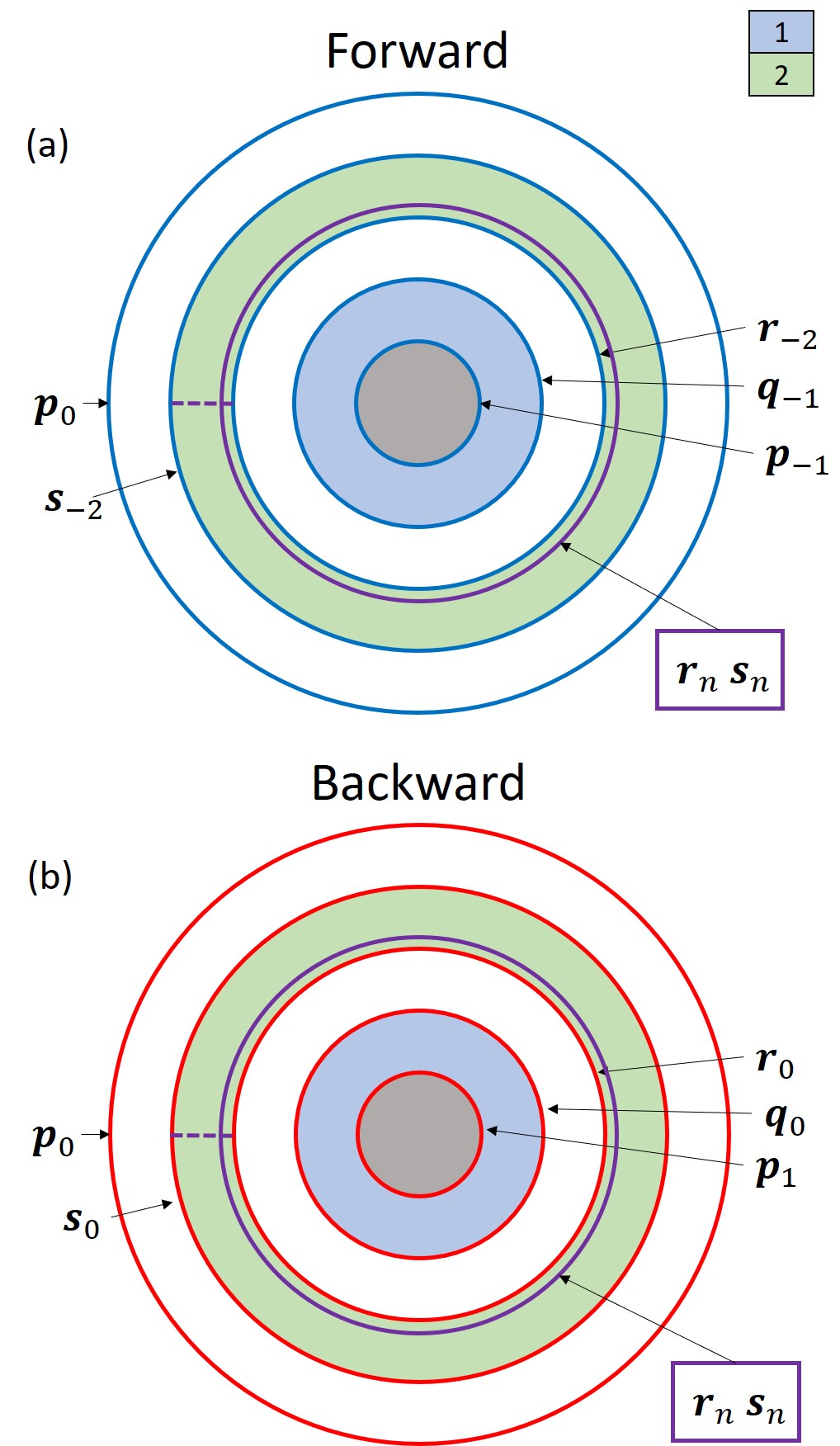}
\caption{Escape time plots for Example 3. (a) The forward escape time plot. (b) The backward escape time plot. $\bm{r}_{n}$ and $\bm{s}_n$ are pseudoneighbor pairs with a obstruction ring (purple) placed between them slightly perturbed from $\bm{r}_n$ toward $\bm{s}_n$.}
\label{Ex3ETP}
\end{figure}

This section explores the case in Fig.~\ref{Ex3Trel}, where it is more beneficial to look at the 2D stable and unstable manifolds extending from the invariant circle $\bm{z}$ instead of from the fixed points. This example is reversible, as discussed in Sec. \ref{Reversibility}, with symmetry operator $S(x,y,z) = (-x,y,-z)$. 

\begin{figure*}
\centering
\includegraphics[width=.9\linewidth]{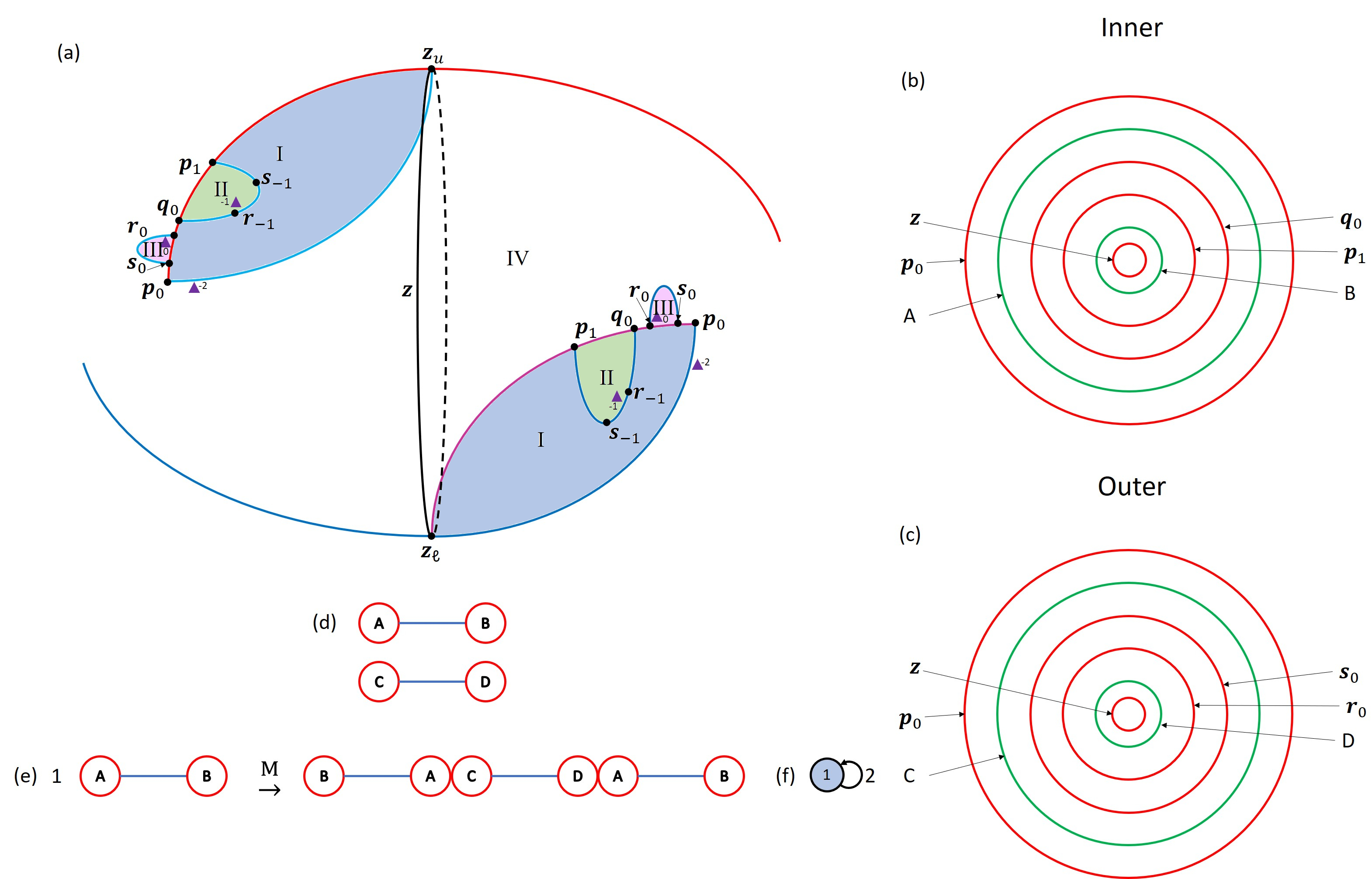}
\caption{(a) The primary division for Example 3. (b) The inner stable division (c) The outer stable division. (d) The two bridge classes. (e) The forward iterate of the one active bridge class. (f) The transition graph of the active bridge class.}
\label{Ex3BridgeIterates}
\end{figure*}

In Fig.~\ref{Ex3Trel} the 2D stable manifold (red) and 1D unstable manifold (cyan) extend from the upper fixed point $\bm{z}_u$ and intersect at the point labeled $\bm{p}_0$ on the left. Similarly the 2D unstable manifold (blue) and 1D stable manifold (magenta) extend from the lower fixed point $\bm{z}_\ell$ and intersect at the point labeled $\bm{p}_0$ on the right. The 1D unstable curve $W^U_{\bm{z}_u}$ lies within the 2D unstable manifold of $\bm{z}$; similarly, the 1D stable curve $W^S_{\bm{z}_\ell}$  lies within the 2D stable manifold of $\bm{z}$. The two points $\bm{p}_0$ exist on a 1D homoclinic intersection curve, which we also denote $\bm{p}_0$, that is formed by the stable and unstable manifolds of the invariant circle. In Fig.~\ref{Ex3Trel}, the stable piece $W^S_{\bm{z}}[\bm{z},\bm{p}_0]$ is an annulus extending from the invariant circle $\bm{z}$ to the homoclinic curve $\bm{p}_0$. The magenta curve at the lower right is twisted by $90^\circ$ with respect to the red curve at the upper left. Note that this twist is topologically trivial and could be removed by untwisting the magenta curve in the counterclockwise direction. The unstable segment $W^U_{\bm{z}}[\bm{z},\bm{p}_0]$ is twisted in the same way. The two pieces $W^U_{\bm{z}}[\bm{z},\bm{p}_0]$ and $W^S_{\bm{z}}[\bm{z},\bm{p}_0]$ intersect at $\bm{z}$ and $\bm{p}_0$ and form a topological torus, enclosing a finite volume. Here, $\bm{p}_0$ has transition number 1 (index number 0) and forms a primary intersection curve. The enclosed volume is a well defined resonance zone. Thus, all boundary curves will lie entirely in $W^S_{\bm{z}}[\bm{z},\bm{p}_0]$.

In Fig.~\ref{Ex3Trel}, $W^U_{\bm{z}}[\bm{p}_{-1},\bm{p}_0]$ is a fundamental annulus. Its first iterate produces two bridges, $W^U_{\bm{z}}[\bm{p}_0,\bm{q}_0]$,  an exterior bundt cake, and $W^U_{\bm{z}}[\bm{q}_0,\bm{p}_1]$, an interior bundt cake. Iterating $W^U_{\bm{z}}[\bm{q}_0,\bm{p}_1]$ forward produces two new interior bundt cakes, $W^U_{\bm{z}}[\bm{q}_1,\bm{r}_0]$ and $W^U_{\bm{z}}[\bm{s}_0,\bm{p}_2]$, as well as one new exterior bundt cake $W^U_{\bm{z}}[\bm{r}_0,\bm{s}_0]$. This trellis could be untwisted by rotating the lower right portion counterclockwise $90^\circ$ and put in a geometric shape that is rotationally invariant about the $x$-axis. This symmetry implies that the topological dynamics could be reduced to a planar map with 1D invariant manifolds. 

Figure~\ref{Ex3ETP} shows the forward and backward ETPs of Fig.~\ref{Ex3Trel}. Since the system is reversible, the forward and backward ETPs have the same pattern of escape domains. From the ETPs we identify $\textbf{r}_n$ and $\textbf{s}_n$ as the sole pseudoneighbor pair and place our obstruction ring slightly perturbed from $\bm{r}_{n}$ toward $\bm{s}_n$. In Fig.~\ref{Ex3Trel} the ring near $\bm{r}_0$ (purple triangles) prevents the bridge $W^U_{\bm{z}}[\bm{r}_0,\bm{s}_0]$ from being pulled through the stable manifold while its backward iterate prevents the bridge $W^U_{\bm{z}}[\bm{q}_0,\bm{p}_1]$ from being pulled through the stable manifold. 

Since we have a well defined resonance zone, the primary division can be constructed as in Sec.~\ref{Example 1}. Figure~\ref{Ex3BridgeIterates}a shows the primary division from which we construct the inner and outer stable divisions seen in Fig.~\ref{Ex3BridgeIterates}b and Fig.~\ref{Ex3BridgeIterates}c. The green circles are the boundary classes used to specify the bridge classes in Fig.~\ref{Ex3BridgeIterates}d. The bridges in Fig.~\ref{Ex3BridgeIterates}a can be broken into two bridge classes, $\llbracket A,B \rrbracket$ representing the interior bundt cakes and $\llbracket C,D \rrbracket$ representing the exterior bundt cake. $\llbracket A,B \rrbracket$ is the only active bridge class and when iterated forward produces two copies of itself and one copy of $\llbracket C,D \rrbracket$ concatenated together as seen in Fig.~\ref{Ex3BridgeIterates}e. The fact that there is no branching in the graph representing the iterate of $\llbracket A,B \rrbracket$ is a consequence of the fact that this system reduces topologically to a 2D map. Compare Fig.~\ref{Ex3BridgeIterates}e to Fig.~\ref{Ex1BridgeIters}a and Fig.~\ref{Ex2Iterates}d. Figure~\ref{Ex3BridgeIterates}f shows the transition graph, with a single active bridge class, which yields a topological entropy $h_{top} = \ln(2)$. 

In this example, the stable and unstable manifolds of the invariant circle are 2D extensions of the standard 1D manifolds of the complete horseshoe in 2D. This is a consequence of the rotational symmetry mentioned above. If we factor out this rotational symmetry we are left with the standard horseshoe in 2D. Another way to see this is that the curves in the upper left of Fig.~\ref{Ex3Trel} form a 2D horseshoe when viewed as 1D invariant manifolds of a 2D map.

\begin{figure*}
\centering
\includegraphics[width=1.\linewidth]{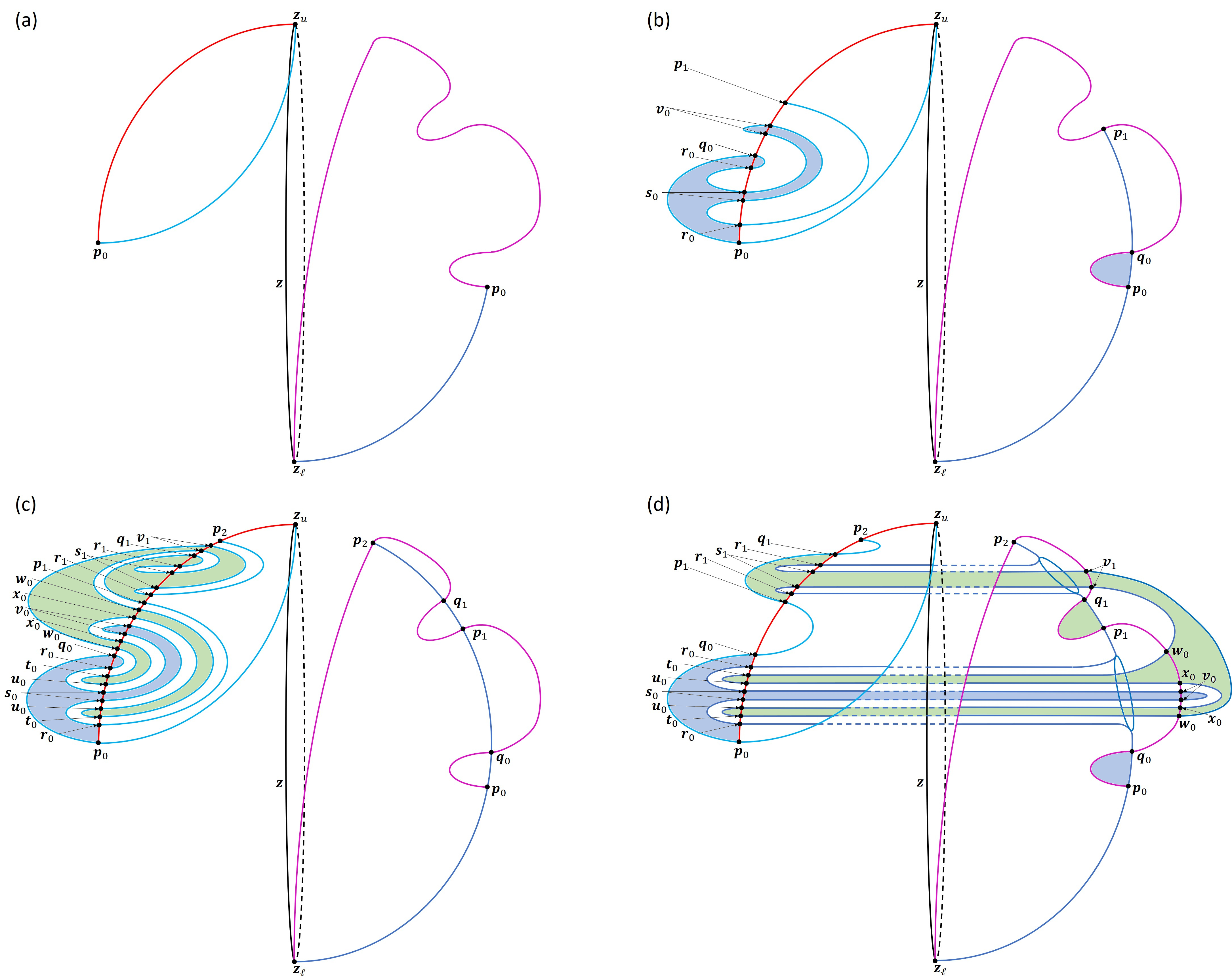}
\caption{(a) The stable and unstable manifolds of the invariant circle up to the primary intersection curve $\bm{p}_0$. (b) The first iterate of the unstable fundamental annulus $W^U_{\bm{z}}[\bm{p}_{-1},\bm{p}_0]$. (c) The second iterate of the unstable fundamental annulus. (d) A trellis topologically identical to (c) but geometrically distorted to be similar to Example 2.}
\label{Ex4Build}
\end{figure*}

\section{Example 4}
\label{Example 4}

\begin{figure*}
\centering
\includegraphics[width=1.\linewidth]{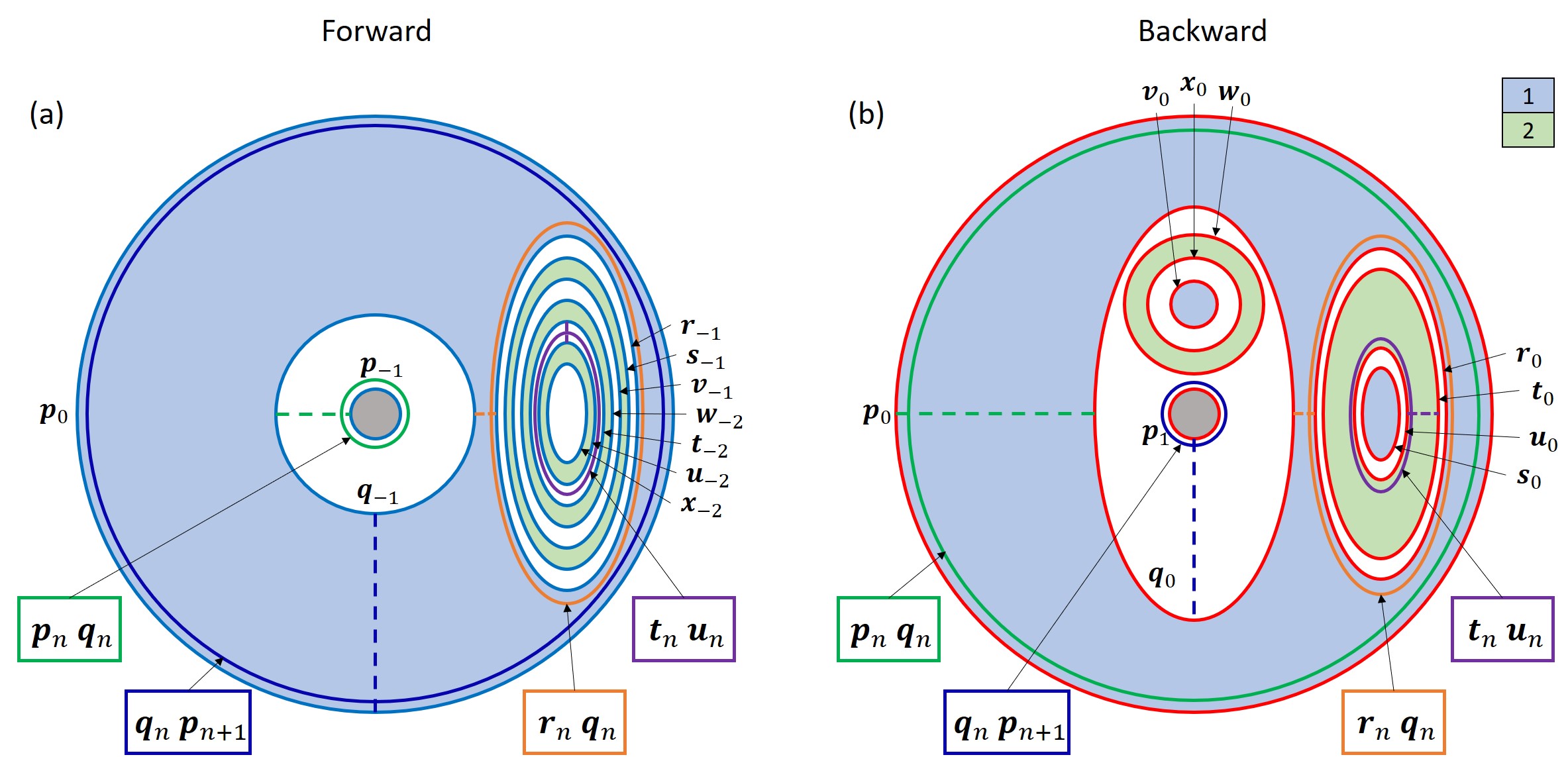}
\caption{The (a) forward and (b) backward capture-time plots (CTPs). Shaded domains represent regions of the fundamental annulus that re-enter the resonance zone after 1 (blue) or 2 (green) iterates. CTPs are used to identify pseudoneighbor pairs in the same way ETPs are used.}
\label{Ex4CTP}
\end{figure*}

We analyzed Example 2 using the 2D stable and unstable manifolds of the fixed points. In Example 3 we showed that we can get a well defined resonance zone if we use the 2D stable and unstable manifolds of the invariant circle. In Fig.~\ref{Ex4Build} we construct an example with a well defined resonance zone based on the manifolds attached to the invariant circle, like Example 3, but incorporating the topological forcing from Example 2. To construct this example, we first suppose that the stable and unstable manifolds of the invariant circle intersect at the primary homoclinic intersection curve $\bm{p}_0$, as seen in Fig.~\ref{Ex4Build}a. As in Example 3 the 2D stable manifold of the upper fixed point intersects the 1D unstable manifold of the upper fixed point at the leftmost point labeled $\bm{p}_0$. In the lower right the 2D unstable manifold of the lower fixed point intersects the 1D stable manifold of the lower fixed point at the rightmost point labeled $\bm{p}_0$. The resulting 2D manifolds form a toroidal resonance zone like Example 3.

In the simplest case, the bridge $W^U_{\bm{z}}[\bm{z},\bm{p}_0]$ iterates forward to form an exterior bridge $W^U_{\bm{z}}[\bm{p}_0,\bm{q}_0]$ and interior bridge $W^U_{\bm{z}}[\bm{q}_0,\bm{p}_1]$. 

Figure~\ref{Ex4Build}b modifies this simple dynamics to match Example 2. We take the bundt cake $W^U_{\bm{z}}[\bm{q}_0,\bm{p}_1]$ and push a small piece of it over to intersect $W^S_{\bm{z}}[\bm{p}_0,\bm{q}_0]$ as shown on the left of Fig.~\ref{Ex4Build}b. This turns the bundt cake $W^U_{\bm{z}}[\bm{q}_0,\bm{p}_1]$ into the tridge $W^U_{\bm{z}}[\bm{q}_0,\bm{r}_0,\bm{p}_1]$ in Fig.~\ref{Ex4Build}b. The remaining piece of $W^U_{\bm{z}}$ attached to $\bm{r}_0$ is pulled back through the stable submanifold forming the exterior bundt cake $W^U_{\bm{z}}[\bm{r}_0,\bm{s}_0]$. We next pull the manifold back through the stable manifold $W^S_{\bm{z}}[\bm{q}_0,\bm{p}_1]$, forming the macaroni $W^U_{\bm{z}}[\bm{s}_0,\bm{v}_0]$, and terminating in an exterior cap $W^U_{\bm{z}}[\bm{v}_0]$. The intersections $\bm{r}_0$ and $\bm{s}_0$ in Example 4 are topologically equivalent to the same intersections in Example 2.  Here we have an additional intersection $\bm{v}_0$ which cuts the cap $W^U_{\bm{z}}[\bm{s}_0]$ in Example 2 into the concatenation of the macaroni $W^U_{\bm{z}}[\bm{s}_0,\bm{v}_0]$ and the cap $W^U_{\bm{z}}[\bm{v}_0]$.

Figure~\ref{Ex4Build}c shows the forward iterate of $W^U_{\bm{z}}[\bm{s}_0,\bm{v}_0]$ and $W^U_{\bm{z}}[\bm{v}_0]$. $W^U_{\bm{z}}[\bm{s}_0,\bm{v}_0]$ maps inertly forward to $W^U_{\bm{z}}[\bm{s}_1,\bm{v}_1]$. The forward iterate of the cap $W^U_{\bm{z}}[\bm{v}_0]$ creates a new intersection $\bm{x}_0$. This requires the forward iterate to be a concatenation of the exterior macaroni $W^U_{\bm{z}}[\bm{v}_1,\bm{w}_0]$, the interior macaroni $W^U_{\bm{z}}[\bm{w}_0,\bm{t}_0]$, the exterior bundt cake $W^U_{\bm{z}}[\bm{t}_0,\bm{u}_0]$, a second interior macaroni $W^U_{\bm{z}}[\bm{u}_0,\bm{x}_0]$, and the exterior cap $W^U_{\bm{z}}[\bm{x}_0]$. In total four new intersections are created: $\bm{w}_0$, $\bm{t}_0$, $\bm{u}_0$, and $\bm{x}_0$. In Example 2, $\bm{t}_0$ and $\bm{u}_0$ are formed by the forward iterate of $W^U_{\bm{z}}[\bm{s}_0]$ (Fig.~\ref{Ex2Build}d) and they have the same topological relationship as exhibited in Fig.~\ref{Ex4Build}c. The curves $\bm{w}_0$ and $\bm{x}_0$ do not occur in Example 2.

\begin{figure*}
\centering
\includegraphics[width=1\linewidth]{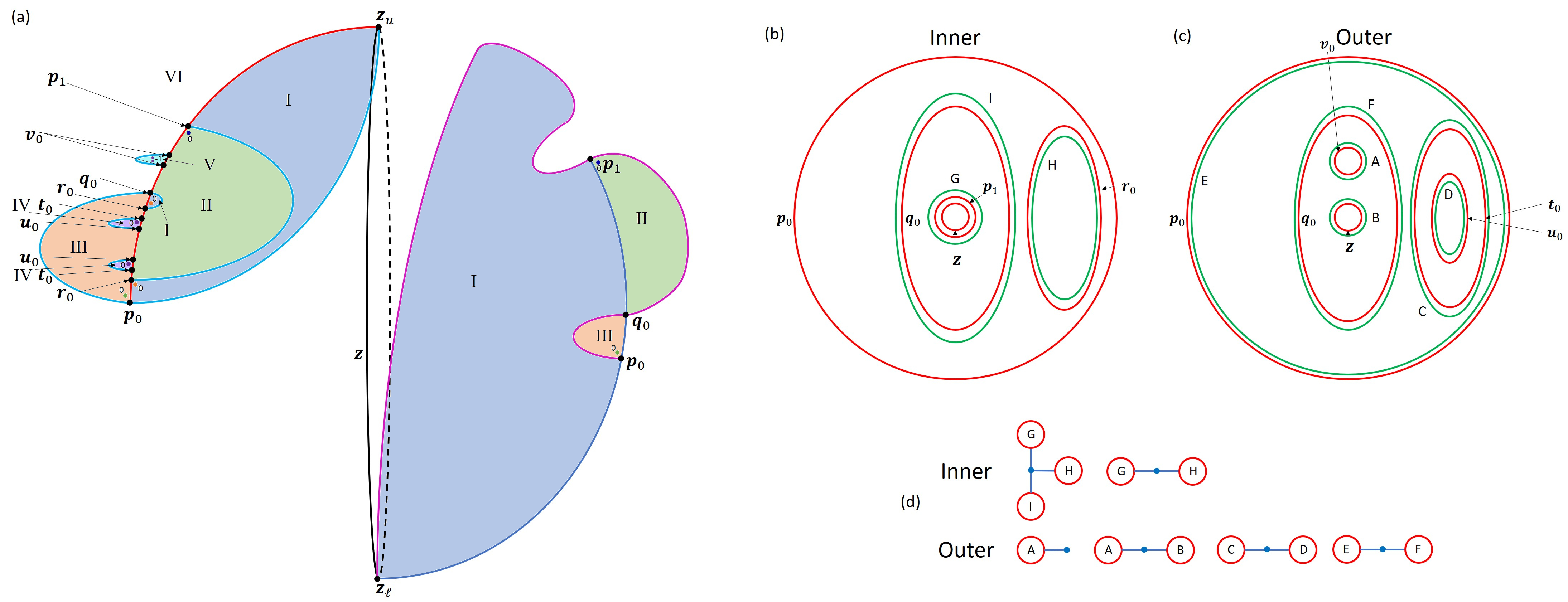}
\caption{(a) The primary division (b) Inner and (c) outer stable primary division. (d) The inner and outer bridge classes for Example 4.}
\label{Ex4PrimaryDiv}
\end{figure*}

Figure~\ref{Ex4Build}d modifies the geometry of the trellis in Fig.~\ref{Ex4Build}c but keeps the topology the same. To accomplish this we start by ``sliding" the intersection curves $\bm{v}_0$, $\bm{x}_0$, and $\bm{w}_0$ along $W^S_{\bm{z}}[\bm{q}_0,\bm{p}_1]$ from the left-hand side of Fig.~\ref{Ex4Build}c to the right-hand side of Fig.~\ref{Ex4Build}d. Similarly, we do the same for $\bm{v}_1$ on $W^S_{\bm{z}}[\bm{q}_1,\bm{p}_2]$. We  adjust the geometry of the tridge $W^U_{\bm{z}}[\bm{q}_0,\bm{r}_0,\bm{p}_1]$ so that it has a tube connecting the unstable manifold on the right-hand side to the curve $\bm{r}_0$ on the left-hand side as in Fig.~\ref{Ex4Build}d. We have drawn Fig.~\ref{Ex4Build}d so that this tube passes behind the ``hole" of the torus that forms the resonance zone. 

The dynamics in Fig.~\ref{Ex4Build}c and Fig.~\ref{Ex4Build}d are topologically identical, but Fig.~\ref{Ex4Build}d now geometrically resembles Fig.~\ref{Ex2Build}d in Example 2. All of the intersection curves between the stable and unstable manifolds in Fig.~\ref{Ex2Build}d of Example 2 are present in Fig.~\ref{Ex4Build}d. However, Fig.~\ref{Ex4Build}d contains extra intersection curves visible on the right-hand side. All of these intersections curves include points on the 1D stable manifold. Hence these curves would always be incomplete in an analysis based solely on the 2D manifolds of the fixed points $\bm{z}_u$ and $\bm{z}_\ell$, as was done in Example 2. 

Note that part of the iterate of the exterior bridge $W^U_{\bm{z}}[\bm{v}_0]$ is inside the resonance zone. This is a case of the recapture of a piece of the unstable manifold that has already escaped. Such recapture is absent from Examples 1 and 3. Additionally, none of the interior bridges of the trellis escape except for the primary bridge $W^U_{\bm{z}}[\bm{z},\bm{p}_0]$. In order to represent the structure of the homoclinic intersections, we use capture-time plots (CTP) instead of escape time plots. See Fig.~\ref{Ex4CTP}. From Fig.~\ref{Ex4CTP} we identify four pseudoneighbor pairs, $[\bm{p}_n, \bm{q}_n]$, $[\bm{q}_n,\bm{r}_n]$, $[\bm{t}_n, \bm{u}_n]$, and $[\bm{q}_n,\bm{p}_{n+1}]$. We place the appropriate obstruction rings in the CTPs: one perturbed from $\bm{q}_n$ toward $\bm{p}_n$ (green), one perturbed from $\bm{u}_n$ toward $\bm{t}_n$ (purple), one perturbed from $\bm{r}_n$ toward $\bm{q}_n$ (orange), and finally one perturbed from $\bm{p}_{n+1}$ toward $\bm{q}_n$ (dark blue). 

\begin{figure*}
\centering
\includegraphics[width=1.\linewidth]{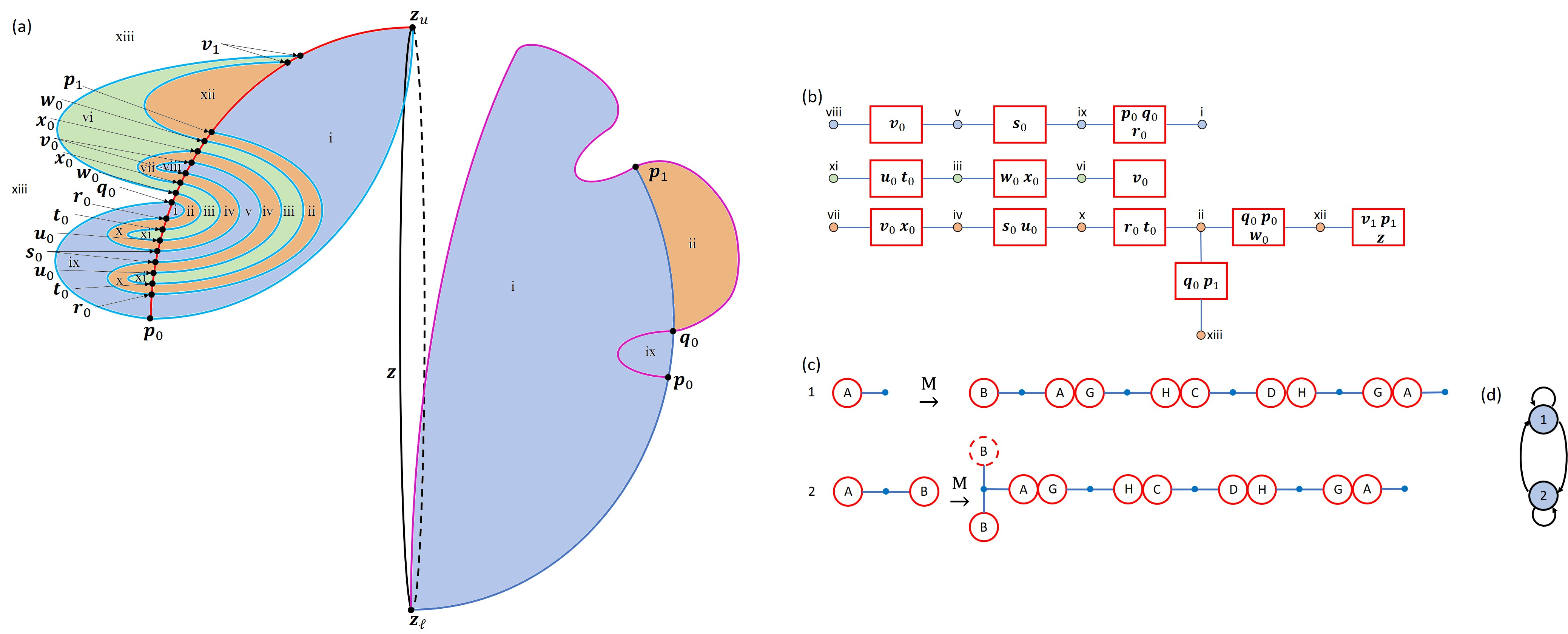}
\caption{(a) The secondary division. (b) The connection graph that shows regions connected across the fundamental stable annulus in (a) (c) The forward iterate of the two active bridge classes. (d) The transition graph for the active bridge classes of Example 4. The transition graph is identical to the transition graph of Example 2. Additionally the iterates of the two bridge classes can be reduced to the iterates in Example 2.}
\label{Ex4SecondaryDiv}
\end{figure*}

Using the information of Fig.~\ref{Ex4CTP}, we construct the primary division in Fig~\ref{Ex4PrimaryDiv}a. The primary division has two interior bridges $W^U_{\bm{z}}[\bm{z},\bm{p}_0]$ and $W^U_{\bm{z}}[\bm{q}_0,\bm{r}_0,\bm{p}_1]$ and three exterior bridges, $W^U_{\bm{z}}[\bm{p}_0,\bm{q}_0]$, $W^U_{\bm{z}}[\bm{t}_0,\bm{u}_0]$ and $W^U_{\bm{z}}[\bm{v}_0]$. The interior bridge $W^U_{\bm{z}}[\bm{q}_0,\bm{r}_0,\bm{p}_1]$ and the first two exterior bridges are inert bridges included by Rule 3 in Sec.~\ref{Example 1}; the primary bridge $W^U_{\bm{z}}[\bm{z},\bm{p}_0]$ and $W^U_{\bm{z}}[\bm{r}_0]$ are included by Rule 2. Using Fig.~\ref{Ex4PrimaryDiv}a we construct the inner and outer stable divisions of $W^S_{\bm{z}}[\bm{z},\bm{p}_0]$ in Fig.~\ref{Ex4PrimaryDiv}b and Fig.~\ref{Ex4PrimaryDiv}c. The inner stable division contains the boundaries for the tridge $W^U_{\bm{z}}[\bm{q}_0,\bm{r}_0,\bm{p}_1]$ and the primary bridge $W^U_{\bm{z}}[\bm{z},\bm{p}_0]$. By inspection of the initial trellis all inner boundary classes are of types $G$ and $H$ shown in Fig.~\ref{Ex4PrimaryDiv}b. The outer stable division is constructed similarly. The outer bridges have boundary classes $A$-$F$ as shown in Fig.~\ref{Ex4PrimaryDiv}c. Examination of Fig.~\ref{Ex4Build}c allows us to identify the bridge classes in Fig~\ref{Ex4PrimaryDiv}d. We have two interior bridge classes, the macaroni $\llbracket G,H \rrbracket$, of which $W^U_{\bm{z}}[\bm{w}_0,\bm{t}_0]$ is a member, and the tridge $\llbracket G,G,H \rrbracket$, of which $W^U_{\bm{z}}[\bm{q}_0,\bm{r}_0,\bm{p}_1]$ is a member. Both of these bridge classes are inert. On the exterior we have inert bridge classes $\llbracket C,D \rrbracket$, of which $W^U_{\bm{z}}[\bm{t}_0,\bm{u}_0] $ is a member, and $\llbracket F,G \rrbracket$, of which $W^U_{\bm{z}}[\bm{p}_0,\bm{q}_0]$ is a member. Finally, we have the active bridge classes $\llbracket A \rrbracket$, including the cap $W^U_{\bm{z}}[\bm{v}_0] $, and $\llbracket A,B \rrbracket$, including the macaroni $W^U_{\bm{z}}[\bm{v}_1,\bm{w}_0]$.

In Fig.~\ref{Ex4SecondaryDiv}a we construct the secondary division following the method outlined in Sec.~\ref{Example 1}. From the secondary division, we construct the connection graph in Fig.~\ref{Ex4SecondaryDiv}b. Together these are used to derive the dynamics of the active bridge classes in Fig.~\ref{Ex4SecondaryDiv}c as done in Example 1. The active bridge class $\llbracket A \rrbracket$ produces one copy of itself and the other active class $\llbracket A,B \rrbracket$. Similarly the active bridge class $\llbracket A,B \rrbracket$ produces one copy of the active classes $\llbracket A \rrbracket$ and $\llbracket A,B,B \rrbracket$. Just like in Example 2 we identify $\llbracket A,B,B \rrbracket$ with $\llbracket A,B \rrbracket$. Fig.~\ref{Ex4SecondaryDiv}d shows the transition graph for the active bridge classes. The transition graph is identical to the transition graph in Example 2 where each active bridge class produces a copy of itself and the other active bridge class. 

Comparing the dynamics between Examples 2 and 4, we see only one point difference,  the addition of the inert macaroni $\llbracket G,H \rrbracket$ in the bridge dynamics of Fig.~\ref{Ex4SecondaryDiv}c relative to Fig.~\ref{Ex2Iterates}. This macaroni is the result of the stable manifold cutting across the cap $W^U_{\bm{z}}[\bm{s}_0]$ and the macaroni $W^U_{\bm{z}}[\bm{s}_1,\bm{t}_0]$ in Example 2. In Example 4 this means that the equivalent to $W^U_{\bm{z}}[\bm{s}_0]$ in Example 2 is $W^U_{\bm{z}}[\bm{s}_0,\bm{v}_0]$ concatenated with $W^U_{\bm{z}}[\bm{v}_0]$. In essence this means the bridge class $\llbracket A \rrbracket$ in Example 2 is the concatenation of bridge classes $\llbracket A \rrbracket$ and $\llbracket G,H \rrbracket$ in Example 4. Additionally $\llbracket A,B \rrbracket$ in Example 2 is the concatenation of $\llbracket A,B \rrbracket$, $\llbracket G,H \rrbracket$, and $\llbracket A \rrbracket$ in Example 4. If we make those substitutions in the bridge dynamics of Example 4, we get bridge dynamics identical to Example 2.

\begin{figure*}
\centering
\includegraphics[width=.9\linewidth]{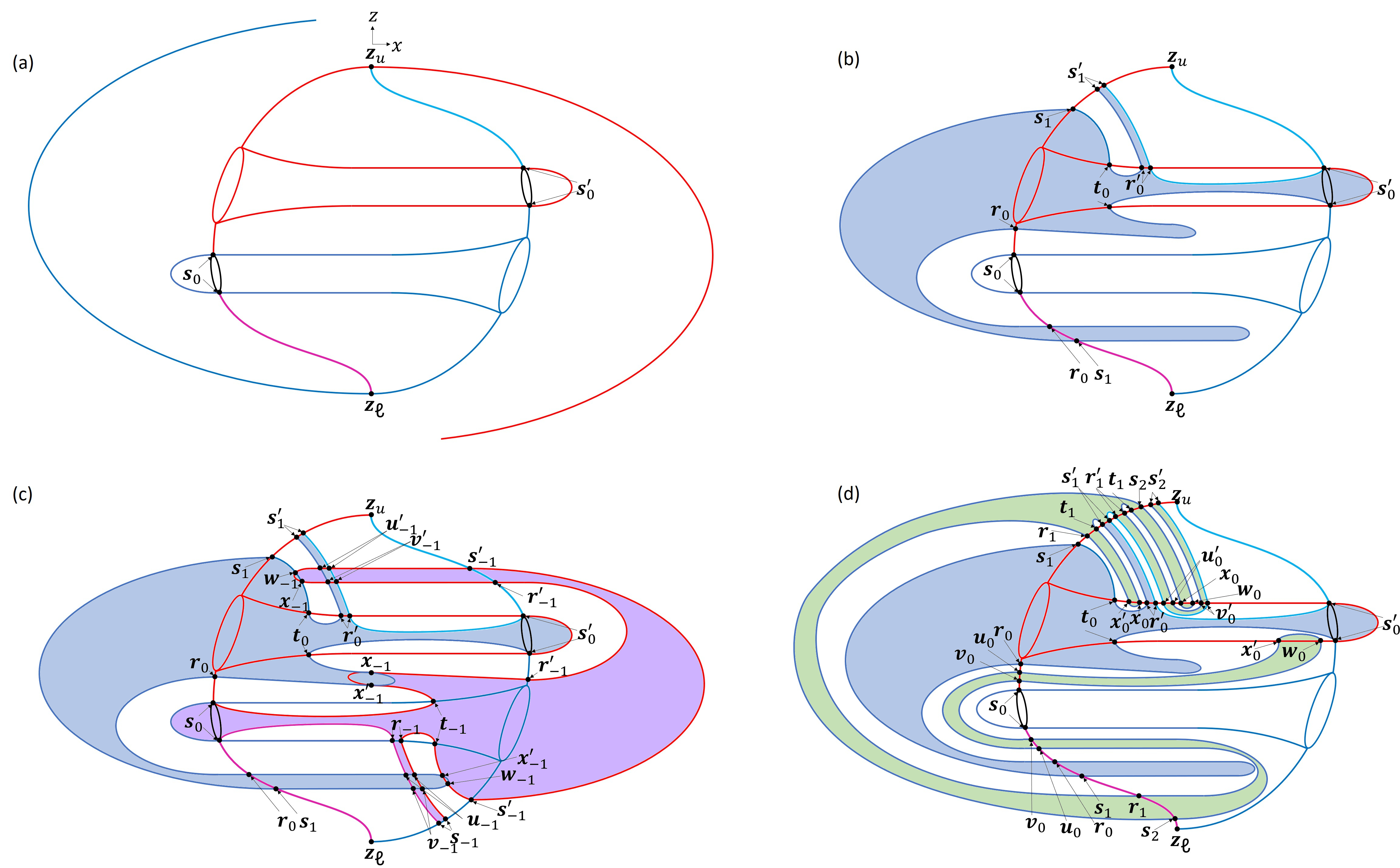}
\caption{(a) The stable and unstable manifolds of the invariant circle up to the intersection curves $\bm{s}_0$ and $\bm{s}^{'}_0$. (b) The first forward iterate of the unstable fundamental annulus $W^U_{\bm{z}}(\bm{s}_{-1},\bm{s}_0]$. (c) The first forward iterate of the unstable fundamental annulus and the first backward iterate of the stable fundamental annulus $W^S_{\bm{z}}[\bm{p}^{'}_{0},\bm{p}^{'}_1)$. (d) The trellis in (c) iterated forward once.}
\label{Ex5Build}
\end{figure*}

\section{Example 5}
\label{Example 5}

Here we analyze the culminating trellis. It lacks an equatorial intersection (like Examples 2-4), its dynamics are fully 3D (like Examples 1, 2 and 4), and the primary bridge class is recurrent (like Examples 1 and 3). Also like Example 3, the system is reversible with $S(x,y,z) = (-x,y,-z)$. Figure~\ref{Ex5Build}a shows the stable manifold of the invariant circle up to the homoclinic curve $\bm{s}_0$ and the unstable manifold up to $\bm{s}'_0$. The unstable manifold reaches across to intersect the stable manifold at $\bm{s}_0$. By symmetry the stable manifold reaches across to intersect the unstable manifold at $\bm{s}'_0$. The union of $W^S_{\bm{z}}[\bm{z},\bm{s}_0,\bm{s}'_0]$ and $W^U_{\bm{z}}[\bm{z},\bm{s}_0,\bm{s}'_0]$ is a topological genus-2 torus which bounds a well defined resonance zone. Note that neither $\bm{s}_0$ nor $\bm{s}'_0$ is a primary intersection curve, according to Sec.~\ref{Invariant manifolds attached to the invariant circle}. As a pair, however, $\bm{s}_0$ and $\bm{s}'_0$ play an analogous role to a single primary intersection curve, since $W^U_{\bm{z}}[\bm{z},\bm{s}_0,\bm{s}'_0]$ and $W^S_{\bm{z}}[\bm{z},\bm{s}_0,\bm{s}'_0]$ only intersect at their boundaries.

We use the intersection curve $\bm{s}'_0$, which is a proper loop on $W^U_{\bm{z}}$, to define the unstable fundamental annulus $W^U_{\bm{z}}(\bm{s}'_{-1},\bm{s}'_0]$. We similarly use $\bm{s}_0$ to define the stable fundamental annulus $W^S_{\bm{z}}[\bm{s}_0,\bm{s}_1)$. We specify that the first iterate of the unstable fundamental annulus produces Fig.~\ref{Ex5Build}b.  This iterate produces an exterior tridge $W^U_{\bm{z}}[\bm{s}'_0,\bm{r}_0,\bm{t}_0]$, an interior tridge $W^U_{\bm{z}}[\bm{t}_0,\bm{r}_0,\bm{s}_1]$, an interior macaroni $W^U_{\bm{z}}[\bm{s}'_1,\bm{r}'_0]$ and two exterior caps $W^U_{\bm{z}}[\bm{s}_1]$ and $W^U_{\bm{z}}[\bm{r}_0]$.  Figure~\ref{Ex5Build}c shows the first backward iterate of the stable fundamental annulus $W^S_{\bm{z}}[\bm{s}_0,\bm{s}_1)$. This backward iterate is obtained by time-reversal symmetry. Notice that the $\bm{r}_0$ and $\bm{r}'_{-1}$ curves are related by the symmetry operator $S$ and therefore $\bm{r}_n$ and $\bm{r}'_n$ are time-reversal-symmetry partners. The curves $\bm{t}_0$ and $\bm{t}_{-1}$ are also related by the symmetry operator $S$ so that the orbit $\bm{t}_n$ is its own time-reversal partner. The primed orbits are always the time-reversal partners of unprimed orbits.

Figure~\ref{Ex5Build}d is obtained by iterating the trellis in Fig.~\ref{Ex5Build}c forward once. The stable component of the trellis is the same as in Fig~\ref{Ex5Build}b. The unstable component of the trellis contains the second iterate of the unstable fundamental annulus $W^U_{\bm{z}}(\bm{s}_{-1},\bm{s}_0]$. 

\begin{figure*}
\centering
\includegraphics[width=1\linewidth]{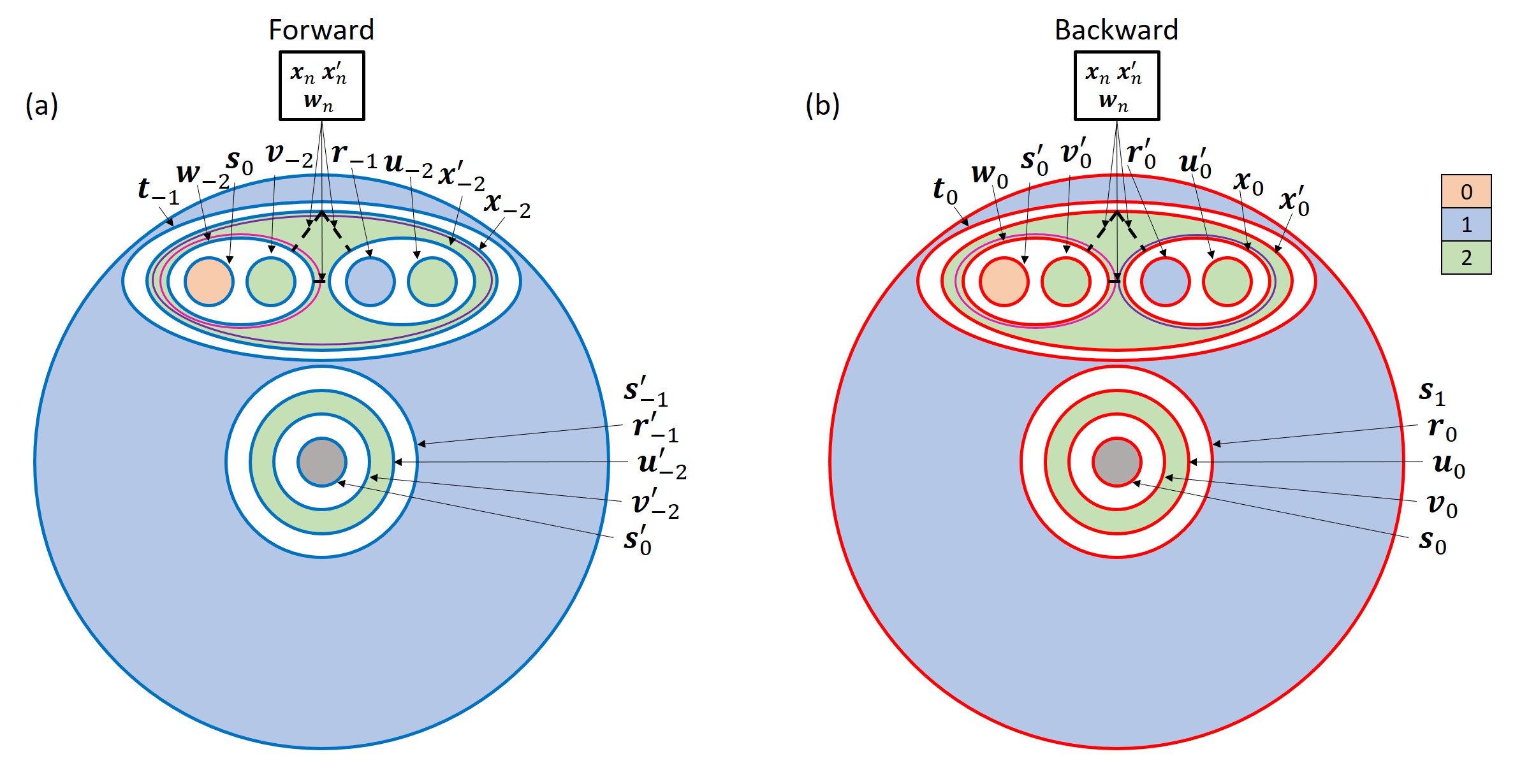}
\caption{The (a) forward and (b) backward ETP for Example 5. The topologies of the ETPs are identical due to time-reversal symmetry. We consider the domains surrounded by $\bm{s}_0$ and $\bm{s}_0^{'}$ to be outside the resonance zone and thus to ``escape" at the zeroth iterate. Every pair formed from the three curves $\bm{x}_n$, $\bm{x}'_n$ and $\bm{w}_n$ is a pseudoneighbor pair. Together they form a ``pseudoneighbor triplet". The triplet can be broken up by two obstruction rings, one perturbed from $\bm{x}_n$ toward $\bm{x}_n^{'}$ and $\bm{w}_n$ (purple), and one perturbed from $\bm{w}_n$ toward $\bm{x}_n^{'}$ and $\bm{x}_n$ (orange). }
\label{Ex5ETP}
\end{figure*}

\begin{figure*}
\centering
\includegraphics[width=1.\linewidth]{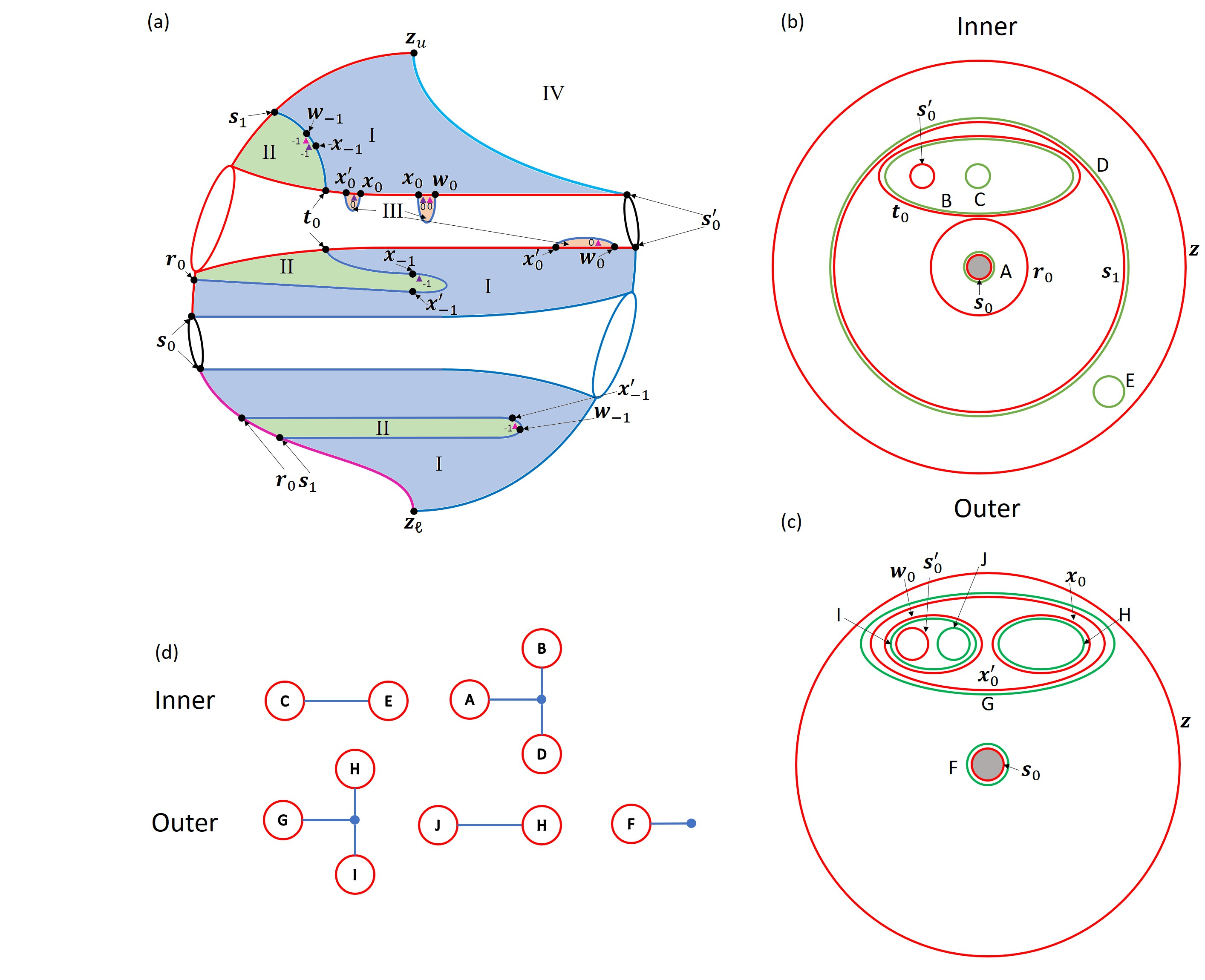}
\caption{(a) The primary division of phase space. (b) The inner and (c) outer stable divisions. (d) The inner and outer bridge classes for Example 5.}
\label{Ex5PrimaryDivision}
\end{figure*}

Figure~\ref{Ex5ETP} contains the forward and backward ETPs. We see the time-reversal symmetry of the forward and backward ETPs because one can be converted to the other by swapping primed and unprimed intersection curves, noting that $\bm{t}_n$ and $\bm{w}_n$ are their own symmetry partners. The escape domains bound by $\bm{s}_0$ in the forward ETP and $\bm{s'}_0$ in the backward ETP escape the resonance zone on the zeroth iterate. This is a result of the caps $W^U_{\bm{z}}[\bm{s}_0]$ and $ W^S_{\bm{z}}[\bm{s'}_0]$ existing outside the resonance zone in Fig.~\ref{Ex5Build}a. We identify pseudoneighbors by looking for curves whose iterates are adjacent in both the forward and backward ETPs. This example has a pseudoneighbor triplet [$\bm{w}_n$, $\bm{x}_n$, $\bm{x'}_n$], i.e., $[\bm{w}_n,\bm{x}_n]$, $[\bm{w}_n,\bm{x'}_n]$ and $[\bm{x}_n,\bm{x'}_n]$ are each pseudoneighbor pairs. For a pseudoneighbor triplet we only need two obstruction rings, one around $\bm{w}_n$ perturbed toward $\bm{x}_n$ (magenta) and one around $\bm{x}_n$ perturbed toward $\bm{x'}_n$ (purple). These two obstruction rings act to uphold the exterior tridge $W^U_{\bm{z}}[\bm{w}_0,\bm{x}_0,\bm{x'}_0]$. See Fig.~\ref{Ex5PrimaryDivision}a.

We construct the primary division in Fig.~\ref{Ex5PrimaryDivision}a using the rules in Sec~\ref{Example 1}. We include the stable portion of the trellis from Rule 1. We include the bridge $W^U_{\bm{z}}[\bm{w}_0,\bm{x}_0,\bm{x'}_0]$ from Rule 2 and the bridges $W^U_{\bm{z}}[\bm{t}_0,\bm{r}_0,\bm{s}_1]$ and $W^U_{\bm{z}}[\bm{z},\bm{s}_0,\bm{s}'_0]$ from Rule 3. From the primary division we obtain the inner and outer stable divisions in Fig.~\ref{Ex5PrimaryDivision}b and Fig.~\ref{Ex5PrimaryDivision}c. The system has two interior bridge classes, the primary tridge $\llbracket A,B,D \rrbracket$ and the macaroni $\llbracket C,E \rrbracket$ shown in Fig~\ref{Ex5PrimaryDivision}d. Figure~\ref{Ex5PrimaryDivision}d contains five bridges classes each of which is represented by bridges present in the trellis in Fig.~\ref{Ex5Build}d. As we will see below, these are the five bridge classes necessary to specify the active bridge dynamics.

\begin{figure*}
\centering
\includegraphics[width=1\linewidth]{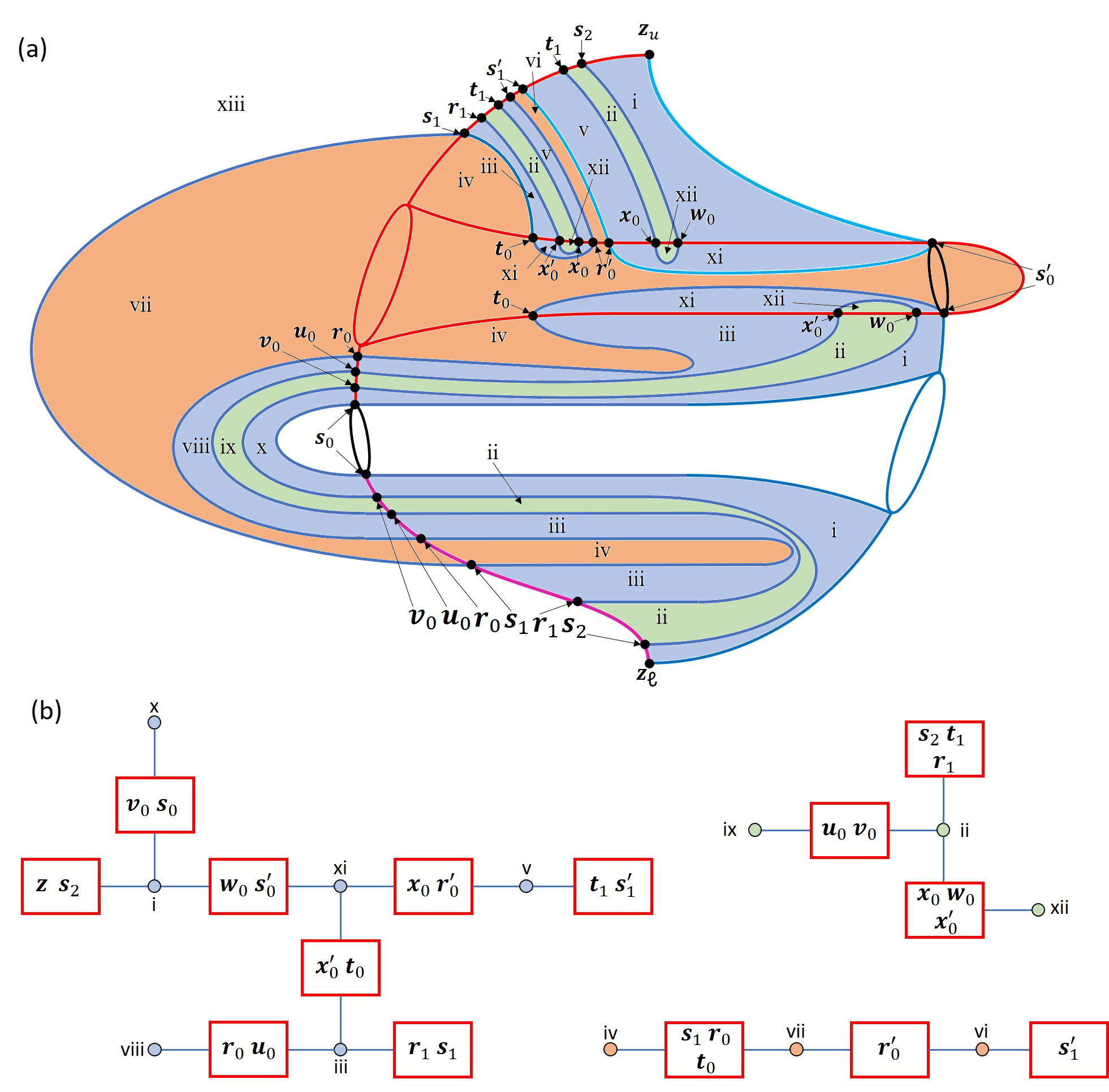}
\caption{(a) The secondary division of phase space for Example 5. (b) The connection graph for the regions of the secondary division. }
\label{Ex5SecondaryDiv}
\end{figure*}

We construct the secondary division in Fig.~\ref{Ex5SecondaryDiv}a based on the rules outlined in Sec.~\ref{Example 1}. The unstable portion of the secondary division contains the iterate of the primary tridge $W^U_{\bm{z}}[\bm{z},\bm{s}_0,\bm{s'}_0]$ and the iterate of $W^U_{\bm{z}}[\bm{t}_0,\bm{r}_0,\bm{s}_1]$ based on Rule 2. The secondary division contains thirteen regions forming three connected components: blue, green and orange as seen in the connection graph of Fig.~\ref{Ex5SecondaryDiv}b.

\begin{figure*}
\centering
\includegraphics[width=1\linewidth]{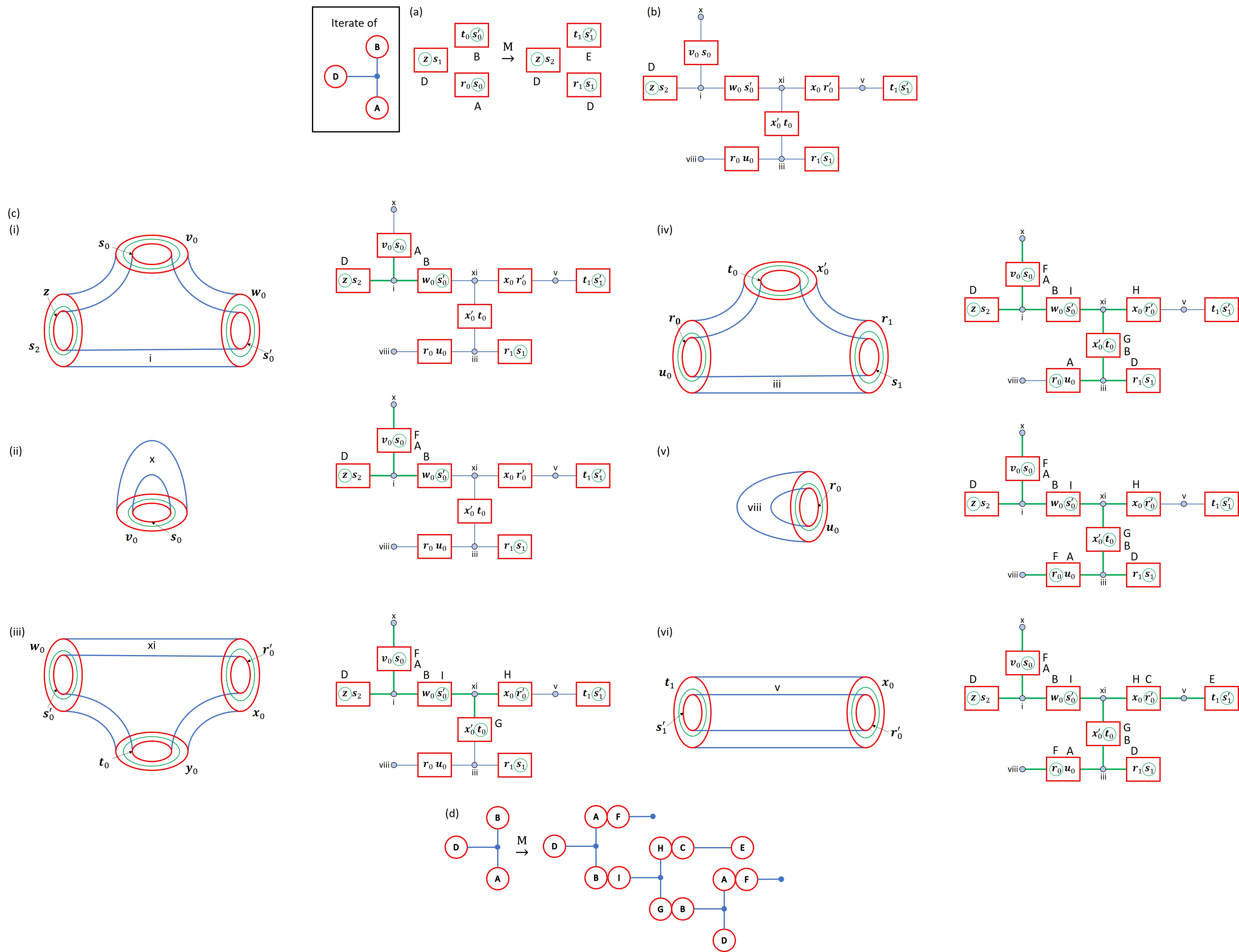}
\caption{ A step-by-step illustration of the process to construct the forward iterate of $\llbracket A,B,D \rrbracket$. (a) We iterate the boundary classes that make up the bridge class to identify where on the secondary division they occur. (b) The component of the connection graph that the forward iterates of the boundary classes lie within. (c) A step-by-step process of identifying the forward iterate of $\llbracket A,B,D \rrbracket$. Each of the regions the forward iterate lies within is shown on the left while the connections being made are shown on the right. Boundary classes are labeled adjacent each box. (d) The concatenation of bridge classes that make up the forward iterate of $\llbracket A,B,D \rrbracket$. }
\label{Ex5ABD_iterate}
\end{figure*}

\begin{figure*}
\centering
\includegraphics[height=1\linewidth]{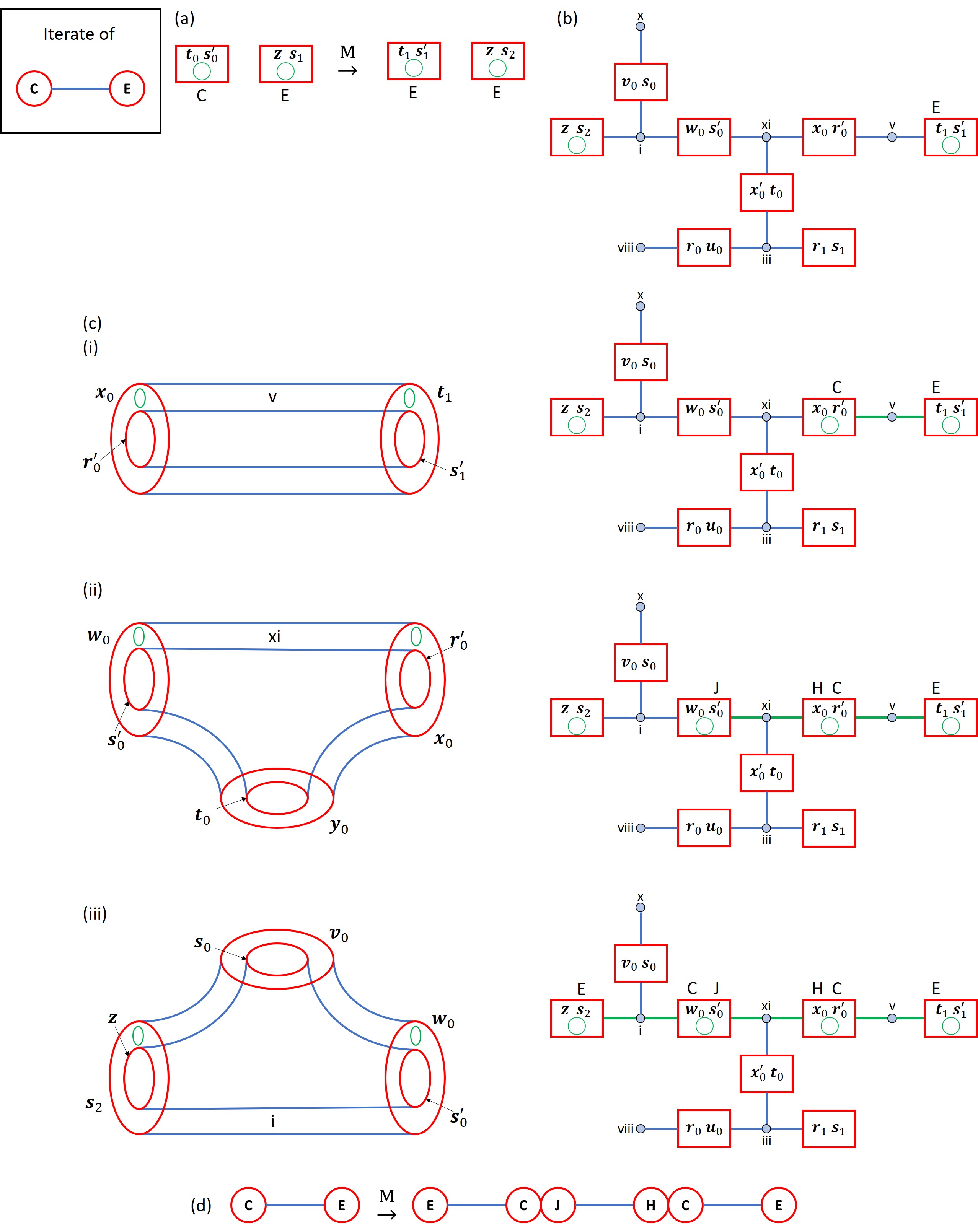}
\caption{ A step-by-step illustration of the process to identify the forward iterate of $\llbracket C,E \rrbracket$. (a) The boundary classes that make up the bridge class and the locations of their forward iterates. Note that in this case the boundary classes do not surround any homoclinic intersections in either the primary or secondary stable divisions. (b)  The component of the connection graph that the forward iterates of the boundary classes lie within. (c) A step-by-step process of constructing the forward iterate of $\llbracket C,E \rrbracket$. (d) The concatenation of the bridge classes that make up the forward iterate of $\llbracket C,E \rrbracket$. }
\label{Ex5CE_iterate}
\end{figure*}

We use the same process as in Example 1 to compute the forward iterates of the interior bridge classes in Fig~\ref{Ex5PrimaryDivision}d. It is easily seen that the three exterior bridge classes are inert. Figure~\ref{Ex5ABD_iterate} computes the iterate of $\llbracket A,B,D \rrbracket$. The iterate of $\llbracket A,B,D \rrbracket$ contains two copies of itself and one copy of the active interior macaroni $\llbracket C,E \rrbracket$. Note that the primary bridge $W^U_{\bm{z}}[\bm{z},\bm{s}_0,\bm{s'}_0]$ belongs to the bridge class $\llbracket A,B,D \rrbracket$ and hence produces copies of itself upon iteration. Figure~\ref{Ex5CE_iterate} computes the iterate of $\llbracket C,E \rrbracket$, which contains three concatenated macaronis: two copies of $\llbracket C,E \rrbracket$ itself and one inert exterior macaroni.

\begin{figure}
\centering
\includegraphics[width=1\linewidth]{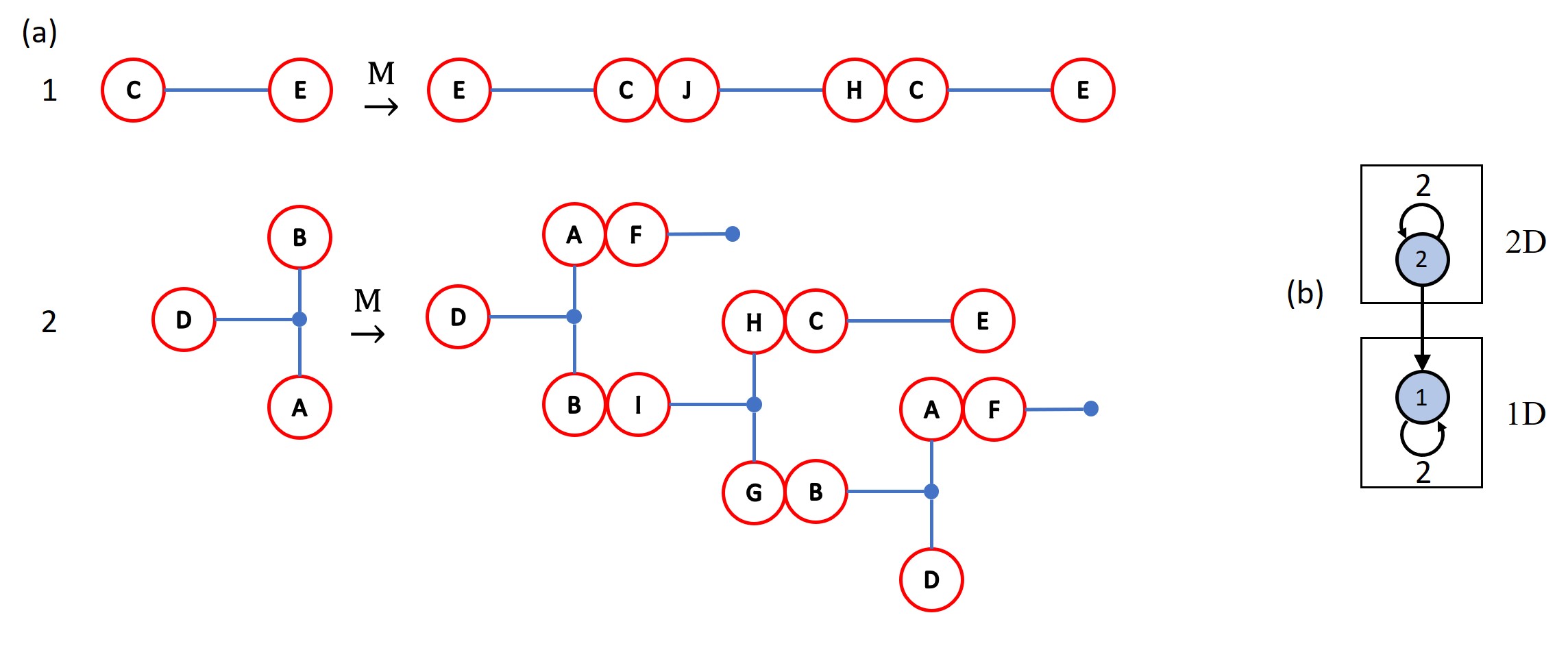}
\caption{ (a) The two active bridge classes in Example 5 and the concatenation of bridge classes that make up their iterates. (b) The transition graph of the two active bridge classes. Note that the forward iterate of bridge class 2 produces copies of itself and bridge class 1 while the forward iterate of bridge class 1 produces only copies of itself. This is an example of a case where the transition graph decomposes into two strongly connected components, one representing 2D dynamics and one representing 1D dynamics.}
\label{Ex5Iterates}
\end{figure}

Figure~\ref{Ex5Iterates}a summarizes the iterates of the two active bridge classes $\llbracket A,B,D \rrbracket$ and $\llbracket C,E \rrbracket$. Note that the dynamics are fully 3-dimensional because there is branching in the forward iterate of $\llbracket A,B,D \rrbracket$ in Fig.~\ref{Ex5Iterates}a. This iterate represents 2-dimensional stretching that is not possible in a 2D map. On the other hand the stretching seen in the forward iterate of $\llbracket C,E \rrbracket$ is 1-dimensional because it contains no branching and is essentially the same stretching seen in 2D maps. We construct the transition graph in Fig.~\ref{Ex5Iterates}b based on iterates of the two active classes. Bridge class 2, i.e. $\llbracket A,B,D \rrbracket$, produces two copies of itself and one of bridge class 1, i.e. $\llbracket C,E \rrbracket$, whereas bridge class 1 produces only two copies of itself. Thus bridge class 2 produces class 1 but not visa versa. This is an example of the phenomenon seen in Ref.~\cite{Smith17} where it was demonstrated that the full transition graph decomposes into two strongly connected components; a strongly connected component is one in which each vertex has a directed path to every other vertex in the component. One strongly connected component corresponds to 2D stretching, and the other strongly connected component corresponds to 1D stretching. The 1D connected component can be reached from the 2D connected component but not visa versa. In the present example, each of these connected components consists of a single vertex. Furthermore both the 1D and 2D connected components produce stretching rates of $\ln{2}$. The topological entropy is the maximum of these two; thus $h = \ln{2}$. In general, the 2D and 1D stretching rates need not be equal. However, in cases with time-reversal-symmetry, like this example, it has been conjectured that they must be equal~\cite{Smith17}.

\section{Conclusion}
\label{Conclusion}

Through a series of topologically specified examples we have demonstrated how HLD can be used to extract symbolic dynamics for systems whose 2D stable and unstable manifolds attached to fixed points do not create a well defined resonance zone. Specifically, we showed in Example 2 that a well defined resonance zone is not strictly necessary to extract some amount of topological forcing. In the remaining three examples we used the 2D stable and unstable manifolds of the invariant circle connecting the two fixed points to construct a well defined resonance zone  and applied HLD to those manifolds. Future work will investigate applying these techniques to numerical examples from the 3D quadratic family of maps. 


\bibliographystyle{elsarticle-num}
\bibliography{MyBibDeskBib.bib}

\end{document}